%% file: thermalizer.tex
\documentclass[%
 reprint,
 nofootinbib,
 amsmath,amssymb,
 aps,
 prx,
]{revtex4-2}

\usepackage[%
    margin=1.8cm,%
    columnsep=0.6cm,%
    marginparwidth=1.7cm,%
    marginparsep=0.1cm%
]{geometry}

\usepackage{mathtools}
\usepackage{amsthm,amsfonts}
\usepackage{bm}
\usepackage{bbm}
\usepackage{dsfont}
\usepackage{physics}
\usepackage{float}

\usepackage[dvipsnames]{xcolor}
\usepackage{graphicx}
\usepackage{subcaption}
\usepackage{array}
\usepackage{booktabs}
\usepackage{enumitem}
\usepackage{comment}
\usepackage[normalem]{ulem}
\usepackage[ruled,linesnumbered]{algorithm2e}
\usepackage[font=small]{caption}

\usepackage{tikz}
\usetikzlibrary{
    arrows.meta,
    positioning,
    calc,
    decorations.pathmorphing,
    decorations.pathreplacing,
    fit,
    backgrounds,
    shapes.geometric,
    patterns,
    matrix
}
\usepackage{pgfplots}
\pgfplotsset{compat=1.18}

\usepackage[autolanguage]{numprint}
\usepackage[
    draft,
    todonotes={textsize=tiny},
    commandnameprefix=always,
]{changes}

\usepackage[%
    colorlinks=true,
    citecolor=teal,
    linkcolor=purple,
    urlcolor=brown,
    linktoc=page
]{hyperref}

\newcommand{\DefCell}[1]{\parbox[t]{0.65\textwidth}{\raggedright #1}}

\newlength{\cellout}
\newlength{\cellin}
\definecolor{cInput}{HTML}{4A90D9}    
\definecolor{cHidden}{HTML}{E8A838}   
\definecolor{cOutput}{HTML}{6BBF6B}   
\definecolor{cForward}{HTML}{2E86C1}
\definecolor{cBackward}{HTML}{C0392B}
\definecolor{cGate}{HTML}{8E44AD}     
\definecolor{cGibbs}{HTML}{27AE60}
\definecolor{cDTM}{HTML}{E67E22}
\definecolor{cAccent}{HTML}{D4A017}
\definecolor{cCond}{HTML}{D63384}     
\definecolor{cGray}{HTML}{7F8C8D}

\definecolor{exSolar}{HTML}{FFF1C8}
\definecolor{exGold}{HTML}{F5D64C}
\definecolor{exYellow}{HTML}{FFB400}
\definecolor{exOrange}{HTML}{FF8400}
\definecolor{exCopper}{HTML}{903001}
\definecolor{exFuchsia}{HTML}{4E012A}
\definecolor{exBrown}{HTML}{1C0101}

\colorlet{cExSolar}{exSolar}
\colorlet{cExGold}{exGold}
\colorlet{cExYellow}{exYellow}
\colorlet{cExOrange}{exOrange}
\colorlet{cExCopper}{exCopper}
\colorlet{cExFuchsia}{exFuchsia}
\colorlet{cExBrown}{exBrown}

\colorlet{extsolar}{exSolar}
\colorlet{extgold}{exGold}
\colorlet{extyellow}{exYellow}
\colorlet{extorange}{exOrange}
\colorlet{extcopper}{exCopper}
\colorlet{extfuchsia}{exFuchsia}
\colorlet{extbrown}{exBrown}

\definecolor{panelBg}{HTML}{FBFAF6}
\definecolor{panelLn}{HTML}{DAD3C5}
\definecolor{boxBg}{HTML}{FFFFFF}
\definecolor{boxLn}{HTML}{D8D2C6}
\definecolor{latLn}{HTML}{C6BEAE}
\definecolor{chipBg}{HTML}{ECE8E1}
\definecolor{chipLn}{HTML}{C7C0B0}
\definecolor{pinLn}{HTML}{B7B0A0}

\colorlet{idealcol}{exCopper}        
\colorlet{hidcol}{exGold}            
\colorlet{implcol}{exOrange}         
\colorlet{hypcol}{exFuchsia}         
\colorlet{naivecol}{exBrown}         
\definecolor{floorcol}{HTML}{8C8079} 
\definecolor{wirecol}{HTML}{B8AEA2}  

\definecolor{vir0}{HTML}{4E012A}
\definecolor{vir1}{HTML}{71131C}
\definecolor{vir2}{HTML}{903001}
\definecolor{vir3}{HTML}{B5440A}
\definecolor{vir4}{HTML}{DA5F00}
\definecolor{vir5}{HTML}{FF8400}
\definecolor{vir6}{HTML}{FF9E00}
\definecolor{vir7}{HTML}{FBBE2E}
\definecolor{vir8}{HTML}{F2D24A}

\pgfplotscreateplotcyclelist{viridis9}{%
  {vir0,mark=*},
  {vir1,mark=*},
  {vir2,mark=*},
  {vir3,mark=*},
  {vir4,mark=*},
  {vir5,mark=*},
  {vir6,mark=*},
  {vir7,mark=*},
  {vir8,mark=*}
}

\colorlet{DarkBlue}{blue!60!green}

\newcommand{\vp}{\varphi}
\newcommand{\Pf}{\tilde{P}^F}
\newcommand{\Pb}{\tilde{P}^B}
\newcommand{\Zf}{Z^F}

\newcommand{\Pmod}{\tilde{P}}
\newcommand{\LVC}{\mathcal{L}^{\mathrm{VC}}}
\newcommand{\LRF}{\mathcal{L}^{\mathrm{RF}}}
\newcommand{\KL}{D_{\mathrm{KL}}}
\newcommand{\EE}{\mathbb{E}}
\newcommand{\softplus}{\mathrm{sp}}
\DeclareMathOperator{\arctanh}{arctanh}
\newcommand{\tk}{\psi}

\newcommand{\Prob}{\mathbf{P}}

\newcommand{\cP}{\mathcal{P}}

\newcommand{\TV}{\ensuremath{\mathrm{TV}}}

\newcommand{\Esamp}{E_{\mathrm{samp}}}

\newcommand{\Eread}{E_{\mathrm{read}}}
\newcommand{\Enb}{E_{\mathrm{nb}}}
\newcommand{\Eclk}{E_{\mathrm{clk}}}

\newcommand{\Erng}{E_{\mathrm{rng}}}
\newcommand{\Eanalog}{E_{\mathrm{ana,supp}}}
\newcommand{\Esram}{E_{\mathrm{SRAM}}}
\newcommand{\Ewrite}{E_{\mathrm{write}}}
\newcommand{\Emac}{E_{\mathrm{MAC}}}

\newcommand{\pL}{p_{\mathrm{L}}}
\newcommand{\Wtri}{W^{(3)}}

\newcommand{\hammcube}{\{-1, 1\}}     
\newcommand{\vs}{\mathbf{s}}

\newcommand{\ind}[1]{\mathbf{1}\!\left\{#1\right\}}
\newcommand{\Bin}{\operatorname{Bin}}

\begin{document}

\title{Thermalizing Stochastic Programs}

\author{Mirko Amico}
\affiliation{Extropic Corporation, San Francisco, California 94111, USA}

\author{Andraž Jelinčič}
\affiliation{Extropic Corporation, San Francisco, California 94111, USA}

\author{Colin Oscar Nancarrow}
\affiliation{Extropic Corporation, San Francisco, California 94111, USA}

\author{Leo Tyrpak}
\affiliation{Extropic Corporation, San Francisco, California 94111, USA}

\author{David Roberts}
\affiliation{Extropic Corporation, San Francisco, California 94111, USA}

\author{Seth Morton}
\affiliation{Extropic Corporation, San Francisco, California 94111, USA}

\author{Dalton Sakthivadivel}
\altaffiliation{CUNY Graduate Center, New York, NY, USA}
\affiliation{Extropic Corporation, San Francisco, California 94111, USA}

\author{Ashwin Gopal}
\affiliation{Extropic Corporation, San Francisco, California 94111, USA}

\author{Guillaume Verdon}
\affiliation{Extropic Corporation, San Francisco, California 94111, USA}

\date{\today}

\begin{abstract}
We present a set of tools for mapping general stochastic programs to thermodynamic hardware designed for energy-efficient stochastic sampling. Given a target stochastic program expressed as a Directed Factor Graph (DFG) of stochastic channels, or equivalently as a Parametrized Stochastic Circuit (PSC), we first introduce a method to approximately compile each factor in the DFG to an Energy-Based Model (EBM) that is native to the hardware. We then analyze how the error of the compiled DFG accumulates from the per-factor errors, and introduce two training refinements, context matching and trajectory-level REINFORCE post-training, which can reduce the residual error left by training each factor in isolation. The \texttt{thermalizers} framework takes a stochastic program expressed in the \texttt{torx} library and replaces its factors with thermodynamic kernels implemented and sampled using the \texttt{thrml} library. We demonstrate it on several example applications, including a market simulator that learns the joint day-to-day dynamics of a panel of financial time series from recorded market history alone, a probabilistic model from mathematical ecology, Gibbs sampling of an EBM the hardware cannot natively express, and a sequential Bayesian design loop over a Gaussian stochastic circuit.
\end{abstract}

\maketitle

\tableofcontents

    \section{Introduction}
\label{sec:intro}

Generative artificial intelligence has found application across a broad range of industries and increasingly supplies expert-level capabilities on-demand. The continued development of artificial intelligence systems with growing capabilities has come with increased energy requirements over the past several years \cite{strubell2019energy,devries2023growing,iea2025energy}, following the discovery of neural scaling laws and the rapid expansion in system sizes and flop throughput of digital hardware \cite{kaplan2020scaling,hoffmann2022chinchilla,sevilla2022compute}. This expansion is already responsible for the revitalization of previously stagnant electricity demand in advanced economies~\cite{iea2025energy}. Today, the bulk of AI training and inference is carried out in a relatively small number of data centers, the largest of which draw power comparable to that of a large city. Realizing artificial intelligence as a general-purpose technology, deployed pervasively rather than consumed as a centrally metered service, requires reducing the energy cost of AI algorithms by orders of magnitude.

Physics-based or `analog' devices, such as photonic \cite{shen2017deep,bandyopadhyay2024single}, stochastic \cite{debashis2020hardware,lockwood2026torx} and neuromorphic \cite{merolla2014truenorth,davies2018loihi,markovic2020physics} chips fall under a broad umbrella of post-digital computing paradigms that show great promise in this direction. Among these, probabilistic computers built from probabilistic-bit (p-bit) arrays \cite{camsari2017,chowdhury2023fullstack} and continuous-variable thermodynamic computers \cite{lockwood2026blueprint, conte2019thermodynamic,coles2023thermodynamic,aifer2024thermodynamic,melanson2025thermodynamic} are the closest relatives of the thermodynamic computers we consider here, discrete stochastic devices that relax to the equilibrium distribution of a programmable energy function through their physical dynamics. Simulations of denoising-based generative algorithms running on such devices \cite{jelincic2025dtm} suggest that non-trivial inference tasks can be executed with as little as $10^{-4}$ or even $10^{-6}$ of the energy required by GPUs.

A general method for executing stochastic programs on this class of hardware has been missing until now. Existing programming models for such devices usually restrict attention to problems the hardware expresses natively. Ising machines minimize quadratic costs programmed directly into their couplings \cite{mohseni2022ising}, and analog Langevin devices perform linear algebra through relaxation to a Gaussian stationary distribution, such that the solution of a linear system is read off as the stationary mean and the matrix inverse as the stationary covariance \cite{aifer2024thermodynamic}. A differentiable programming framework over probabilistic modes and mixtures thereof broadens the continuous-variable repertoire considerably, up to and including deep neural network inference \cite{lockwood2026blueprint}. Invertible logic encodes a Boolean relation in the ground states of a synthesized energy function \cite{camsari2017}. However, these constructions are either tied to a single target fixed in advance or confined to continuous-variable equilibrium dynamics, and none of them extends to the execution of a general stochastic program.

An equilibrium distribution of an energy function is an undirected object with no notion of input or output. A \emph{stochastic program}\footnote{We define here an object distinct from what appears in the unrelated field of \emph{stochastic programming} in mathematical optimization.}, on the other hand, is a directed composition of conditional distributions, each of which maps a set of input variables to a set of output variables. Any general execution method must therefore translate between the two representations, and the errors introduced by this translation can accumulate over the course of the program. The perspective we adopt is that the energy-based model underlying the hardware supplies the instruction set of a probabilistic computer, whose single primitive is the Gibbs update of its p-bits under a programmable energy.

In this manuscript, we show how to execute programs written in the framework of Ref.~\cite{lockwood2026torx}, which describes a stochastic program as a directed composition of local stochastic operations, or factors, on thermodynamic hardware. When a subset of the variables of a thermodynamic computer is held fixed, the device samples from the conditional distribution of the remaining variables, so that a single choice of energy function realizes a probabilistic transition kernel from the clamped variables to the free variables. We call this object a thermodynamic kernel. The thermodynamic kernel is the basic instruction from which our programs are assembled. By tuning the thermodynamic kernel's parameters appropriately, we can sample from a broad class of probabilistic transition kernels even when the device is restricted in the distributions it can natively support. Our method, which we call variational compilation, adjusts the energy function of each kernel to minimize the relative entropy between the kernel and the local operation it must reproduce, one factor of the program at a time.

We detail several methods to obtain the kernel parameters, analyze how the compilation errors of individual kernels compound or dissipate through the composed program, and describe post-training procedures, at the level of full trajectories, that reduce the residual error of the executed program. We demonstrate the framework by training EBMs to approximate the local rules of a biased random walk, a birth-death process, the day-to-day dynamics of a panel of financial time series learned from recorded market history, a Gibbs sweep for an EBM with interactions not native to the hardware, and a Gaussian stochastic circuit. All demonstrations are executed in software, based on the device model of Ref.~\cite{jelincic2025dtm}.

The manuscript is organized as follows. Section \ref{sec:background} introduces the hardware platform, thermodynamic hypergraphical models, which serve as the compilation substrate, and directed factor graphs, which we use to describe generic stochastic programs. Section \ref{sec:training} develops the training framework, from the variational compilation of a single factor to the choice of input distribution, pre-compiled kernel libraries, trajectory-level REINFORCE post-training of the composed program, organized around an error-propagation bound that connects per-factor, trajectory, and readout errors. Section \ref{sec:demos} demonstrates the framework on the examples listed above. Section \ref{sec:discuss} closes with a discussion of the framework and open questions. The appendices collect the derivations and implementation details omitted from the main text.

\begin{figure}[tbp]
    \centering
    \includegraphics[width=\columnwidth]{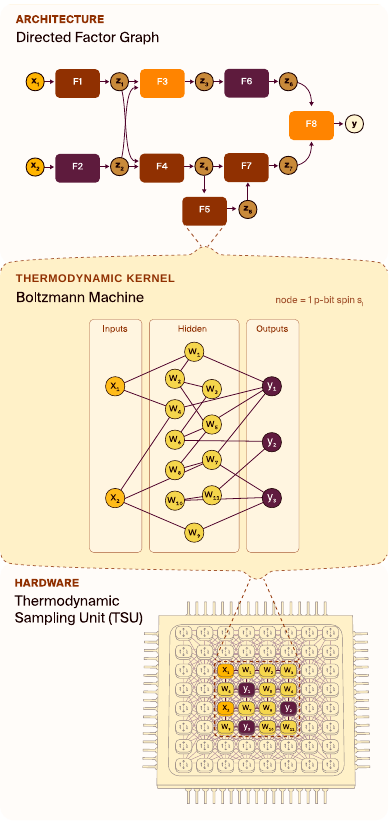}
    \caption{The framework operates on three levels. \emph{(top)} At the highest level of abstraction, the stochastic program is a directed factor graph of Markov kernels, or equivalently a parametrized stochastic circuit \cite{lockwood2026torx}, passing wire variables $z_\ell$ from the program's inputs to its output. \emph{(middle)} Each factor of the DFG is compiled to a thermodynamic kernel, a conditional distribution obtained from an EBM by clamping the input spins $x$ (orange), marginalizing hidden spins $w$ (yellow) and normalizing over the output spins $y$ (maroon). \emph{(bottom)} The compiled kernel is laid out spatially on a region of the thermodynamic hardware, a Thermodynamic Sampling Unit (TSU), whose physical dynamics samples it natively.}
    \label{fig:overview}
\end{figure}

\section{Background and preliminaries}
\label{sec:background}

In this Section we define the objects used in the rest of the manuscript. Thermodynamic Hypergraphical Models (THMs) are defined in \S\ref{sec:ebm}. THMs are a special class of energy-based models and provide an abstract representation of possible hardware substrates. The hardware platform that we compile to in the examples of \S\ref{sec:demos} is described in \S\ref{sec:hardware}. To represent information processing from inputs to outputs, we use Directed Acyclic Graphs (DAG) of stochastic factors.
In the literature \cite{frey2003unifying}, DAGs of stochastic factors are known as directed factor graphs; we define them in \S\ref{sec:dfg}.
An equivalent framework for representing stochastic programs is the one of parametrized stochastic circuits. These are equivalent to DAGs of stochastic factors and described briefly in \S\ref{sec:psc}. More details on the PSC framework are given in the companion paper~\cite{lockwood2026torx}.  
Most models in this manuscript will be structured as a DFG at the top level, where each factor inside the DFG is a stochastic kernel, which can then be mapped to thermodynamic hardware via a THM. Figure~\ref{fig:overview} gives a visual overview of the framework and how the different objects presented in this Section interact with each other.

\subsection{Thermodynamic hypergraphical models}\label{sec:ebm}

An \emph{energy-based model} is a family of probability distributions of the form
\begin{equation}\label{EQN:boltzmann_dist}
p_\vp(s)
=
\frac{1}{Z_\vp}e^{-E_\vp(s)},
\end{equation}
referred to as the Boltzmann distribution, where
\[
Z_\vp
=
\sum_{x\in\mathcal X}e^{-E_\vp(x)}
\]
is a normalizing constant called the partition function. A \emph{Thermodynamic Hypergraph Model} is an energy-based model defined on a finite hypergraph \(G=(V,\mathcal E)\),
\begin{equation}
E_{\vp}(s)
=
-\sum_{e\in \mathcal E}J_e\prod_{v_i\in e}s_i
\qquad
s\in\{-1,1\}^{n},
\end{equation}
where $\vp=\{J_e\}_{e\in \mathcal E}$ and $n = |V|$. The \emph{order} of the model is $\max_{e\in \mathcal E}|e|$.
Note that in the machine-learning literature Boltzmann machines usually take values in $\{0,1\}$; one translates to the bipolar setting above with the mapping \(s_i\mapsto2s_i-1\).

We will sample from this Boltzmann distribution using Gibbs sampling~\cite{glauber1963,geman1984,gonzalez2011parallel}, wherein we define a Markov chain that resamples spins according to
\begin{equation}\label{eq:gibbs-cond}
    p_\vp(s_i=1\mid s_{\setminus i})=\sigma\left(2\sum_{e \ni i}J_e\prod_{j\in e\setminus\{i\}}s_j\right),
\end{equation}
where \(s_{\setminus i}\) denotes the state of all spins other than \(s_i\). The equilibrium distribution of this chain can be shown to yield Eq.~\eqref{EQN:boltzmann_dist}.
The \texttt{thrml} software library \cite{thrml} allows for seamless definition and sampling of THMs.

A THM of order \(2\) is also called a Boltzmann machine (BM),
\begin{equation}\label{eq:ising}
    E_\vp(s_1,\ldots,s_n)=-\sum_{\{i,j\}\in \mathcal E}J_{i,j}s_is_j-\sum_ih_is_i.
\end{equation}
We emphasize this as most of our examples only use pairwise interactions, yet we keep the framework general in order to facilitate future developments. A chromatic partition of \(V\) assigns spins to blocks in which no two vertices share a hyperedge ~\cite{gonzalez2011parallel}. The spins of a block are conditionally independent given the rest of the configuration, so a block update, in which each spin of the block draws from its neighborhood conditional~\cite{glauber1963} resamples the whole block at once. We call a step of resampling all spins from a block, or just one spin, a Gibbs update.

In order for a THM to represent a Markov kernel, we impose a certain structure on it. Specifically, we partition the graph's vertices \(V=V_{\text{in}}\cup V_{\text{hidden}}\cup V_{\text{out}}\).
We call the input nodes \(x\in V_{\text{in}}\), the hidden nodes \(w \in V_{\text{hidden}}\) and the output nodes \(y \in V_{\text{out}}\) so the full state is \(s=(x,w,y)\in \{-1,1\}^n\) with \(n = n_\mathrm{in} + n_h + n_\mathrm{out}\). Note that unlike Restricted Boltzmann Machines (RBMs) where hidden nodes represent one color block and visible nodes represent the other color block, we make no such assumptions. The input-hidden-output partition need not relate to the graph coloring used for block-Gibbs sampling.
During inference, we are given input \(x\) and would like to sample \(y\) (or sometimes the opposite), the hidden nodes \(w\) being used only to enrich the hypothesis class of our models.

We can write the joint distribution over all node types as
\begin{align*}
    p_\vp(x,w,y)=\frac{1}{Z_\vp}e^{-E_\vp(x,w,y)}.
\end{align*}
We get a joint distribution over \((x,y)\) by marginalising over hidden nodes,
\begin{align*}
    p_\vp(x,y)=\sum_w p_\vp(x,w,y).
\end{align*}
On hardware this corresponds to discarding the values of the hidden nodes when reading out the final node configuration. Further marginalizations over \(x\) and \(y\) yield the distributions \(p_\vp(y)\) and \(p_\vp(x)\), respectively.

Concretely, we would like to sample from the forward distribution
\begin{align}
    \Pf(y \mid x;\vp) &= p_\vp(y \mid x) = \frac{p_\vp(x, y)}{p_\vp(x)} \notag
    \\&=\frac{1}{Z^F(x; \vp)}\sum_we^{-E_\vp(x,w,y)},
\end{align}
where \(Z^F(x; \vp)=\sum_{w,y}e^{-E_\vp(x,w,y)}\) is the normalizing constant for a fixed \(x\).
To sample \(y\) conditioned on an input state \(x\), we first clamp the input nodes \(x\), then run Gibbs updates on \((w,y)\) and then read \(y\).

We will also use the backward distribution,
\begin{align}\label{eq:backward}
    \Pb(x \mid y;\vp) &= p_\vp(x \mid y) = \frac{p_\vp(x, y)}{p_\vp(y)}.
\end{align}
The backward distribution can be sampled analogously, with the roles of \(x\) and \(y\) interchanged.
We suppress the parameter subscript \(\vp\) when it is not needed, writing \(\Pf(\cdot\mid x)\).
It is the forward \(\Pf\) that is optimized at compile time to match a target conditional, a procedure we expand on in \S\ref{sec:compile}. The backward $\Pb$ is the Bayes inverse of $\Pf$ against the implicit prior $p_\vp(x) \propto \Zf(x)$ that the construction imposes on $x$.

The forward $\Pf$ does not fully determine the backward $\Pb$. For any real-valued function
$\lambda$ of the input alone, the shifted energy
\begin{equation}\label{eq:gauge_energy}
  E_{\vp,\lambda}(x,w,y) := E_\vp(x,w,y) - \lambda(x)
\end{equation}
defines conditionals $\Pf_\lambda$ and $\Pb_\lambda$ through the same construction. Because $e^{\lambda(x)}$ depends only on $x$, it factors out of the sums over $w$ and $y$ in the forward normalization and cancels,
\begin{equation}\label{eq:gauge_fwd}
  \Pf_\lambda(y \mid x) \;=\; \Pf(y \mid x).
\end{equation}
The forward is invariant under this shift. The backward is not,
\begin{equation}\label{eq:gauge_bwd}
  \Pb_\lambda(x \mid y) \;\propto\; e^{\lambda(x)}\,\sum_w e^{-E_\vp(x,w,y)},
\end{equation}
which corresponds to reweighting the implicit prior to $p_{\vp,\lambda}(x) \propto e^{\lambda(x)}\,\Zf(x)$. Forward compilation thus determines an equivalence class of energies related by shifts supported on the input. We can fix a representative of this equivalence class by choosing the function $\lambda$, which leaves $\Pf$ unchanged. This is just a convention when considering a single conditional but it starts to matter once forward-compiled conditionals are tiled together. Appendix~\ref{sec:backward-compilation} goes into more details on how the backward can be compiled.

\subsection{Hardware platform}
\label{sec:hardware}

Extremely energy efficient thermodynamic computing architectures can be assembled using electronic circuits of subthreshold CMOS transistors whose per-site effective conductance and temperature are voltage-programmable~\cite{freitas2026neatrn}. An example implementation of this concept, which we refer to as Z1, was described in Ref.~\cite{jelincic2025dtm}. This hardware would support pairwise Ising~\cite{mohseni2022ising} energies $E(s) = -\frac{1}{2}s^\top J s - h^\top s$ on a fixed planar topology. The Z1 topology is a sparse, locally connected, 2-colorable graph, where each node supports two different couplings per edge (one per direction; see below for details). In particular, we have a set of nodes $V$ (of size $n = |V|$) connected with a set of edges $\mathcal E$. The graph is mostly regular with degree~16, and is locally connected: the nodes are laid out on a 2D plane and connected only to nearby nodes (for more details see \S\ref{subsec:graph_arch}). Since the graph is 2-colorable, there exist sets $V_1$ and $V_2$ such that $V_1 \cup V_2 = V$, $V_1 \cap V_2 = \varnothing$, and for all $\{u,v\} \in \mathcal E$ we have $|\{u,v\} \cap V_1| = |\{u,v\} \cap V_2| = 1$. For each $v \in V$ we define the set of its neighbors as $N(v) \coloneqq \big\{ u \in V \,|\, \{u,v\} \in \mathcal E \big\}$. Each node $v$ has a bias $h_v$ and couplings $\mathbf{J}_v \coloneqq \big(J_v^u \big)_{u \in N(v)}$, and at any given time has a state $s_v \in \hammcube$. We write $\vs \coloneqq \big( s_v \big)_{v \in V}$, $\mathbf{J} \coloneqq \big(\mathbf{J}_v \big)_{v\in V}$, $\mathbf{h} = \big(h_v \big)_{v\in V}$, and $\vp \coloneqq (\mathbf{h}, \mathbf{J})$. In practice, these nodes are an array of p-bits~\cite{camsari2017,camsari2019pbits,borders2019factorization,aadit2022sparse,jelincic2025dtm,kaiser2021probabilistic,chowdhury2023fullstack,patel2024pass} initialized in some initial configuration and interact according to a programmed set of couplings and biases.

We define the update probability of a node $v \in V$ as
\begin{equation}
\label{eq:node_update_dist}
    p_v(\vs) \coloneqq \sigma \Big( 2 \, h_v + 2 \sum_{u \in N(v)} s_u J_v^u \Big),
\end{equation}
the analog of Eq.\eqref{eq:gibbs-cond} for this setting. The chip performs block Gibbs sampling~\cite{glauber1963,geman1984,gonzalez2011parallel} by first resampling every node $v \in V_1$, setting $s_v = 1$ with probability $p_v(\vs)$ and $s_v = -1$ otherwise. Since this depends only on the states of the nodes in $V_2$, all nodes in $V_1$ can be updated in parallel. The same is then done for $V_2$. These two block updates constitute one Gibbs iteration. The Z1 chip will perform Gibbs sampling extremely rapidly and cheaply: the entire chip has $\sim$250,000 nodes, each Gibbs iteration is estimated to cost approximately $3 \times 10^{-10}\,\mathrm{J}$, and the chip will perform between $10^6$ and $10^7$ Gibbs iterations per second.

Note that $J_v^u$ need not equal $J_u^v$. When they are equal for every edge, the model is a Boltzmann machine. In that case, writing $J_{uv} \coloneqq J_u^v = J_v^u$ for each edge, we can define an energy function analogous to Eq.~\eqref{eq:ising}
\begin{equation}
\label{eq:bm_energy}
    E(\vs; \vp) \coloneqq - \sum_{v \in V} s_v h_v \, - \, \sum_{\{u,v\} \in \mathcal E} s_v s_u J_{uv}
\end{equation}
such that the Gibbs sampling Markov chain converges to the stationary distribution
\begin{equation}
\label{eq:bm_stat_dist}
    p_\vp(\vs) = \frac{1}{Z_{\vp}} e^{-E(\vs; \vp)} \quad \text{where} \; Z_{\vp} = \sum_{\vs' \in  \hammcube^n} e^{ -E(\vs'; \vp)}.
\end{equation}

BMs can be trained using Contrastive Divergence (CD)~\cite{hinton2002}, a sampling-based algorithm that requires very little computation beyond the sampling itself, making it particularly well suited to our hardware. In each training step, CD compares two sets of statistics: a \emph{positive phase} that lowers the energy of states consistent with the training data, and a \emph{negative phase} that raises the energy of states the model currently favors.

If some edges are asymmetric (i.e.,\ $J_v^u \neq J_u^v$), the Gibbs sampling dynamics no longer obeys detailed balance. The Markov chain still has a stationary distribution (provided all couplings are finite), but this distribution generally does not admit a closed-form expression, and it is unclear how to train such a model. A viable learning rule for the asymmetric case could be very impactful, but we consider this unlikely given that the problem appears intractable. For this reason, the rest of this document focuses on the symmetric case, i.e.,\ on Boltzmann machines.

\subsubsection{Architecture}
\label{subsec:graph_arch}
\begin{figure*}[ht]
    \centering
    \includegraphics[width=\textwidth]{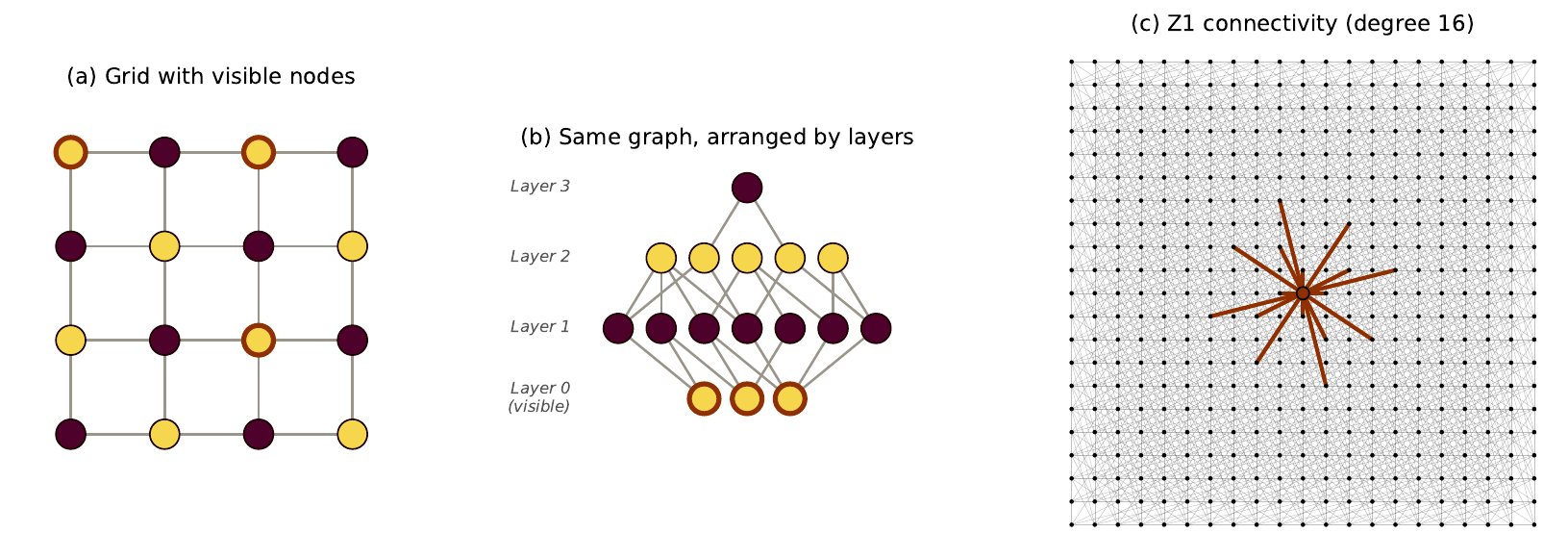}
    \caption{\textbf{Z1 architecture.} \textbf{(a)}~A small nearest-neighbor grid with randomly chosen visible nodes (red outlines). Nodes are colored in a bipartite pattern; all nodes of one color can be updated simultaneously during Gibbs sampling. \textbf{(b)}~The same graph rearranged into layers based on graph distance from the nearest visible node, revealing an emergent deep Boltzmann machine structure. \textbf{(c)}~Z1 connectivity on a $21 \times 21$ grid: all edges from a single highlighted node are shown in red.}
    \label{fig:z1_arch}
\end{figure*}

Nodes in the Z1 chip are arranged on a 2D grid (each node has two integer-valued coordinates). A node at position $(x, y)$ with connection rule $(a, b)$ is connected to the four nodes at $(x+a, y+b)$, $(x-b, y+a)$, $(x-a, y-b)$, $(x+b, y-a)$. All nodes in Z1 have connection rules $(1, 0)$, $(2, 1)$, $(2, 3)$, $(4, 1)$, giving degree~16 (except at grid boundaries where some edges fall outside the grid). All edges are short, with the longest having Euclidean length $\sqrt{17} \approx 4.1$ grid units. A visual representation of the device connectivity for a $21 \times 21$ grid is shown in Figure~\ref{fig:z1_arch}, panel (c).

When using a grid of $L^2$ nodes as a BM for $d$-dimensional data, we randomly designate $d$ of the $L^2$ nodes as visible (one per data dimension); the remaining $L^2 - d$ nodes become hidden (latent variables). The hidden nodes' effective depth is determined by their graph distance to the nearest visible node, giving rise to an emergent layer structure akin to a deep Boltzmann machine (Fig.~\ref{fig:z1_arch} panel (b)).

\subsubsection{Operation}
In addition to performing Gibbs sampling (with fixed couplings, biases and temperature), the chip supports other operations:
\begin{enumerate}
    \item \emph{Readout:} reading the current state of some nodes and sending them to the host. This costs about as much energy as $10^2$--$10^3$ Gibbs iterations.
    \item \emph{Coupling flashing:} receiving a new set of couplings and biases from the host. This is significantly more expensive than readout.
    \item \emph{Clamping:} setting some nodes into ``clamped'' mode so they are not updated during Gibbs sampling. About as expensive as coupling flashing.
\end{enumerate}

As long as coupling re-flashing is infrequent (no more than about once per second), the chip can sample from Boltzmann distributions with extraordinary speed and energy efficiency. A more detailed breakdown of the energy cost of the hardware operations is given in Appendix~\ref{sec:z1-energy}.

\subsection{Directed factor graphs}
\label{sec:dfg}

Having reviewed how an EBM/THM can be used as a stochastic kernel in \S\ref{sec:ebm}, we can now construct a directed acyclic graph of stochastic kernels, also known as a DFG~\cite{frey2003unifying, lockwood2026torx}.

A DFG is a directed acyclic graph where each node holds a stochastic kernel and each directed edge carries some random variable sampled as an output of one stochastic kernel to be then used as input to a subsequent stochastic kernel. This structure is similar to that of directed graphical models~\cite{bishop2006pattern}.

A DFG hosts kernels $P_1, \ldots, P_L$. Each kernel $P_\ell$ is a function $P_\ell(y \mid x; \phi_\ell)$ of an input $x$ and an output $y$, parametrized by \(\phi_\ell \in \Phi_\ell \subseteq \mathbb{R}^{d_\ell}\). A kernel is a conditional distribution over its outputs, implying
\begin{align}
    P_\ell(y \mid x; \phi_\ell)\geq0,\quad\sum_y P_\ell(y \mid x; \phi_\ell)=1.
\end{align}
We shall sometimes omit the dependence on \(\phi_\ell\) when convenient. We call \(P_j\) a \emph{parent} of \(P_\ell\) if there is a directed edge from \(P_j\) to \(P_\ell\), i.e.,\ the output of \(P_j\) serves as an input to \(P_\ell\), and we write $\mathrm{pa}(\ell)$ for the set of $P_\ell$'s parents.
We assume a \emph{topological ordering}: all parents appear before their children, so if \(P_j\) is a parent of \(P_\ell\), then \(j<\ell\).
Writing $z_\ell$ for the random variable on the wire leaving kernel $\ell$ (with $z_0$ denoting the DFG's global input, carried by edges with dangling sources) and $z_{\mathrm{pa}(\ell)} \coloneqq (z_j)_{j \in \mathrm{pa}(\ell)}$ for the tuple of kernel $\ell$'s parent wires, the full DFG defines a joint distribution over all of its wires by ancestral sampling along the DAG:
\[
    P^\mathrm{DFG}_\phi \big( z_1, \ldots, z_L \mid z_0 \big) \coloneqq \prod_{\ell=1}^L P_\ell \big( z_\ell \mid z_{\mathrm{pa}(\ell)}; \phi_\ell \big).
\]
We call the tuple $z_{0:L} = (z_0, \ldots, z_L)$ of all wire variables a \emph{trajectory} of the DFG. When stating losses and gradient identities it is convenient to treat the DAG as a chain, $\mathrm{pa}(\ell) = \{\ell - 1\}$, which loses no generality: relabel the kernels topologically and let each wire carry forward, unchanged, every earlier variable that a later kernel still needs, so that each kernel reads only the wire before it. Implementations keep the original DAG; the chain form is purely a notational device, and we use it throughout \S\ref{sec:training}.

The DFG could also have designated outputs, which could be any subset of the DFG's variables \( z_\mathrm{out} \coloneqq \big( z_s \big)_{s \in S}\) for some \(S \subset \{1, \ldots, L \}\). This then means that the DFG itself becomes a kernel if we marginalize out all \(z_\ell\) for \(\ell \not\in S\):
\[
P^\mathrm{out} (z_\mathrm{out} \mid z_0 ; \phi) = \sum_{( z_\ell )_{\ell \not\in S}} P^\mathrm{DFG}_\phi \big( z_1, \ldots, z_L \mid z_0 \big).
\]
More generally, we call any map $R$ from trajectories to some observable a \emph{readout} of the DFG, and write $R_\# P$ for the distribution of $R(z_{0:L})$ when $z_{0:L} \sim P$; reading out a designated output is the special case $R(z_{0:L}) = z_\mathrm{out}$.
Conversely, any kernel can also be written as a DFG of several factors. Therefore, each DFG has infinitely many equivalent DFGs. The exact choice of DFG representation we use to represent some stochastic program is therefore largely dependent on what we want to convey and what kinds of kernels we consider as our basic building blocks. As we will see in the next section, the granularity we choose for our DFG (do we use a few complicated factors or many simple ones) can greatly influence how and how well the DFG gets variationally compiled.

While we say that a kernel abstractly represents a conditional distribution \( P_\ell(y \mid x; \phi_\ell) \), we do not always assume that we have access to a closed form expression for this conditional distribution. In some cases we might only be able to efficiently sample from the conditional distribution, or we might only have access to samples from the joint trajectory distribution \( ( z_1, \ldots, z_L ) \sim P^\mathrm{DFG}_\phi \). However, as we will see in the next section, the type of information we have about the kernels being compiled determines the type of compilation method we can use for them.

\subsection{Parametrized stochastic circuits}
\label{sec:psc}
\begin{figure}[tbp]
    \centering
    \includegraphics[width=1.1\columnwidth]{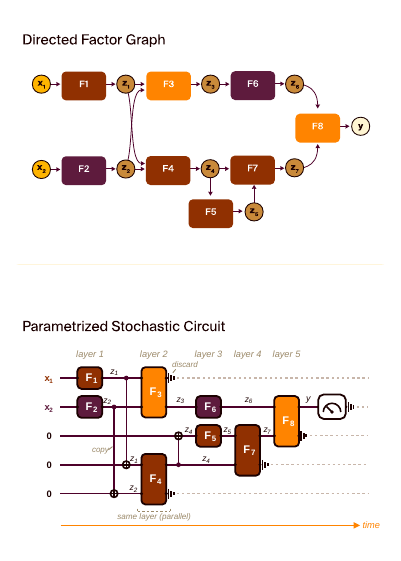}
    \caption{Example of the equivalence between the directed factor graph view and the parametrized stochastic circuits view of a given stochastic program. PSCs organize operations more explicitly than a DFG, collecting them in layers that run simultaneously, which makes PSCs more natural when considering how a program is executed on hardware. On the other hand, DFGs hide some of these details away in favor of a clearer view of the relationship between operations. This resembles algorithmic reasoning more closely. Despite their graphical differences, the two frameworks are equivalent.}
    \label{fig:dfg_psc_equivalence}
\end{figure}

An architecture-level companion object to the DFG is a \emph{Parametrized Stochastic Circuit}, introduced by the \texttt{torx} framework~\cite{lockwood2026torx}.
A PSC is a DFG whose kernels are grouped into ordered layers of
disjoint-support gates for parallel execution. 
The two describe the same class of stochastic programs and serve the same goal of characterizing the set of algorithms that can be run on our hardware.
PSCs correspond more closely to how circuits are designed in quantum computers while DFGs are inspired by graphical models.
Different audiences might prefer a different way to visualize identical frameworks.
We show the equivalence in Figure~\ref{fig:dfg_psc_equivalence}.

PSCs are hardware-agnostic by design and the same logical circuit can compile to different backends: a software simulator, the XTR-0 test chip (see \cite{lockwood2026torx}), or Z1. 
Within a PSC, wires may carry a discrete or a continuous state and all stochasticity is localized to the gate kernels.
Kernels are built up by composing smaller kernels, with the atomic kernels called elementary gates.
Examples of such gates include \(PNOT(p)\) which acts on one bit and flips its state with probability \(p\).
Gates acting on different bits combine into a single layer by tensor product, for example in Fig.~\ref{fig:dfg_psc_equivalence} the first layer is \(F_1\otimes F_2\) acting on bits \(x_1,x_2\) independently.

\section{Training}
\label{sec:training}

Training turns a stochastic program into a hardware-executable form. Its input is a \emph{target} program: a DAG of kernels in the sense of \S\ref{sec:dfg}, whose kernels we label $P_\ell$, $\ell = 1, \ldots, L$ (in the chain form introduced there), together with a distribution $q_0$ for the program's global input $z_0$. From the target, we construct a \emph{model} program on the same DAG by replacing every kernel with a thermodynamic kernel $\Pf_\ell(\cdot \mid \cdot\,; \vp_\ell)$, a conditional of a THM that the hardware samples natively, and we optimize the parameters $\vp = (\vp_1, \ldots, \vp_L)$ so that executing the model program reproduces the behavior of the target (Fig.~\ref{fig:overview}).

More formally, the target DFG and the model DFG each define a probability distribution over trajectories $z_{0:L}$, respectively given by
\begin{align}
    \label{eq:target_traj}
    P(z_{0:L}) \;=\; q_0(z_0) \prod_{\ell=1}^{L} P_\ell(z_\ell \mid z_{\ell-1}) \quad \text{(target)}, \\[1mm]
    \Pmod_\vp(z_{0:L}) \;=\; q_0(z_0) \prod_{\ell=1}^{L} \Pf_\ell(z_\ell \mid z_{\ell-1}; \vp_\ell) \quad \text{(model)}.
    \label{eq:model_traj}
\end{align}
We are normally interested in minimizing $D\big( P \,\big\|\, \Pmod_\vp \big)$ for some distributional distance $D$. In the rest of this paper our distance of choice is usually the Kullback-Leibler (KL) divergence $\KL$ or the Total-Variation (TV) distance $\big\| \cdot \big\|_\TV$.

In some cases we might not care about the joint distribution of the entire trajectory, but might only be interested in the behavior of a ``readout" $Y \coloneqq R(z_{0:L})$ for some function $R$. Namely, if we define $R_\# P(y) \coloneqq P(R^{-1}(y))$ to be the pushforward of $P$ under $R$ (where $R^{-1}$ stands for the pre-image of $R$), then we say our goal is to minimize $D\big( R_\# P \,\big\|\, R_\# \Pmod_\vp \big)$. Even though it is often impossible to directly minimize $D\big( R_\# P \,\big\|\, R_\# \Pmod_\vp \big)$, we later provide justification why our proposed methods are expected to bring down this objective indirectly.

We write $q_\ell$ and $\tilde q_\ell$ for the marginal distributions of the wire variable $z_\ell$ under the two laws. Throughout, a tilde marks quantities of the compiled model, while plain letters are used for the target.

Training proceeds in two stages. \emph{Variational compilation} fits each kernel to its own target conditional, independently of all the others (\S\ref{sec:compile}). Its main degree of freedom is the distribution of the inputs each kernel trains under (\S\ref{sec:input_distributions}), and standard kernels can be compiled once and reused across programs (\S\ref{sec:libraries}). \emph{REINFORCE post-training} then trains all kernels jointly, against a single objective on the executed program (\S\ref{sec:reinforce}). Both stages can run with the hardware itself in the loop, which absorbs part of the execution error into the trained parameters.

How closely must each kernel match its own target for the deployed program as a whole to be accurate? The section is organized around the key Eq.~\eqref{eq:mitigation_chain} below, which answers this by relating three levels of error: the error of any readout of the program (\S\ref{sec:dfg}), the error of its whole trajectory distribution, and the errors of its individual kernels. Let
\begin{equation}\label{eq:J_def}
    J_\ell(x; \vp_\ell) \;\coloneqq\; \KL\big( P_\ell(\cdot \mid x) \,\big\|\, \Pf_\ell(\cdot \mid x; \vp_\ell) \big)
\end{equation}
measure the mismatch of kernel $\ell$ and its target at a single input $x$, and let $\varepsilon_\ell(\vp_\ell) \coloneqq \EE_{x \sim q_{\ell-1}}[\, J_\ell(x; \vp_\ell) \,]$ be its average over the inputs the target program produces. Finally, let $\mu_\ell$ denote the distribution of the inputs under which kernel $\ell$ is trained and let $c_\ell \coloneqq \max_x\, q_{\ell-1}(x) / \mu_\ell(x)$ be the largest factor by which the target's input marginal exceeds it. As derived in Appendix~\ref{sec:kernels-chains-appendix}, we obtain
\begin{widetext}
\begin{equation}\label{eq:mitigation_chain}
    \underbrace{\KL\big( R_\# P \,\big\|\, R_\# \Pmod_\vp \big)}_{\text{readout error}}
    \;\le\;
    \underbrace{\KL\big( P \,\big\|\, \Pmod_\vp \big)}_{\text{trajectory error}}
    \;=\;
    \sum_{\ell=1}^{L} \varepsilon_\ell(\vp_\ell)
    \;\le\;
    \sum_{\ell=1}^{L} c_\ell\, \EE_{x \sim \mu_\ell}\big[ J_\ell(x; \vp_\ell) \big].
\end{equation}
\end{widetext}
The middle equality is the chain rule of the KL divergence: under the true input marginals, $\mu_\ell = q_{\ell-1}$, the trajectory error is the sum of the per-kernel errors. This sum is the variational-compilation objective (\S\ref{sec:compile}). It splits over kernels, which is why we train each kernel independently, and it identifies the target input marginals as the canonical training distribution (\S\ref{sec:input_distributions}).

The right-most inequality bounds the per-kernel losses when a different measure is used for the input distribution (as opposed to $\mu_\ell = q_{\ell-1}$ in the middle). If the kernel is trained under a generic $\mu_\ell$, such as the uniform distribution, each per-kernel loss enters the bound inflated by the factor $c_\ell \ge 1$. Compiling under a generic input distribution therefore still controls the trajectory error, which makes it possible to compile standard kernels into reusable libraries before their target program is known (\S\ref{sec:libraries}), with $c_\ell$ quantifying the error incurred from compiling without access to information about the target.

On the left-most side is the data-processing inequality: full-trajectory training provably helps, even if the quantity we ultimately care about is the pushforward $R_\# \Pmod_\vp$ of our model under some readout $R$. However, this bound can be loose, because errors from different kernels can cancel in the readout law while remaining fully visible at the trajectory level. No per-kernel objective can exploit such cancellations and this is where trajectory-level REINFORCE post-training comes into play (\S\ref{sec:reinforce}).

Equation~\eqref{eq:mitigation_chain} shows that trajectory error grows at worst linearly in program depth. This cannot be improved in general, but it is often pessimistic. Applying the same kernel to two different input distributions can never increase the total-variation distance between them, and kernels that de-correlate inputs, such as the mixing dynamics of a sampler, shrink it at a geometric rate. For programs composed of such kernels, residuals injected early are washed out rather than accumulated, and the deployed error saturates at a depth-independent floor, the per-kernel residual amplified by the target's relaxation time~\cite{mitrophanov2005}. Appendix~\ref{app:inputs_model} makes both regimes precise, and the meta-EBM demonstration of \S\ref{sec:demos_meta} measures the floor against an exactly known target.

\subsection{Variational compilation}
\label{sec:compile}

\paragraph*{The thermodynamic kernel.}
As explained in \S\ref{sec:ebm}, clamping the input region of a THM and letting the remaining spins thermalize realizes a valid Markov kernel from inputs to outputs. We take this as our compilation primitive to execute target conditionals on thermodynamic hardware. Each kernel is realized by a region of the substrate hosting input spins $x$, hidden spins $w$, and output spins $y$, equipped with an energy $E_\vp(x, w, y)$. The only requirement on $E_\vp$ is that the substrate thermalizes $p_\vp(s) \propto e^{-E_\vp(s)}$ natively. It is convenient to collect the Boltzmann weight of the hidden spins into the \emph{affinity}
\begin{equation}\label{eq:kernel_def}
    \tk_\vp(x, y) \;\coloneqq\; \sum_{w} e^{-E_\vp(x, w, y)},
\end{equation}
the unnormalized joint distribution over inputs and outputs. The thermodynamic kernel is the Markov kernel obtained by conditioning over an input \(x\) and normalizing,
\begin{equation}\label{eq:forward}
    \Pf(y \mid x; \vp) = \frac{\tk_\vp(x, y)}{\Zf(x; \vp)}, \quad \Zf(x; \vp) \coloneqq \sum_{y'} \tk_\vp(x, y').
\end{equation}
It is sampled by clamping $x$, Gibbs-sampling $(w, y)$, and reading out $y$. The same programming also exposes the backward conditional \eqref{eq:backward} under the swapped clamping pattern. However, compiling the forward leaves the backward undetermined, since rescaling the affinity by any function of the inputs alone changes $\Pb$ while leaving $\Pf$ untouched. Any use of the backward direction therefore involves an extra choice to fix this degree of freedom, developed in Appendix~\ref{sec:backward-compilation}.

\paragraph*{The loss.}
Variational compilation minimizes the trajectory-level KL divergence of \eqref{eq:mitigation_chain},
\begin{equation}\label{eq:compile_loss}
    \LVC(\vp) \;\coloneqq\; \KL\big( P \,\big\|\, \Pmod_\vp \big)
    \;=\; \sum_{\ell=1}^{L} \varepsilon_\ell(\vp_\ell).
\end{equation}
Each summand depends on the parameters of one kernel only, so the kernels are trained independently of one another, in parallel. Factors that recur in the program may share one set of parameters, whose gradient contributions then accumulate across occurrences. Minimizing $\varepsilon_\ell$ is equivalent to maximizing log likelihood,
\begin{equation}\label{eq:compile_mle}
    \min_{\vp_\ell}\;
    \EE_{x \sim q_{\ell-1}}\, \EE_{y \sim P_\ell(\cdot \mid x)}\big[ -\log \Pf_\ell(y \mid x; \vp_\ell) \big].
\end{equation}

\paragraph*{The gradient.}
Differentiating \eqref{eq:compile_mle} gives
\begin{equation}\label{eq:compile_grad}
    \nabla_{\vp_\ell} \varepsilon_\ell
    =
    \EE_{x \sim q_{\ell-1}}\, \EE_{y \sim P_\ell(\cdot \mid x)}\big[ {-\nabla_{\vp_\ell} \log \Pf_\ell(y \mid x; \vp_\ell)} \big].
\end{equation}
In \eqref{eq:compile_grad}, the expectations are taken with respect to the target only, which supplies the input $x$ and the output $y$. Knowledge about the target determines how they are evaluated. The integrand is a function of the model only, and the structure of the model determines how it is computed. We note that we can have the same kernel applied many times in the same directed factor graph. This is enforced through weight sharing. For example if we were to specify that \(P_{\ell_1},P_{\ell_2},P_{\ell_3}\) are the same kernel then we enforce weight sharing \(\vp_{\ell_1}=\vp_{\ell_2}=\vp_{\ell_3}\) and so in taking the gradient we would get a sum of terms with contributions coming from the same kernel appearing at different places in the DFG.

\paragraph*{Target specification.}
The target kernel may be specified in three ways, ordered by decreasing access:
\begin{enumerate}[label=(\alph*), leftmargin=*]
    \item \emph{Explicit conditional.} $P_\ell(y \mid x)$ is available in closed form and its output space is small enough to enumerate. The expectations in \eqref{eq:compile_grad} are then computed exactly, as weighted sums. A Trotter gate given as a small stochastic matrix, as in \S\ref{sec:demos_random_walk}, is of this kind.
    \item \emph{Conditional sampler.} We cannot enumerate the probabilities $P_\ell(y\mid x)$, but we can draw $y \sim P_\ell(\cdot \mid x)$ at any prescribed input $x$. The expectations are estimated by Monte Carlo.
    \item \emph{Trajectory data.} We possess recorded trajectories $z_{0:L} \sim P$ but cannot query the kernels at inputs of our choosing. This is the case when a dataset is available, e.g.,\ learning the transition kernel of an observed system from its history, with many (yesterday, today) pairs and no way to re-run yesterday. The market simulator of \S\ref{sec:demos_market} fits in this category. The data supply pairs $(x, y) = (z_{\ell-1}, z_\ell)$ from the joint distribution $q_{\ell-1}(x)\, P_\ell(y \mid x)$, on which both expectations are estimated.
\end{enumerate}
For the purpose of computing the gradient \eqref{eq:compile_grad}, cases (b) and (c) coincide. Both supply input--output pairs from the same joint, and the estimator does not care whether a pair was drawn on demand or read from a dataset. The difference between them reappears in \S\ref{sec:input_distributions}, where case (c) fixes the input distribution, which rules out model-context matching.

\paragraph*{Model structure.}
On the model side there are two possibilities.
\begin{enumerate}[label=(\arabic*), leftmargin=*]
    \item \emph{Tractable kernel.} $\Pf_\ell(y \mid x; \vp_\ell)$ is an explicitly computable, differentiable function of $\vp_\ell$, and the integrand of \eqref{eq:compile_grad} is evaluated exactly by automatic differentiation. Compilation is then ordinary deterministic optimization, with no sampling on the model side. Tractability reaches much further than few-spin kernels. Enumerating the output spins costs $2^{n_{\mathrm{out}}}$ terms, and hidden spins need not be enumerated at all when they can be summed out analytically, which is possible in closed form whenever they form an independent set. The \emph{analytic thermodynamic kernel} (ATK, Appendix~\ref{app:atk}) generalizes this approach. In some cases the fit is itself closed-form, requiring no gradient descent at all (Appendix~\ref{sec:pnot}).

    \item \emph{Intractable kernel.} If the kernel is too large to explicitly compute, the gradient is estimated by sampling the kernel itself. Differentiating $\log \Pf_\ell$ through \eqref{eq:kernel_def} and \eqref{eq:forward} turns the integrand of \eqref{eq:compile_grad} into a difference of two expectations (Appendix~\ref{app:vc_details}),
    \begin{multline}\label{eq:cd_identity}
        -\nabla_\vp \log \Pf(y \mid x; \vp)
        \;=\;
        \EE_{(w, y') \sim p_\vp(\cdot,\, \cdot \mid x)}\big[ \Phi_\vp(x, w, y') \big] \\
        \;-\; \EE_{w \sim p_\vp(\cdot \mid x, y)}\big[ \Phi_\vp(x, w, y) \big],
    \end{multline}
    where $\Phi_\vp \coloneqq -\nabla_\vp E_\vp$ is the vector of the THM's couplings \(J_e\) for each hyperedge, one spin product $\prod_{i \in e} s_i$ per hyperedge $e$ and one spin value $s_i$ per spin, read directly off a spin configuration. Both terms are expectations of $\Phi_\vp$ under thermalizations of the kernel, each estimated by a Gibbs run of $K$ sweeps on the substrate or its simulator, and the two runs differ only in the role each group of spins plays. The input spins are clamped (to $x$) in both, the hidden spins thermalize freely in both, and the output spins thermalize freely in the first term (the \emph{negative phase}) but are clamped to the target's sample $y$ in the second (the \emph{positive phase}). This is the contrastive-divergence estimator of Boltzmann-machine learning~\cite{ackley1985,hinton2002,carreira2005cd}, in conditional form. Its bias is governed by the mixing time of the clamped kernel, which grows as training sharpens the kernel, the mixing--expressivity tradeoff (MET)~\cite{jelincic2025dtm}. The adaptive correlation penalty of Ref.~\cite{jelincic2025dtm} bounds the mixing time throughout training and keeps the estimator reliable.
\end{enumerate}
Appendix~\ref{app:vc_details} assembles the concrete loss and gradient estimator in each of the possible combinations of target specification and model structure, and collects further considerations on the compilation objective.

\begin{figure}[t]
\centering
\resizebox{\columnwidth}{!}{%
\begin{tikzpicture}[>=Stealth, line cap=round,
   classlbl/.style={font=\scriptsize, text=black!70},
   srclbl/.style={font=\scriptsize},
   mathlbl/.style={font=\small}]

\fill[cAccent!16, draw=cAccent!45, line width=0.8pt]
  plot[smooth cycle, tension=0.7] coordinates
  {(0.6,3.2) (1.6,5.1) (3.6,5.35) (5.6,4.8) (7.05,3.35)
   (6.8,1.6) (5.2,0.8) (3.2,0.7) (1.4,1.2) (0.5,2.2)};
\fill[cAccent!30, draw=cAccent!60, line width=0.8pt]
  plot[smooth cycle, tension=0.7] coordinates
  {(1.0,3.1) (1.7,4.5) (3.3,4.6) (4.4,4.0) (4.78,3.0)
   (4.4,2.0) (3.3,1.5) (1.8,1.6) (1.05,2.4)};
\fill[cAccent!46, draw=cAccent!75, line width=0.8pt]
  plot[smooth cycle, tension=0.7] coordinates
  {(1.5,3.05) (2.0,3.9) (2.9,4.0) (3.38,3.5) (3.47,3.0)
   (3.3,2.4) (2.8,2.1) (2.0,2.2) (1.55,2.6)};

\node[classlbl] at (2.45,2.62) {$n_h{=}0$};
\node[classlbl] at (3.15,4.28) {$n_h{=}1$};
\node[classlbl] at (1.85,4.98) {$n_h \ge k^\ast$};

\node[star, star points=5, star point ratio=2.3, fill=cBackward,
      draw=white, line width=0.6pt, minimum size=0.42cm, inner sep=0pt]
      (tgt) at (6.1,3.0) {};
\node[mathlbl, text=cBackward, right=2pt of tgt] {$P_\ell$};

\coordinate (phi0) at (3.47,3.0);   
\coordinate (phi1) at (4.78,3.0);   
\coordinate (phih) at (4.20,3.0);   

\begin{scope}
  \fill[cExOrange!18] (phih) ellipse (0.36 and 0.30);
  \foreach \p in {(3.95,3.12),(4.42,3.10),(4.05,2.84),(4.38,2.86),(4.20,3.22),(4.20,2.74)}
     \fill[cExCopper!75] \p circle (0.025);
\end{scope}

\draw[cInput, line width=1.3pt, ->] (phih) -- (phi1);
\draw[cBackward, line width=1.3pt, dashed, ->] (phi1) -- (5.98,3.0);
\draw[cBackward!50, line width=0.8pt, dashed, ->]
   (phi0) to[bend left=22] (5.95,3.08);

\fill[cInput] (phih) circle (0.055);
\draw[cBackward, line width=1pt, fill=white] (phi1) circle (0.06);
\fill[black!70] (phi0) circle (0.05);

\node[mathlbl, anchor=south east] at ([yshift=1pt]phih) {$\hat\vp$};
\node[mathlbl, anchor=south west] at ([yshift=1pt]phi1) {$\vp^\ast$};
\node[srclbl, text=cBackward!70, anchor=south] at (4.90,3.58) {$\epsilon_0$};
\node[srclbl, text=cBackward, anchor=north]    at (5.35,2.92) {$\epsilon_1$};

\node[srclbl, text=cAccent!75!black, align=center] at (6.05,1.6)
   {$\epsilon\!\to\!0$\\[-1pt]\scriptsize (target reachable)};
\draw[cAccent!70!black, line width=0.6pt, ->] (6.05,1.9) -- (tgt);

\node[srclbl, text=cInput, anchor=south] at (3.7,5.55) {optimization residual};
\draw[cInput!55, line width=0.5pt] (3.9,5.40) -- (4.49,3.05);

\node[srclbl, text=cBackward, anchor=west, align=left] at (6.55,4.5)
   {hypothesis-class\\limitation};
\draw[cBackward!55, line width=0.5pt] (6.55,4.45) -- (5.40,3.05);

\node[srclbl, text=cExCopper, anchor=north] at (3.9,0.40) {sampling bias (MET)};
\draw[cExCopper!60, line width=0.5pt] (4.0,0.66) -- (4.20,2.66);

\end{tikzpicture}%
}
\caption{Expressivity hierarchy and the three sources of compilation residual. Each hypothesis class (the set of forward conditionals the kernel can realize at a fixed hidden-spin count $n_h$ and topology) is nested inside the next. The example target $P_\ell$ lies outside the $n_h{=}0$ and $n_h{=}1$ classes. The achieved parameter $\hat\vp$ has three sources of residual errors (\S\ref{sec:compile}). The \textcolor{cInput}{optimization residual} is the gap from $\hat\vp$ to the true in-class minimizer $\vp^\ast$. The \textcolor{cBackward}{hypothesis-class limitation} lower-bounds the attainable accuracy $\epsilon_1$, it is the irreducible gap from $\vp^\ast$ to the target, fixed by $n_h$, topology, and the coupling and bias caps. The \textcolor{cExCopper}{sampling bias} of unmixed phases displaces the achieved point within the class.}
\label{fig:expressivity_hierarchy}
\end{figure}
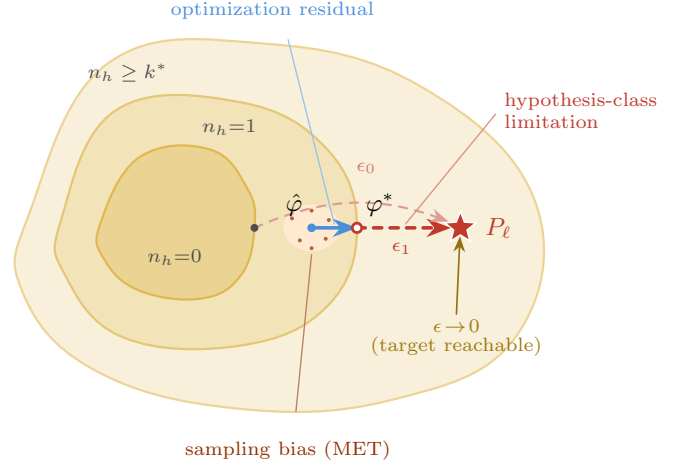

\paragraph*{Sources of residual.}
Independently of the method used for computing the gradient, compilation ends with each kernel matching its target only up to a residual $J_\ell(\cdot\,; \hat \vp_\ell)$. Figure~\ref{fig:expressivity_hierarchy} separates the three sources that contribute to it: the \emph{hypothesis-class limitation} set by the substrate's topology, the hidden-spin count, and the coupling caps; the \emph{optimization residual} of finite gradient tolerance; and, in the intractable case, the \emph{sampling bias} of unmixed phases.

\paragraph*{Error accumulation.}
The per-kernel residual is what the composed program accumulates, and for that accounting it is convenient to pass from the relative entropy $J_\ell$ in which we compile to total variation, in which Markov kernels are non-expansive. The per-kernel discrepancy is measured by
\begin{equation}
  \eta_\ell \;=\; \sup_x \big\| \hat P_\ell(\cdot \mid x) - P_\ell(\cdot \mid x) \big\|_\TV, \label{eq:error-defn}
\end{equation}
the worst-case total-variation error of the deployed conditional $\hat P_\ell$ against its target, and the accumulated error by the discrepancy $\hat\delta_\ell = \| \hat q_\ell - q_\ell \|_\TV$ between the marginals of the deployed and target programs at wire $\ell$. REINFORCE post-training (\S\ref{sec:reinforce}) attacks this accumulation directly. A large class of targets, however, is error-robust before any mitigation. This robustness is measured in terms of the {\it contraction coefficient} $\rho(P_\ell)$ of each kernel, which satisfies
\begin{equation}
  \big\| P_\ell\, (p - q) \big\|_\TV \;\le\; \rho(P_\ell)\, \| p - q \|_\TV \label{eq:contraction-coeff}
\end{equation}
for any pair $p,q$ of input distributions. When this coefficient is small, inherited error is washed out geometrically rather than passed along, and $\hat\delta_\ell$ saturates at the depth-independent floor of Eq.~\eqref{eq:deployed_floor}, the per-kernel residual amplified by the target's relaxation time, instead of accumulating indefinitely. Rapidly mixing MCMC chains are the canonical case; the meta-EBM demonstration of \S\ref{sec:demos_meta} compiles exactly such a target and measures its floor against this prediction.

\subsection{Context-matching input distributions}
\label{sec:input_distributions}

The expectation over inputs in the gradient \eqref{eq:compile_grad} is independent of the underlying input distribution $\mu_\ell$ under which a kernel is trained. This is a degree of freedom of variational compilation, and the error accounting in \eqref{eq:mitigation_chain} shows its cost. If every kernel matched its target exactly, the input distribution would not matter. Any $\mu_\ell$ with full support finds the same optimum. However, with the constraints in parameters and topology imposed by realistic hardware architectures, the ability of a kernel to express a given target may be limited. In this case, the input distribution determines how a kernel's parameters are optimized to spend their limited accuracy.

The error a kernel contributes to the \emph{deployed} circuit is its error on the inputs the deployed circuit actually feeds it, and those are distributed as $\tilde q_{\ell-1}$, not $q_{\ell-1}$: upstream residuals drift the input stream away from the target's, most sharply by leaking probability onto inputs the target never produces, where a kernel trained under target inputs was free to be arbitrarily wrong. (The conservation leakage in the random-walk demonstration of \S\ref{sec:demos_random_walk} is exactly this failure mode).

There are three possible choices for $\mu_\ell$:
\begin{enumerate}[label=(\roman*), leftmargin=*]
    \item \emph{target inputs}, $\mu_\ell = q_{\ell-1}$: the weights of the exact decomposition in \eqref{eq:mitigation_chain}, under which the sum of per-kernel losses \emph{is} the trajectory error. This is the default;
    \item \emph{model inputs}, $\mu_\ell = \tilde q_{\ell-1}$: each kernel is trained under the input marginal it receives inside the compiled circuit. These differ from the target input by the accumulated error over the preceding kernels;
    \item \emph{generic inputs}, e.g.,\ uniform: agnostic to the context of the rest of the program. This is justified when the destination program is not yet known (\S\ref{sec:libraries}).
\end{enumerate}
Re-optimizing an already-compiled kernel under the first two is called \emph{target-context matching} (target-CM) and \emph{model-context matching} (model-CM), respectively.

\paragraph{target-context.}
In case of (c) Trajectory data, the data supply pairs whose inputs are already distributed as $q_{\ell-1}$, the distribution already matches the context and no other choice of $\mu_\ell$ can improve the KL bound of \eqref{eq:mitigation_chain}. In the cases of (a) explicit conditional and (b) conditional sampler, inputs for the re-optimization of the kernels are drawn from the target conditional. In principle these can be produced by rolling the target program forward, which would seem to make per-kernel training cost grow with depth. However, training already materializes samples of that kernel's \emph{target output} so target rollouts accumulate as a by-product of training itself. Appendix~\ref{app:input_distributions} gives training schedules that exploit this, together with their guarantees.

\paragraph{model-context.} Model inputs weight each input by how often the deployed circuit visits it. Appendix~\ref{app:input_distributions} makes the comparison precise: target inputs already guarantee that per-kernel residuals are never amplified in propagation, but the depth-independent floor at which the error of a contracting program settles is set by the error each kernel injects on the inputs it receives, and only model inputs control that quantity.
Model-context matching requires querying the target conditional at model-generated inputs, so it is available in cases (a) explicit conditional and (b) conditional sampler only. In case (c) trajectory data it is ruled out entirely, and trajectory-level REINFORCE (\S\ref{sec:reinforce}) remains the only stage that trains the program under the inputs it produces. This is the situation of the market simulator of \S\ref{sec:demos_market}. The inputs themselves are produced by rolling out the compiled circuit, on the hardware if available, in which case they carry the upstream execution error too and the re-optimization trains against it as well.

To improve performance of the compiled kernels, one would train under target inputs from the start: in case (c) trajectory data they are automatic and in cases (a) explicit conditional and (b) conditional sampler they are free. Optionally, switch to model inputs late in training, for two reasons. Early on, residuals are dominated by optimization slack rather than by capacity, so the two objectives barely differ where it matters. Also, the model-input objective moves whenever upstream kernels update, so chasing it from the start spends capacity on input distributions that will not survive training. Generic inputs enter only through pre-compiled libraries, repaired by context matching after placement (\S\ref{sec:libraries}). REINFORCE post-training comes last (\S\ref{sec:reinforce}). The random-walk demonstration of \S\ref{sec:demos_random_walk} deliberately runs the full pipeline (generic inputs, then model-CM, then REINFORCE) to expose how much the input distribution alone matters. Whenever possible, however, one should always aim to train using target inputs.

\subsection{Pre-compiled kernel libraries}
\label{sec:libraries}

A target conditional that recurs across applications, such as the stochastic gates in \texttt{torx}'s PSC library \cite{lockwood2026torx} or a conditional that is particularly common across stochastic programs, can be compiled once and reused. In order to do so, one would compile such targets under a generic input distribution $\mu_\ell$, such as the uniform one. The rightmost bound of \eqref{eq:mitigation_chain} guarantees that after placement into a specific program, the kernel's contribution to the trajectory error is at most $c_\ell\, \EE_{x \sim \mu_\ell}[J_\ell(x; \vp_\ell)]$, where for uniform $\mu_\ell$ on $n_{\mathrm{in}}$ input spins $c_\ell = 2^{n_{\mathrm{in}}} \max_x q_{\ell-1}(x)$. Context matching after placement of the kernel within the DAG representing a stochastic program can repair the mismatch by re-allocating the kernel's capacity towards inputs that are more relevant in the context of the program it belongs to. Which kernels merit a library, and whether generic input distributions other than uniform transfer better across applications, we leave as future work. Pre-compiled gate sets play an analogous role in quantum computing, where quantum programs are generally constructed out of known basic gates.

\subsection{Trajectory-level post-training via REINFORCE gradient}
\label{sec:reinforce}

The training stages in \S\ref{sec:compile}--\S\ref{sec:libraries} focused on matching each kernel against its target conditional, discarding the context from the DFG they are embedded in. Therefore, the trajectory error such training can reach is bounded below by the sum $\sum_\ell \varepsilon_\ell$ of per-kernel residuals in~\eqref{eq:mitigation_chain}. Trajectory-level REINFORCE instead cares only about how well the readout (so only a subset of the DFG's variables) satisfies a given objective. Training uses all compiled kernels jointly against a single global objective, which optimizes the entire program rather than individual kernels. Each thermodynamic kernel then receives the outputs of its ancestors and can be driven to compensate for the errors therein. We obtain such a joint training rule from the REINFORCE, or score-function, gradient~\cite{williams1992,glynn1990}, which only requires the ability to sample from each compiled kernel and to evaluate a scalar objective, and does not require a gradient propagated through the discrete samples.

The model program is executed forward as in \eqref{eq:model_traj}, $z_{0:L} \sim \Pmod_\vp$, giving the loss
\begin{equation}\label{eq:rf_loss}
    \LRF(\vp_1, \ldots, \vp_L) \;=\; \EE_{z_{0:L} \sim \Pmod_\vp}\big[ F(z_{1:L}) \big].
\end{equation}
$F$ need not depend on every variable, it can be, and often is, a function of the terminal $z_L$ value.

Let $\Phi_{\vp_\ell} = -\nabla_{\vp_\ell} E_{\vp_\ell}$ be factor $\ell$'s vector of sufficient statistics, as in \eqref{eq:cd_identity}, evaluated at the factor's clamped input $z_{\ell-1}$ and at the hidden and output spins $(w_\ell, z_\ell)$ it samples. Applying the score-function identity factor-by-factor gives
\begin{multline} \label{eq:conditional_reinforce_grad}
    \nabla_{\vp_\ell} \LRF = \EE \bigg[ F(z_{1:L})\Big( \Phi_{\vp_\ell}(z_{\ell-1}, w_\ell, z_\ell)
    \\[1mm]
    - \EE_{(w', y') \sim p_{\vp_\ell}(\cdot,\,\cdot \mid z_{\ell-1})}\big[ \Phi_{\vp_\ell}(z_{\ell-1}, w', y') \big] \Big) \bigg],
\end{multline}
where the outer expectation is over executions of the model program, with each factor's hidden spins $w_\ell$ recorded alongside its output, and the inner expectation is conditional on the sampled parent $z_{\ell-1}$. Appendix~\ref{app:reinforce} gives an unbiased estimator of \eqref{eq:conditional_reinforce_grad} that costs only one extra ``reference'' sample per factor, and explains several important details to be mindful of when using this method in practice. Just like variational compilation, the only model-side information that REINFORCE requires are the spin products $\Phi$ computed from thermalized states. However, unlike contrastive divergence (which is one way of executing VC), which needs the two clamping patterns to compute the gradient (known as the positive and the negative phase), REINFORCE only requires inputs to be clamped with all hidden and output nodes being free (just like the negative phase of CD).

What should $F$ be? If $F$ is held fixed while $\vp$ varies, the minimizer of $\LRF$ drives $\Pmod_\vp$ onto $\arg\min F$, collapsing it to a trivial distribution. A useful $F$ must therefore stand in for a \emph{distributional} objective, and converting a distributional target into a per-sample reward is a standard problem in generative modeling. There are two routes. The first is adversarial: take $F = D_\phi$, the score of a discriminator with parameters $\phi$ trained against the program's own samples, as in a generative adversarial network (GAN)~\cite{goodfellow2014,arjovsky2017wgan}; the min--max game between the compiled program and the discriminator turns the per-sample score into a distributional distance, the natural choice when the target is a high-dimensional distribution available only through data. The second applies when the quantity of interest is a differentiable functional of a readout law. For the squared error $\mathcal{D}(\vp) = \lVert m_\vp - t \rVert^2$ between a vector of readout statistics $m_\vp = \EE_{z_{0:L} \sim \Pmod_\vp}[f(z_L)]$ and a target value $t$, the chain rule supplies the effective per-sample reward
\begin{equation}\label{eq:f_eff}
    F_{\mathrm{eff}}(z_L) \;=\; 2 \sum_i \big( m_{\vp, i} - t_i \big)\, f_i(z_L),
\end{equation}
with $m_\vp$ estimated by the minibatch mean and held fixed under differentiation (a stop-gradient): REINFORCE run with the reward \eqref{eq:f_eff} descends $\mathcal{D}$ exactly (Appendix~\ref{app:reinforce}). The biased-random-walk benchmark of \S\ref{sec:demos_random_walk} follows this route, Eqs.~\eqref{eq:rw_leaf_loss} and \eqref{eq:f_eff}, as does the post-training of the market simulator (\S\ref{sec:demos_market}).

If this rule can train the whole program jointly, why compile and context-match each kernel first? Because the score-function gradient can have very high variance. Equation~\eqref{eq:conditional_reinforce_grad} uses only the scalar value $F(z_{1:L})$, not the gradient of $F$, and only at the samples the model itself draws. If those samples lie where $F$ is uninformative, say at uniformly low reward, the gradient carries almost no signal. REINFORCE becomes effective only once the model is already good enough that its samples reach the region where $F$ discriminates, and supplying a good starting model is exactly what variational compilation and context matching are for.

This division of labor, a cheap local objective to bring the model up to competence followed by a global reward to refine it, is also the recipe behind much of the recent leap in large language models, where the second stage is reinforcement learning from human feedback (RLHF)~\cite{christiano2017,stiennon2020,ouyang2022}. The resemblance is structural: a language model generating a sequence is a chain of weight-tied conditional samplers (a DFG in the sense of \S\ref{sec:dfg}) pre-trained on next-token prediction, a local objective in which every position is supervised by its own target, the role per-kernel compilation plays here; RLHF then optimizes a reward defined on the whole sequence, which no per-token target can express, and it works only because pre-training has already made the samples good enough for that reward to be informative. Trajectory-level REINFORCE is the counterpart of that post-training step for compiled stochastic programs, and the market simulator of \S\ref{sec:demos_market} realizes the full recipe on a compiled program of the same shape: a weight-tied kernel applied day after day, compiled from recorded pairs and then post-trained with a reward on whole rollouts.

\section{Demonstrations}
\label{sec:demos}

This section demonstrates the framework on a variety of examples that highlight different capabilities available when executing stochastic programs on thermodynamic hardware. The biased random walk of \S\ref{sec:demos_random_walk} shows the compilation and error mitigation methods. The birth--death process of \S\ref{subsec:Eco} compiles an interacting particle system to analytic thermodynamic kernels. The market simulator of \S\ref{sec:demos_market} learns the transition kernel of a panel of financial time series from recorded data alone, an example of a target where only trajectory data is available, and post-trains the compiled program with trajectory-level REINFORCE. The meta-EBM of \S\ref{sec:demos_meta} samples a non-native three-body Ising energy by variationally compiling its Gibbs conditionals. The Gaussian stochastic circuit of \S\ref{sec:demos_gaussian} compiles a hierarchical Gaussian model in closed form and runs a measurement-selection loop on it. 
All our experiments are executed using the \texttt{thrml} Python library with THMs simulating realistic hardware by matching the topology of the Z1 thermodynamic computer and, except where noted, the hardware caps $|J| \leq J_\mathrm{max}$, $|h| \leq h_\mathrm{max}$.
We will describe the underlying representation of each of our examples as a directed factor graph, write out the kernels we are compiling and which method was used for training.

\subsection{Biased random walk}
\label{sec:demos_random_walk}
\begin{figure*}[t]
    \centering
    \begin{subfigure}[b]{0.36\textwidth}\centering
        \includegraphics[width=\linewidth]{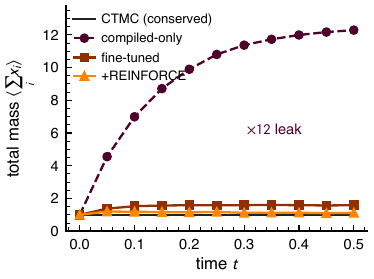}
        \caption{}\label{fig:rw-mass}
    \end{subfigure}\hfill
    \begin{subfigure}[b]{0.36\textwidth}\centering
        \includegraphics[width=\linewidth]{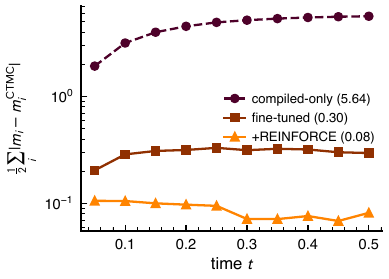}
        \caption{}\label{fig:rw-tv}
    \end{subfigure}\hfill
    \begin{subfigure}[b]{0.255\textwidth}\centering
        \includegraphics[width=\linewidth]{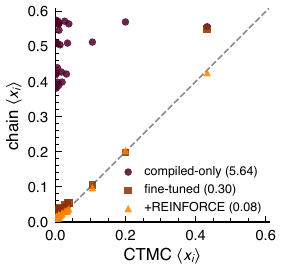}
        \caption{}\label{fig:rw-parity}
    \end{subfigure}
    \caption{%
        \textbf{Error mitigation on the $5\times5$ biased random walk.} A single particle starts at site $(0,0)$ and evolves under the Trotterized biased-walk circuit of Eq.~\eqref{eq:asym_swap} ($M=10$ macro steps, $\delta t = 0.05$, $\gamma = 2$), compiled per gate into atomic kernels with $n_h = 1$. All panels compare the analytic CTMC, the compiled-only chain, the chain after model-context matching (\S\ref{sec:input_distributions}), and the chain after trajectory-level REINFORCE on top of context matching (\S\ref{sec:reinforce}); markers (circle/square/triangle) are shared across panels. \textbf{(a)}~Total occupancy mass $\mathbb{E}[\sum_i z_i]$ versus time. The CTMC conserves particle number (flat at $1$) while the compiled chain experiences leakage out of the single-particle sector. \textbf{(b)}~Half-$\ell_1$ occupancy error between the per-site occupancy and the CTMC marginal (log scale) versus time;  the error falls by nearly two orders of magnitude across the pipeline. \textbf{(c)}~Final-time per-site occupancy of each chain against the CTMC reference, one point per site. Conservation leakage appears as the compiled-only points fanning above the diagonal, context matching and REINFORCE collapse them onto it.
    }
    \label{fig:random_walk}
\end{figure*}
Consider the stochastic dynamics of a single particle hopping on a graph $\mathcal{G} = (\mathcal V, \mathcal E)$, where the vertices $\mathcal V$ are the possible locations of the particle and the edges $\mathcal{E}$ the transport links between vertices.
For our example we will take the graph to be an \(L\times L\) two-dimensional torus.
The occupation probabilities $p_i(t)$ with $i \in \mathcal V$ evolve under the master equation,
\begin{equation}\label{eq:master}
  \frac{dp_j}{dt} \;=\; \sum_{i} Q_{ji}\, p_i ,
\end{equation}
with solution $p(t) = e^{Qt} p(0)$ where $Q$ is the generator of this continuous-time Markov chain (CTMC). For the random walk, $Q_{ji}$ is the rate from source site $i$ to destination site $j$,
\begin{equation}\label{eq:walk_generator}
  Q_{ji} \;=\; \gamma\,\sigma(a_j - a_i)\,\mathbbm{1}_{\{(i, j) \in \mathcal E\}}, \quad Q_{ii} \;=\; -\!\sum_{j \neq i} Q_{ji},
\end{equation}
with $\sigma(u)=1/(1+e^{-u})$ the logistic function, $a_i$ a per-site logit, and $\gamma$ the overall transition rate. Detailed balance with respect to $\pi_i \propto e^{a_i}$ holds so the walk converges to the stationary state, $\pi_i = e^{a_i}/\sum_k e^{a_k}$.

\begin{figure}
\centering
\includegraphics[width=1\columnwidth]{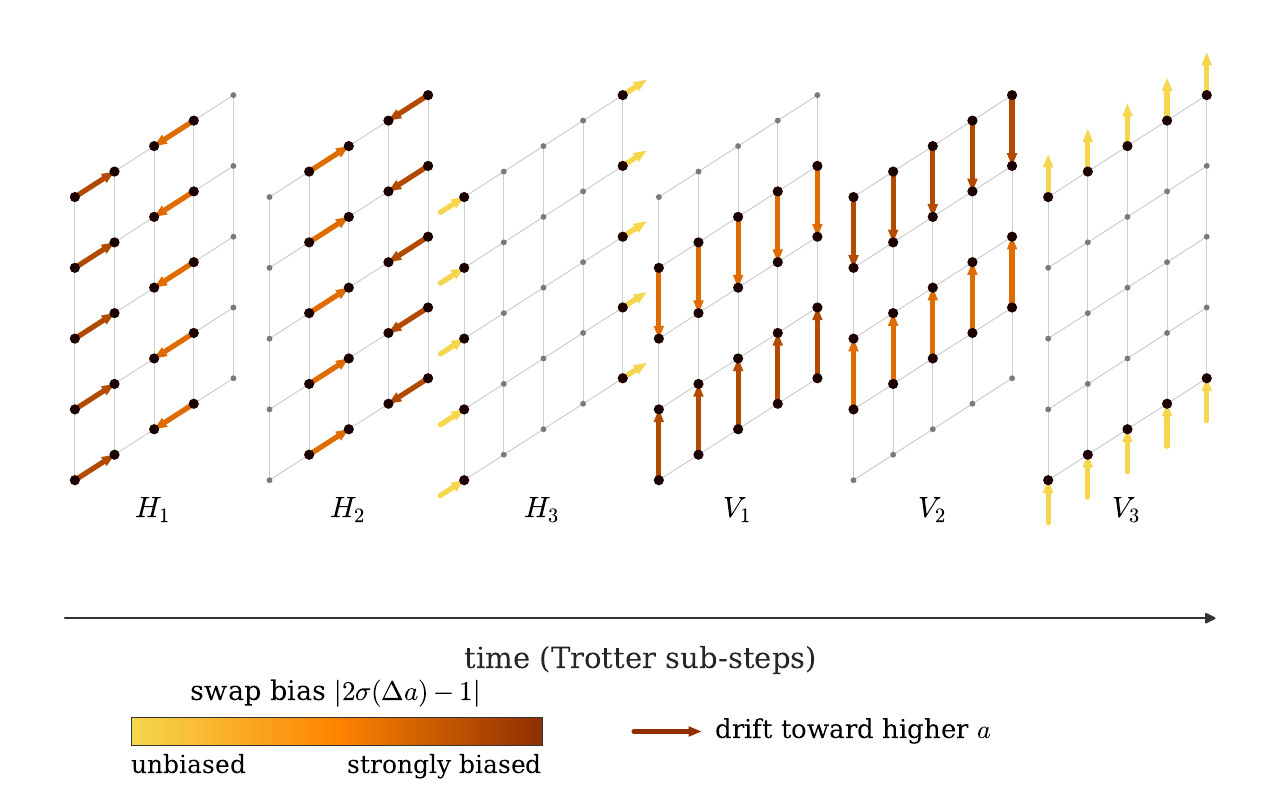}
\caption{Biased-random-walk circuit on the $5\times5$ torus over one macro Trotter step. Each tilted slice is one Trotter sub-step, the three horizontal families $H_1$, $H_2$, $H_3$ followed by the three vertical families $V_1$, $V_2$, $V_3$ with arrows marking the \texttt{PAsymSwap} gates active in that sub-step. Arrow color encodes the swap bias $|2\sigma(\Delta a)-1|$. The periodic classes $H_3$ and $V_3$ are drawn as split arrows at the boundary.}
\label{fig:brw_circuit}
\end{figure}

One can rewrite the solution of the master equation~\eqref{eq:master} by discretizing time into $M$ steps of size $\delta t = t/M$,
\begin{equation}\label{eq:time_slicing}
  p(t) \;=\; e^{Qt}\,p(0) \;=\; \bigl(e^{Q\delta t}\bigr)^{M} p(0),
\end{equation}
the single-step propagator $e^{Q\delta t}$ is a stochastic matrix and the dynamics is now a discrete-time Markov chain on $\mathcal V$. The propagator is, however, a dense $|\mathcal V| \times |\mathcal V|$ matrix that couples all sites at once and does not correspond to a local operation. Locality enters through the generator, each rate in~\eqref{eq:walk_generator} involves a single edge, so $Q$ decomposes as a sum of two-site terms $Q = \sum_{(i,j) \in \mathcal E} Q_{(i,j)}$, with each $Q_{(i,j)}$ acting
only on the occupancies of the two sites it connects. Trotterization~\cite{trotter1959product} exploits this decomposition to approximate the single-step propagator by a product of local factors,
\begin{equation}\label{eq:trotter}
  e^{Q\delta t} \;\approx\; \prod_{\mathcal A} e^{Q_{\mathcal A} \delta t},
  \qquad Q_{\mathcal A} = \sum_{(i,j) \in \mathcal A} Q_{(i,j)},
\end{equation}
where the edges are partitioned into color classes $\mathcal A_1,\ldots,\mathcal A_c$, shown in Figure~\ref{fig:brw_circuit}, such that no two edges in a class share a vertex. Within a class the two-site factors act on disjoint pairs and commute, so each $e^{Q_{\mathcal A} \delta t}$ is an exact product of two-site gates applied in
parallel, and the only error in~\eqref{eq:trotter} is the $O(\delta t^2)$ per-step Trotter error from the non-commutativity of different classes, accumulating to $O(t\,\delta t)$ over the
trajectory.

For small $\delta t$ the Trotter factor $e^{Q_{(i,j)}\delta t}$ can be expanded as $e^{Q_{(i,j)}\delta t} = I + Q_{(i,j)}\delta t + O(\delta t^2)$, identity plus the
rates $Q_{(i,j)}$. Therefore, to first order in $\delta t$ we can write each Trotter factor as a $4 \times 4$ stochastic matrix acting on the occupation of adjacent vertices, assuming one-hot encoding for the state. Particle conservation makes $00$ and $11$ fixed, and the biased hop lives in the middle block. We take this stochastic matrix as the target conditional
\begin{align}\label{eq:asym_swap}
  P_{(i,j)} =
  \begin{pmatrix}
    1 & 0 & 0 & 0 \\
    0 & 1 - p_{ji} & p_{ij} & 0 \\
    0 & p_{ji} & 1 - p_{ij} & 0 \\
    0 & 0 & 0 & 1
  \end{pmatrix},
\end{align}
with the hop probabilities
\begin{equation}\label{eq:asym_swap_prob}
  p_{ij} \;=\; \gamma\,\sigma(a_j - a_i)\,\delta t .
\end{equation}
This is the asymmetric stochastic swap gate \texttt{PAsymSwap} in the \texttt{torx} library~\cite{lockwood2026torx}. Each \texttt{PAsymSwap} is variationally compiled into a thermodynamic kernel with $n_\mathrm{in} = n_\mathrm{out} = 2$ and $n_h=1$. In this case, the median compiled total variation distance for the \texttt{PAsymSwap} gates is 0.096.

The overall directed factor graph representation for a random walk of duration $t$ is then a chain of $\ell=1,\dots,M$ color class kernels $P_\ell=\prod_{\mathcal A}\prod_{(i,j) \in \mathcal A}P_{(i,j)}$, each factor $\ell$ involves a product over all the edges of the lattice, grouped by their corresponding color. We note that applying \(P_{(i,j)}\) is an approximation to the dynamics of the random walk for time \(\delta t\), with an error coming from the non-commutativity and another from the approximation in \eqref{eq:asym_swap}. Each kernel \(P_i\) takes in as input the value of the vertices adjacent to edge \(e_i\) and outputs their updated values. We write the compiled DFG as $\tilde{P}^\mathrm{DFG} = \tilde{P}_1^F\tilde{P}_2^F\cdots\tilde{P}_M^F$.
The kernels \(\tilde{P}^F_{(i,j)}(y\mid x;\vp_i)\) are parametrized by \(\vp\) and trained under a uniform input distribution with loss,
\begin{align*}
    \mathcal L^{VC}(\vp)=\mathbb{E}_{x\sim\mu}\left[D_{KL}\left(P_{(i,j)}(\cdot\mid x)\mid\mid\tilde{P}^F_{(i,j)}(\cdot\mid x;\vp)\right)\right],
\end{align*}
where \(\mu\) is the uniform distribution over the \(2\) input bits.
The example is small enough that we take gradients explicitly and train with gradient descent as in the case of a tractable kernel.

We then apply model-context matching to the compiled circuit and re-optimize each gate under the input marginal it receives when the compiled circuit itself is rolled out. These input marginals are highly non-uniform: gates near the initial site see most of their mass in the one-particle sector, while gates far from the initial site see almost exclusively the $(0,0)$ input configuration. Model-context matching re-allocates each gate's limited capacity towards the inputs that matter most, the ones the deployed circuit actually visits. On top of the model-context-matched circuit we run trajectory-level REINFORCE (§\ref{sec:reinforce}) on the endpoint occupancies: the objective is the squared error between the final-time per-site occupancy
$m_{\vp,i}=\EE[\,(z_L)_i\,]$ and the analytical CTMC marginal
$m^\star_i = (e^{QT}p(0))_i$,
\begin{equation}\label{eq:rw_leaf_loss}
  \mathcal{L}^{\mathrm{RF}}_{\mathrm{rw}}(\vp)
  \;=\; \sum_i \bigl(m_{\vp, i} - m^\star_i\bigr)^2 ,
\end{equation}
optimized with the effective per-sample reward \eqref{eq:f_eff}, here with $f_i(z_L) = (z_L)_i$ and $t = m^\star$.

For the experiment we take a square lattice of side $L=5$ and the logits, $a_i=  2\sin \left[ 2 \pi \left(\frac{2x_i+y_i}{L}+0.2\right) \right] +0.75\cos \left[ 2 \pi \left(\frac{x_i-2y_i}{L}-0.4\right) \right] $, where $x_i, y_i \in [0, \cdots, L-1]$ are the horizontal and vertical coordinates of each site, respectively. Two errors accumulate over the $6M =60$-layer trajectory. The Trotter error is $O(t\,\delta t)$ from the non-commutativity of edge factors, present in the analytical target circuit itself. On top of that, every gate carries a compilation residual that sums across paths, accounted for by the trajectory-KL identity~\eqref{path-kl-chain-eq}. For capped finite-energy kernels, the dominant artifact is single-particle conservation leakage. The compiled gate conditionals assign small non-zero probability to transitions outside the one-particle sector, and over many layers the realized marginal accumulates multi-particle support.

Figure~\ref{fig:random_walk} compares the analytical CTMC, the compiled-only chain, the context-matched chain, and the chain after REINFORCE on top of context matching, on a $5 \times 5$ torus with $M = 10$ macro steps. Reported results are measured over $4096$ chains, with each compiled layer sampled after $K=30$ block-Gibbs sweeps. We quantify the quality of the results with three diagnostics: the total occupancy mass, the final-time half-$\ell_1$ occupancy error $\tfrac{1}{2}\sum_i |m_{\vp,i}-m_{\mathrm{CTMC},i}|$, and the final-time per-site occupancy scatter. The compiled marginals systematically over-shoot the CTMC at distant sites and under-shoot near the initial site, the signature of accumulated conservation leakage. Stacking context matching and REINFORCE reduces the final-time half-$\ell_1$ occupancy error from $5.64$ (compiled-only) to $0.30$ (context matched) to $0.08$ (context matched $+$ REINFORCE), nearly two orders of magnitude. This quantity coincides with the total variation between site-occupancy distributions when particle number is conserved, but under leakage it is bounded below by half the mass discrepancy, $\tfrac12\,\big|\mathbb{E}[\sum_i (z_L)_i]-1\big|$. The compiled-only occupancies exceed the CTMC marginal at every site (Fig.~\ref{fig:random_walk}c), so the error $5.6 = (12.2 - 1)/2$ is a pure conservation leak, half the excess mass of panel (a), and the error mitigation techniques are vital in recovering expectation values near the ideal ones.

\input{figures/eco_local_updates}
\subsection{Birth-death Process}
\label{subsec:Eco}
We detail a particular local interacting particle system with rich dynamics and compile its forward conditional distributions to the \emph{Analytic Thermodynamic Kernels} of Appendix~\ref{app:atk}.

Consider a finite, two-dimensional grid on a torus, on which the vertices support a finite carrying capacity $C$ of each of two species: rabbits and foxes. Each time-step, there is an on-site reaction process in which a rabbit has some chance to reproduce and some chance to be eaten by a fox, while a fox has some chance to starve and some chance to reproduce in the event that it ate a rabbit. Then there is a diffusion-like process, in which each animal crosses to a neighboring site on the grid with minute probability $p$. This is effected in a series of two-site migrations tiled across the grid, multiple times per time-step\footnote{We are here describing a Trotter discretization of continuous-time diffusion, but the continuous-time process is incidental and unconsidered.}. Transitions that would otherwise increment a site beyond its carrying capacity are stipulated to simply not occur\footnote{This simplifies the eschatology.}. A diagram representing the local conditional probabilistic updates that this ecosystem comprises may be found in Figure~\ref{fig:local-updates}.

This model has two absorbing states. In the absence of foxes, since rabbits do not starve, any number of rabbits will eventually reproduce to fill the ecosystem. If rabbits are scarce and there are many foxes who consume them at a fast rate, then the rabbits will go extinct and the foxes will follow as they all starve. The predation rate $d_p$ that parametrizes the likelihood of the on-site $(r,f)\to (r-1,f+1)$ and $(r,f) \to (r-1,f)$ processes determines the statistical proportion of these eventualities for a given initial state.

As the conditional distributions in question are given analytically and supported on small state spaces (we take $C = 15$) we can minimize the objective~\eqref{eq:compile_loss} with gradient descent by choosing a small thermodynamic kernel and evaluating its marginal distribution in closed form. Kernels with only four and eight visible nodes are capable of reproducing the conditional distributions of the migration and reaction processes respectively to a worst-case total variation distance of $\sim 0.01$. The Z1-topology kernels to which we compile the conditional distributions of this ecological model appear in Figure~\ref{fig:atk}.

The forward conditionals are best understood through the diagrams provided in Figure~\ref{fig:local-updates}, augmented by the understanding that events which would otherwise populate a site with a greater number of animals than it has capacity resolve uneventfully: animals stay put rather than dying from overcrowding. We keep all parameters fixed save $d_p\in\{4,\ldots,12\}.$ The only other parameters of note are capacity $C = 15$ and ecosystem size $L = 16.$ Remaining parameter values are incidental and collected in Table~\ref{fig:EcoTable}.

The forward conditional of the on-site birth-death process is
\begin{widetext}
\begin{equation}
\label{eq:Reaction}
P_{\mathrm{reaction}}(r',f'\mid r,f)=\frac{1}{B}
\begin{cases}
b_r\,r\,\ind{r<C} & \underset{\text{rabbit birth}}{(r',f')=(r+1,\,f)},\\[4pt]
m_f\,f & \underset{\text{fox death}}{(r',f')=(r,\,f-1)},\\[4pt]
d_p \, \eta\,\ind{r\ge1,\,f\ge1,\,f<C}
  & \underset{\text{fertile predation}}{(r',f')=(r-1,\,f+1)},\\[4pt]
  d_p\,\ind{r\ge1,\,f\ge1}\bigl(1-\eta\,\ind{f<C}\bigr)
  & \underset{\text{sterile predation}}{(r',f')=(r-1,\,f)},\\[4pt]
0 & \text{otherwise.}
\end{cases}
\end{equation}
\end{widetext}
where
\[
B=b_r\,r\,\ind{r<C}+d_p\,\ind{r\ge1,\,f\ge1}+m_f\,f .
\]
and $P_{\mathrm{reaction}}(r',f'\mid r,f)=\ind{(r',f')=(r,f)}$ for $B=0$.
We specify the forward conditional of the migration process as follows. Rabbits ($n=r$) and foxes ($n=f$) migrate independently, with directional migration proposals between sites $i$ and $j$
\[
D=n_{i\to j}-n_{j\to i}
\]
where $n_{i\to j}\sim\Bin(n_i,p)$ and $n_{j\to i}\sim\Bin(n_j,p)$ are independent. The capacity-constrained net change realized in light of the proposal's effect on the present occupants is
\begin{equation}
\label{eq:Migration}
\Delta n=\min\{\max\{D,n_i - C\},C - n_j\}
\end{equation}
which implicitly defines the forward conditional $P_\mathrm{migration}(\Delta n|n_i,n_j).$

\begin{table}[h]

\centering
\small
\renewcommand{\arraystretch}{1.25}
\begin{tabular}{@{}lll@{}}
\toprule
Symbol & Quantity & Value \\
\midrule
$L$          & lattice size ($L\times L$ torus)               & $16$ \\
$C$          & per-species site capacity               & $15$ \\
$p$          & migration probability               & $0.05$ \\
$b_r$        & rabbit birth rate                       & $1.0$ \\
$d_p$        & predation rate                   & $\{4,\dots,12\}$ \\
$\eta$       & fertility rate   & $0.5$ \\
$m_f$        & fox death rate                          & $0.5$ \\
\bottomrule
\end{tabular}
\caption{Parameter values for forward conditionals ~\eqref{eq:Reaction} and ~\eqref{eq:Migration}}
\label{fig:EcoTable}
\end{table}

\begin{figure}[!t]
    \centering
    \includegraphics[
        width=\columnwidth,
        height=0.82\textheight,
        keepaspectratio,
        trim=8mm 0 0 0
    ]{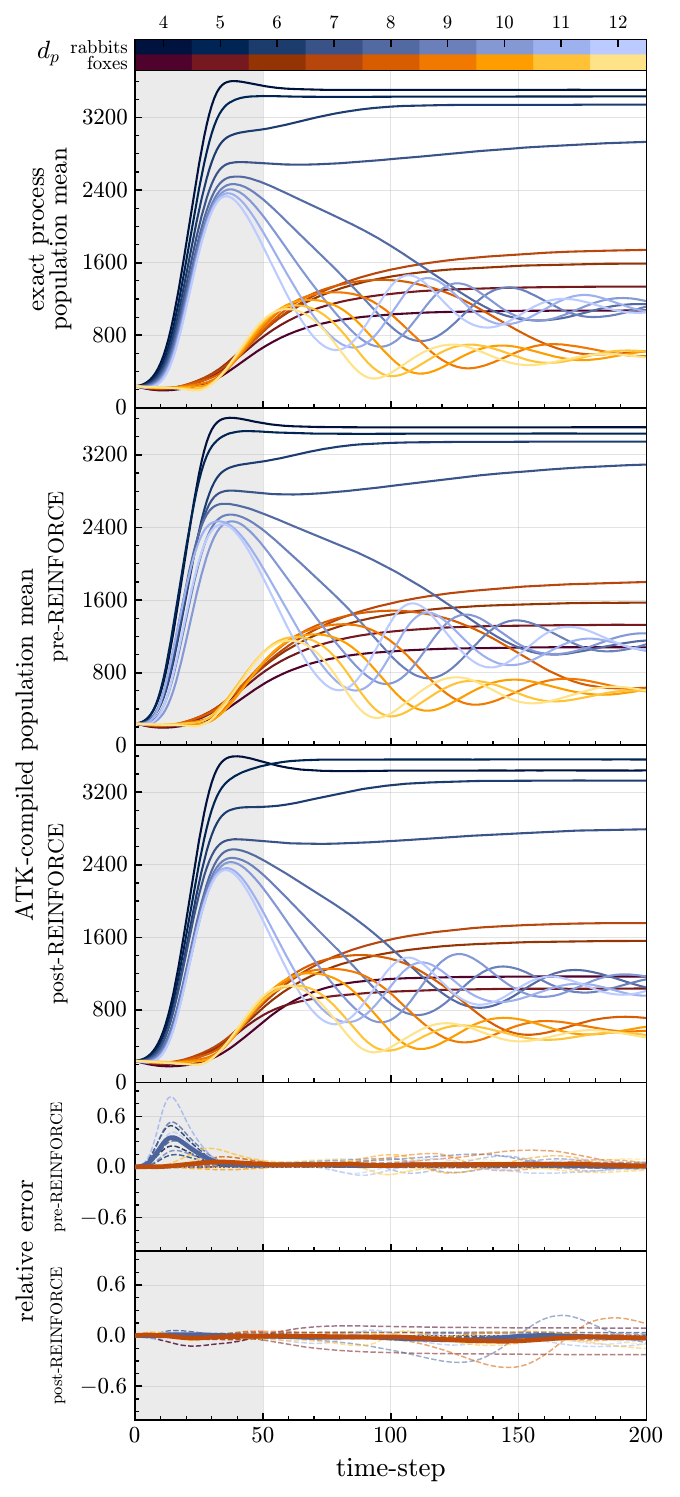}
    \caption{Mean population dynamics across 1000 simulations of the predator-prey ecosystem model. Rabbit population is shown in blue and foxes in orange. The legend at the top of the plot indicates the value of $d_p$ chosen for the reaction process~\eqref{eq:Reaction}. REINFORCE training was carried out only over the time-span shown in gray. Heavy lines in the plots of relative error represent the average over all values of $d_p$; individual relative errors are plotted as dashed lines.}
    \label{fig:CompiledEcoPopRelativeError}
\end{figure}

In Figure~\ref{fig:CompiledEcoPopRelativeError} we show averaged population dynamics characteristic of metastable cohabitation evident in both the exact process and the compiled process. The initial state of the system consists of an otherwise empty environment containing a $4\times4$ central region fully occupied by rabbits and foxes, in total $4\times4\times15 = 240$ of each. The populations display a boom-and-bust cycle for $d_p \in \{8,\ldots,12\}$. The objective for REINFORCE training, performed over the first $T_\mathrm{RF} = 50$ time-steps was
$$\LRF(\vp) = \sum_{t=1}^{T_\mathrm{RF}} \big\| \EE_{z_{0:T_\mathrm{RF}}\sim\Pmod_\vp}[N_t] - N^*_t \big\|^2$$
with $N_t= \left(\sum_i r_i(t),\sum_i f_i(t)\right)$ the total populations of rabbits and foxes at time-step $t$ and $N^\star_t$ the exact-process average over $1000$ seeds.

REINFORCE training substantially improved the quality of the compiled model's population mean for $t \in \{1,\ldots, T_\mathrm{RF}\}$, restoring the monotonic relationship between $d_p$ and average total rabbit population in the early phase of the dynamics. Shortly thereafter, more significant departures emerge from the exact process than were evident in the pre-REINFORCE compilation, suggesting that `reward hacking' poses a risk even with models built from a small number of unique kernels.

\subsection{Market simulator}
\label{sec:demos_market}

This demonstration compiles a stochastic program whose target kernel belongs to a real system known only through its recorded history: the day-to-day movement of financial markets. The data is a panel of $N = 14$ daily series covering the major asset classes (US and international stocks, government bonds, corporate credit, exchange rates, gold, oil, copper, and a volatility index) over $4{,}802$ trading days from 2007 to 2026 (Appendix~\ref{app:market_task}). Writing $x_t$ for the panel's state on day $t$, the target is the one-day transition kernel $P(x_t \mid c_t)$, where the context $c_t=(x_{t-B},\ldots,x_{t-1})$ collects the $B$ preceding days. This is target case (c) trajectory data of \S\ref{sec:compile}. History supplies a few thousand (context, next day) pairs, and we cannot query the target kernel at an input of our choosing. Each series contributes one number per trading day, its \emph{move}, the day's change of the log price (Appendix~\ref{app:market_task}). 

\begin{figure*}[!t]
\centering
\includegraphics[width=0.92\textwidth]{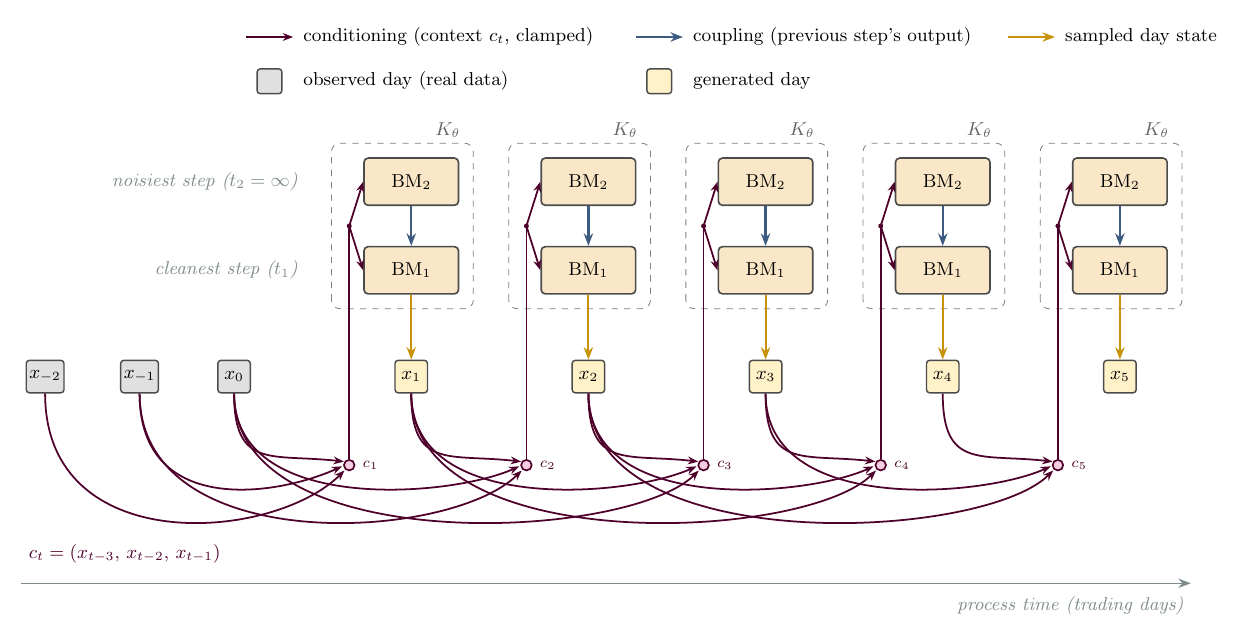}
\caption{\textbf{The market simulator as a directed factor graph.} Grey squares are real observed days seeding the rollout; green squares are generated days. Each dashed box is one application of the one-day kernel $K_\theta$, weight-tied across all days: a two-step conditional TCM (Appendix~\ref{app:tcm}) whose two BMs are both conditioned on the clamped context $c_t$ (pink), the noisier $\mathrm{BM}_2$ drafts the next day's state, and the cleaner $\mathrm{BM}_1$ refines that draft, which it receives through the denoising coupling (blue). Each sampled day (green arrows) enters the contexts of the $B$ following days. The drawing shows $B = 3$ context days for readability; the model uses $B = 5$.}
\label{fig:market_dfg}
\end{figure*}

Markets are close to \emph{efficient}: any dependable pattern in the direction of a day's moves would be traded away, leaving move direction essentially unpredictable from public history~\cite{fama1970efficient}. We therefore do not attempt prediction. What survives efficiency is the joint distribution of the moves, whose robust regularities are known as the stylized facts of asset returns~\cite{cont2001empirical}: same-day moves are strongly correlated across assets, calm and turbulent days cluster in time (volatility clustering), and extreme moves hit many assets at once (markets crash together). A model that reproduces these facts is a \emph{market simulator}: a generator of synthetic market histories, used in finance for stress-testing and risk estimation (Appendix~\ref{app:market_task}). The kernel is accordingly judged on the statistics of its long rollouts, not on any single prediction.

Each day is discretized into $x_t \in \{-1,1\}^{56}$, \(4\) bits per asset class, encoding \(8\) quantiles of the market movement (see Appendix~\ref{app:market_disc}). The kernel is compiled to a conditional \emph{Thermodynamic Consistency Model} (TCM, Appendix~\ref{app:tcm}). A TCM is a chain of two BM kernels, both conditioned on the clamped $B = 5$-day context, in which the first samples a draft of the next day and the second refines it. The simulator is the rollout of Fig.~\ref{fig:market_dfg} (Appendix~\ref{app:market_kernel}): starting from a real $B$-day window, the kernel samples the next day, the sampled day is appended to the context while the oldest day drops out, and the kernel is applied to the updated context. The same kernel, with the same parameters, acts on every day, so the program is a weight-tied chain over process time. We call this a feedback rollout, since after the first $B$ days every context the kernel sees is assembled from its own outputs. 

The target conditional ranges over $2^{56}$ outputs, and the kernels that compile it carry over a thousand hidden spins per step, so its conditional cannot be explicitly normalized, the intractable-model case (2) of \S\ref{sec:compile}. Each step is trained by conditional CD with the adaptive correlation penalty on the $\approx 3{,}300$ recorded pairs (Appendices~\ref{app:market_arch} and~\ref{app:market_vc}). This per-kernel compilation is the first of two training stages.

Here there is no target conditional to compare against to calculate execution error as the target is accessed only through recorded pairs. Therefore, we judge the compiled program from its readouts, under the same conditions that will be used in deployment: 256 rollouts of $1{,}200$ days each, seeded with real held-out windows, after which the kernel sees only its own outputs. Three error terms, one per stylized fact, measure performance: the error in the matrix of same-day sign correlations between assets, the error in the autocorrelation of the panel-wide volatility, defined as the day's move magnitude averaged over the 14 series, and the error in the frequency of joint extreme down-moves. We compare results across different models (single BM, TCM, TCM+REINFORCE) by leveraging two training-free baselines. The first, \emph{IID day-resampling}, draws whole historical days with replacement, which reproduces every same-day statistic of the training data by construction but carries no structure in time. The second, \emph{independent per-series Markov chains}, fits one eight-state transition matrix per series by counting on the training split, which captures the dynamics of each series but renders the series independent of one another. Each error term is then divided by the same error evaluated on the baseline that lacks the corresponding structure, the volatility term by the day-resampler and the correlation and tail terms by the Markov chains, so each baseline scores one on its own term and a ratio below one means the compiled model captures structure that baseline cannot represent. Definitions, the data split, and a sanity check (the simulator must not invent the predictability that market efficiency rules out) are given in Appendix~\ref{app:market_eval}.

\begin{table}[!t]
\caption{\textbf{Stylized-fact errors of long feedback rollouts against held-out real data:} the three normalized error terms and their sum, the composite of Eq.~\eqref{eq:market_composite} (256 rollouts of $1{,}200$ days; lower is better). Each term is normalized by the training-free baseline built to fail it, so that baseline's entry is $1$ by definition. The learned kernels beat both baselines on the composite, and REINFORCE post-training improves it by a further $27\%$.}
\label{tab:market_results}
\footnotesize
\begin{tabular}{lcccc}
\toprule
 & \multicolumn{3}{c}{error terms} & \\
\cmidrule(lr){2-4}
System & corr. & vol. & tail & composite \\
\midrule
indep.\ Markov chains & 1.000 & 0.961 & 1.000 & 2.961 \\
iid day-resampling & 0.176 & 1.000 & 0.475 & 1.651 \\
single conditional BM & 0.239 & 0.483 & 0.095 & 0.818 \\
conditional TCM (stage 1) & 0.242 & 0.378 & 0.047 & 0.667 \\
TCM + REINFORCE (stage 2) & 0.202 & 0.202 & 0.082 & \textbf{0.486} \\
\bottomrule
\end{tabular}
\end{table}

\begin{figure}[t]
\centering
\includegraphics[width=0.9\columnwidth]{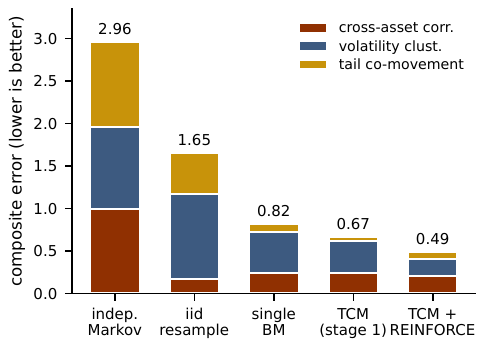}
\caption{\textbf{Composite stylized-fact error, stacked by its three terms} (the numbers of Table~\ref{tab:market_results}; lower is better).}
\label{fig:market_bars}
\end{figure}

\begin{figure*}[!t]
    \centering
    \includegraphics[width=\textwidth]{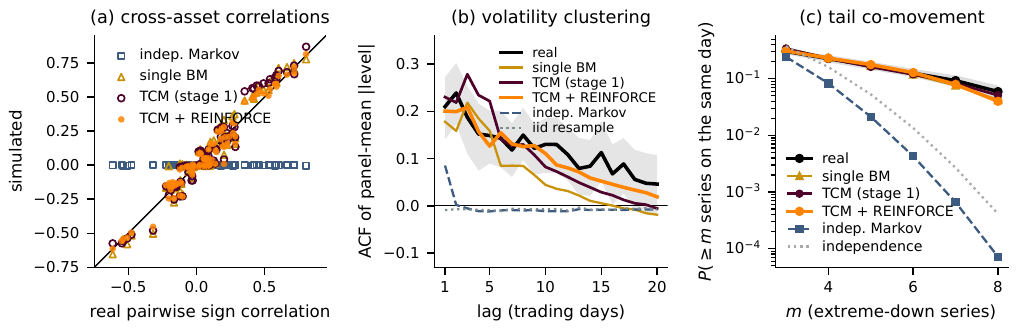}
    \caption{\textbf{Rollout statistics against held-out real data} (256 rollouts of $1{,}200$ days).\textbf{(a)}~Same-day sign correlations of all 91 asset pairs, simulated vs.\ real: the compiled kernels lie on the diagonal, while the independent-Markov baseline carries none of this structure (iid resampling matches this panel by construction and is omitted). \textbf{(b)} Autocorrelation of the panel-wide volatility, the day's move magnitude averaged across the 14 series. Both training-free baselines are flat beyond lag 1. Stage 1 overshoots at short lags and decays too fast while REINFORCE post-training pulls the curve closer to the real one. Grey bands are bias-corrected 95\% block-bootstrap intervals on the real statistics. The errors are given in the 2nd column of table \ref{tab:market_results}. \textbf{(c)}~Probability that at least $m$ of the 14 series post an extreme down-move on the same day (log scale): extreme moves arrive together, orders of magnitude more often than independent series would allow (dotted), and the compiled kernels reproduce this.}
    \label{fig:market_results}
\end{figure*}

The resulting scores are reported in Table~\ref{tab:market_results} and Fig.~\ref{fig:market_bars}. Moreover, Fig.~\ref{fig:market_results} shows the statistics behind them. The compiled two-step TCM reaches a composite error of $0.667$, below both baselines on the composite. Its noisiest step run alone, corresponding to a single conditional BM used as a complete model of the kernel (Appendix~\ref{app:tcm_conditional}), scores $0.818$. This means that the TCM setup (two BMs chained together) provides a meaningful improvement. The TCM also holds its quality far better when trained past its best point (Appendix~\ref{app:market_addl}). The dominant error from the model in stage 1 (Fig.~\ref{fig:market_results}b) is the volatility autocorrelation.

The variationally compiled TCM is optimized to approximate the simulator trajectory wise, but has no guarantees of reproducing the stylized facts. We post-train the whole weight-tied chain on 20-day rollouts with REINFORCE, using the linearized-functional reward of Eq.~\eqref{eq:f_eff} to match a vector of window statistics chosen to parallel the three scored axes (Appendix~\ref{app:market_rg}). The composite error falls from $0.667$ to $0.486$, and most of the error reduction comes from the volatility clustering (Fig.~\ref{fig:market_bars}). The reward sees only short rollouts of the training split, 20 generated days per trajectory, while the evaluation scores feedback rollouts of 1,200 days on held-out data. The improvement therefore reflects generalization along two axes that the training signal does not contain: from 20-day windows to 1,200-day rollouts, and from the training split to held-out data. Since the reward statistics were chosen to parallel the scored terms, the claim is only that the improvement transfers, not that post-training improved anything that wasn't already specified as part of the reward.

\subsection{Meta-EBMs: sampling beyond hardware-native energies}
\label{sec:demos_meta}

In this example, we focus on sampling from stationary distributions corresponding to energies that are not natively expressible by the hardware substrate. We refer to this construction as a meta-EBM. Rather than compiling the target EBM and then sampling it, a meta-EBM is the compilation of the Gibbs sampling program that samples it. The Gibbs sampling algorithm decomposes into local, detailed-balanced update kernels, each kernel being a target conditional that can be compiled to any hardware substrate using variational compilation. In this way, the hardware is able to sample from any family of distributions that arise as stationary measures of compiled Markov chains (the formal construction is collected in Appendix~\ref{app:meta_ebm}).

Take as target EBM any thermodynamic hypergraphical model (THM, \S\ref{sec:ebm}), namely a distribution $p_\vp(x) \propto e^{-E_\vp(x)}$ on $x \in \{-1,+1\}^d$ whose energy is a sum of local terms on an interaction hypergraph, cf.\ Eq.~\eqref{EQN:boltzmann_dist}. We write $\pi \equiv p_\vp$ for this target. Its MCMC sampling algorithm can be unrolled to a DFG. In particular, the elementary Gibbs update resamples a block $A$ from its exact conditional while holding its complement $A^c$ fixed, yielding the target conditional
\begin{equation}\label{eq:meta_conditional}
    p\!\left(x_{A} \mid x_{\partial A}\right)=\frac{e^{-E(x_{A},\, x_{A^c})}}{\sum_{x_A'} e^{-E(x'_A,\, x_{A^c})}}
\end{equation}
where $\partial A$ denotes the Markov blanket of $A$, i.e. the set of nodes that share a hyperedge with a node in $A$. This target conditional is itself a small normalized EBM whose energy is the target energy with $x_{A^c}$ clamped. Each such update is in detailed balance with $\pi$ and thus leaves $\pi$ invariant. The conditional is then compiled to a thermodynamic kernel (\S\ref{sec:compile}), with $x_{A^c}$ playing the role of the clamped input and the resampled block $x_A$ playing the role of the output.

The compiled kernels alone do not yet form a sampler, since we must specify the order in which each kernel is sampled. This choice is the sampling schedule. The schedule does not affect correctness, because each kernel leaves $\pi$ invariant and a composition of $\pi$-invariant kernels is again $\pi$-invariant, so any schedule that keeps visiting every site converges to the correct stationary law under the usual irreducibility and ergodicity conditions. However, one must be careful when resampling sites in parallel, since a simultaneous update is a valid Gibbs move only when neither site lies in the blanket of the other. This can be avoided by using a chromatic sampling schedule. Properly coloring the interaction hypergraph (\S\ref{sec:ebm}) guarantees that same-color sites share no hyperedge and have disjoint blankets, so an entire color class can be sampled in parallel while the update remains exact. Compilation itself is schedule-independent, since each kernel is compiled once from its blanket conditional and the same compiled kernels run under any schedule, making the schedule a runtime choice rather than a property of the couplings.

We demonstrate the meta-EBM construction for a target EBM with three-body interactions compiled to the Z1 substrate, a pairwise Ising lattice with fixed topology
$\mathcal{E}_{\mathrm{Z1}}$,
\begin{equation}\label{eq:native}
  E_{\text{native}}(s)
  \;=\;
  -\sum_i h_i\, s_i \;-\; \tfrac{1}{2}\!\!\sum_{(i,j)\in\mathcal{E}_{\mathrm{Z1}}}\!\! J_{ij}\, s_i s_j.
\end{equation}
The target is
\begin{multline}\label{eq:target3body}
  E(x) = -\sum_n W_n\, x_n
  \;-\; \tfrac{1}{2}\sum_{(m,n)} W_{mn}\, x_m x_n
  \\
  \;-\; \tfrac{1}{3!}\sum_{(n,m,m')} \Wtri_{nmm'}\, x_n x_m x_{m'}.
\end{multline}
The sums run over ordered tuples of distinct sites, with fully symmetric couplings. In \eqref{eq:logit} below, they run over the tuples that contain $n$. The energy contains terms of three orders: a field $W_n$, a pairwise coupling $W_{mn}$, and a three-body coupling $\Wtri_{nmm'}$. The target is non-native to this substrate in the order of interactions and its connectivity. Every coupled pair in $\mathcal{E}_{\mathrm{Z1}}$ sits at a lattice offset $(dx,dy)$ with $dx+dy$ odd, so the lattice is a chessboard, and every Z1-realizable Ising model is an RBM-like \emph{bipartite} model that cannot host a dense interaction graph. In addition to these limitations, every programmable weight in (\ref{eq:native}) has finite dynamic range, $|J_{ij}|\le J_{\max}$ and $|h_i|\le h_{\max}$.

\begin{figure*}[ht]
  \centering
  \includegraphics[width=\textwidth]{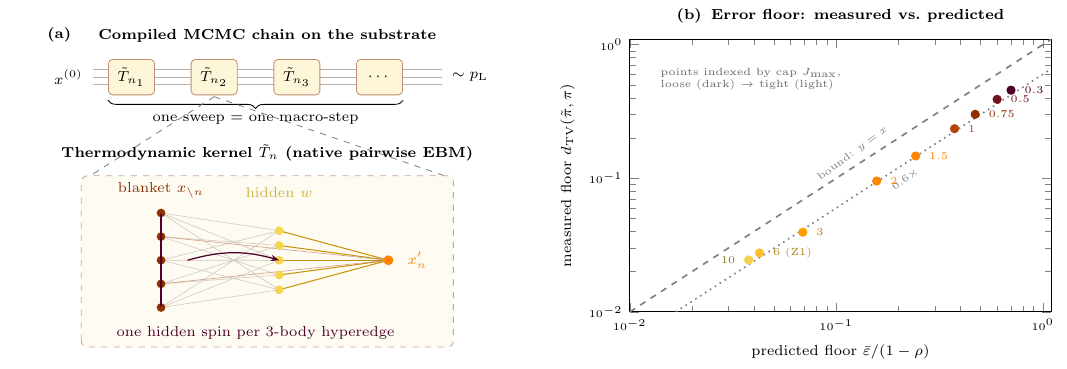}

  \caption{\textbf{A meta-EBM circuit and its un-mitigated error
  floor.} \emph{(a)} The meta-EBM circuit: a sparse three-body Ising target is sampled by single-site Gibbs, whose per-site updates $K_n$ are compiled into thermodynamic kernels $\tilde K_n$ and composed into an MCMC chain. The zoomed-in inset shows one kernel, a native pairwise EBM with a single hidden layer (here up to $n_h=8$), emulating a three-body Gibbs update. \emph{(b)} Measured stationary error of the compiled chain against the parameter-free floor $\bar\varepsilon/(1-\rho)$ predicted by the perturbation bound \eqref{eq:deployed_floor}, one point per coupling cap $J_{\max}$. The measured floor is the large-$t$ plateau of the exact marginal error $\|\tilde\pi_t- \pi_t\|_\TV$, reached within $t\approx3$ sweeps, with each sweep being a pass over all the color classes. Data is from the enumerable $d=12$ target, which has a contraction coefficient $\rho_0\approx 0.28$ (estimated using exact diagonalization). The kernels are compiled with the cap as the only substrate constraint; the connectivity floor of the Z1 lattice (Appendix~\ref{app:meta_kernel}) is not part of this measurement.}
  \label{fig:results_banner}
\end{figure*}

In the example, we run single-site Gibbs sampling, the case of \eqref{eq:meta_conditional} in which every block shrinks to a single node $A=\{n\}$. Each target kernel is then a conditional Bernoulli distribution, $K_n\!\left(x_n \,\middle|\, x_{\partial n}\right)\propto e^{\theta_n\!\left(x\right)\, x_n}$,
whose half log-odds follows from differentiating \eqref{eq:target3body} at
$x_n = \pm 1$,
\begin{multline}\label{eq:logit}
  \theta_n(x)
  \;=\;
  W_n
  \;+\; \tfrac{1}{2}\sum_{(m,n)} W_{mn}\, x_m
  \\
  \;+\; \tfrac{1}{3!}\sum_{(n,m,m')} \Wtri_{nmm'}\, x_m x_{m'}.
\end{multline}
The substrate kernel realizes this conditional with a pairwise Ising energy on the blanket inputs $x$, hidden spins $w \in \{-1,+1\}^{n_h}$, and one output $y$. Marginalizing the hidden layer gives a logistic regression in which the forward half log-odds carries one additive softplus feature per hidden spin,
\begin{multline}\label{eq:kernel_logit}
  \theta_y(x)
  \;=\;
  \underbrace{J_{xy}^\top x + h_y}_{\text{affine in } x}
  \\
  + \sum_{a=1}^{n_h}
  \tfrac{1}{2}\Bigl[
    \softplus \bigl(-2(\alpha_a^\top x - \beta_a)\bigr)
    -
    \softplus \bigl(-2(\alpha_a^\top x + \beta_a)\bigr)
  \Bigr],
\end{multline}
where $\alpha_a$ is the input-to-hidden coupling of spin $a$ and $\beta_a$ its coupling to the output. The full derivation is given in Appendix~\ref{app:meta_kernel}. This is the pattern of the closed-form PNOT example of Appendix~\ref{sec:pnot}, in which affine log-odds require no hidden spins while each nonlinearity requires one. Matching \eqref{eq:kernel_logit} to the target logit \eqref{eq:logit} then proceeds term by term from lowest to highest order. The constant $W_n$ is absorbed exactly by the output bias $h_y$ and the linear coefficients $\tfrac{1}{2}W_{mn}$ by the direct input--output couplings $J_{xy}$, with no hidden spins involved. Each bilinear term is reproduced by one hidden spin acting as a soft product gate, obtained by aligning its input projection $\alpha_a$ with the pair $(x_m, x_{m'})$ and driving the coupling large, which sharpens the softplus difference in \eqref{eq:kernel_logit} toward a ReLU. The contribution of the spin then converges to the product $x_m x_{m'}$, with a residual that decays exponentially in the coupling magnitude. Distinct hidden spins do not interact and their contributions add by superposition, so one hidden spin per three-body hyperedge touching site $n$ reproduces the entire conditional \eqref{eq:logit}, up to $n_h = 8$ per site for the sparse target used here (Fig.~\ref{fig:results_banner}a). On a fully connected substrate with unbounded couplings the compilation is exact and the loss infimum is zero.

The Z1 hardware substrate introduces two error sources. One is the finite dynamic range of the coupling $|J| \le J_{\max}$. Here we treat this value as a free parameter and sweep different choices of it. The other source of error is the hardware topology. On the bipartite chessboard, a spin that is simultaneously a pairwise neighbor and a three-body partner of site $n$ cannot keep both its affine edge and its bilinear edge, leaving an irreducible per-site residual that persists at any value of the cap (the placement argument is given in Appendix~\ref{app:meta_kernel}). The following numerics isolate the first source: the kernels are compiled over a fully connected spin set with only the imposed dynamic-range constraint, so $J_{\max}$ is the single tunable knob of the comparison, and every deviation from ideal behavior is attributable to the cap. The connectivity residual would also contribute but is not simulated here.

We consider a three-body Ising target on $d = 12$ spins, small enough that the full $2^{12}$-state distribution is enumerable and every quantity can be computed exactly. 18 pairwise couplings and 20 three-body hyperedges are drawn uniformly at random from the pairs and triples of the 12 sites, with the fields and coupling magnitudes sampled i.i.d.\ from $\mathcal{N}(0,\,0.6^2)$. The per-site kernels are compiled variationally under a uniform input measure $\mu_\ell = \mathcal{U}$, with no context matching.

We sweep the cap from $J_{\max}=10$ down to $J_{\max}=0.3$, where the coupling cap badly clips the conditional logits, and at each setting we run the compiled chain to stationarity and measure its exact marginal error against the ideal chain for the logical target $\pL \equiv p_\vp$ (measurement definitions are collected in Appendix~\ref{app:meta_numerics}). Figure~\ref{fig:results_banner}b shows the result. For each compiled chain, the layer-wise error saturates within $t \approx 3$ sweeps and stays flat thereafter, instead of growing linearly in depth as the naive budget of \eqref{eq:deployed_unrolled} would allow. The plateau itself matches a parameter-free prediction, the floor
\begin{equation}\label{eq:meta_floor}
    \tilde{\delta}_t \leq \frac{\bar\varepsilon}{1-\rho_0}
\end{equation}
of \eqref{eq:eta_floor}, with $\tilde\delta_t \equiv \|\tilde{q}_t - q_t\|_\TV$ denoting the marginal error after the $t$th Gibbs sweep. Here the per-step residual $\bar\varepsilon$ is the worst-case conditional total variation of the compiled kernels, $\bar\varepsilon = \max_\ell \eta_\ell$, and $\rho_0 \equiv \rho(P)$ is the contraction coefficient of the ideal sweep $P$, defined in Eq.~\eqref{eq:dobrushin}. The measured stationary error tracks this prediction across the full cap sweep, spanning over a decade of error, from $0.46$ at the most aggressive cap down to $2.4\times10^{-2}$ at $J_{\max}=10$, at fixed $\rho_0 = 0.28$. The measured floor sits consistently at $\approx 0.6\times$ the bound, which makes the bound a reasonably tight upper envelope rather than an order-of-magnitude estimate.

The same sweep exposes the degradation mechanism analyzed in Appendix~\ref{app:inputs_model}. As $J_{\max}$ tightens, the floor climbs not because the chain mixes worse, since $\rho$ is fixed by the target and not by the substrate, but because the per-step residual $\bar\varepsilon$ rises as the cap clips sharper and sharper logits. The cap thus traces out the entire error budget of the meta-EBM along a single axis, and the perturbation bound converts the per-step budget into a bound on the stationary sampling bias. At the maximum cap $J_{\max}=10$ the observable-level picture agrees, with per-site expectations of the compiled chain tracking ideal Gibbs (mean single-site error $\approx 5\times10^{-3}$) and relaxing onto the exact stationary values within roughly three sweeps. Full per-site detail, including uniformity across equilibrium occupancies, is given in Appendix~\ref{app:meta_numerics}.

\subsection{Gaussian stochastic circuits}
\label{sec:demos_gaussian}

Linear-Gaussian models appear throughout applied probability, from spatial statistics~\cite{rue2005} and Kalman filtering~\cite{kalman1960} to Bayesian optimization~\cite{shahriari2016} and the simulation of stochastic differential equations~\cite{sarkka2019}. Here the target is a Gaussian stochastic circuit, a stochastic circuit in which every gate is a linear-Gaussian conditional, so the joint law of all wires is multivariate Gaussian. The graphical models literature calls this object a Gaussian Bayesian network \cite{shachter1989gaussian, koller2009probabilistic}. We compile it to the thermodynamic hardware of \S\ref{sec:hardware}. Since the hardware operates on binary variables, we discretize each real-valued variable of the process and encode it in a group of spins. The discretization couples all pairs of spins that encode interacting variables, so the coupling graph of the compiled energy is denser than the hardware lattice, and running the program requires a minor embedding~\cite{choi2008minor, choi2011}, in which each logical spin is carried by a chain of strongly coupled p-bits.

\begin{figure*}[t!]
\centering
\begin{minipage}[c]{0.22\textwidth}
\centering
\includegraphics[height=4.05in]{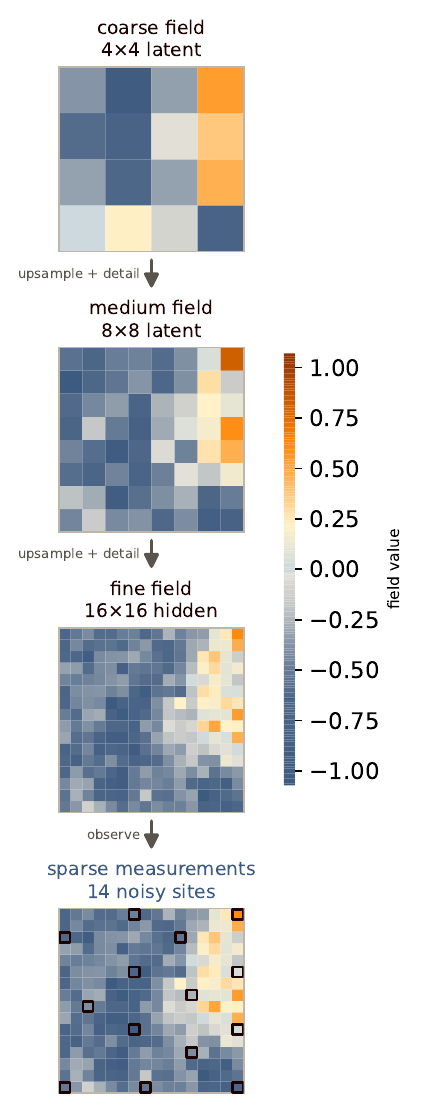}
\end{minipage}\hfill
\begin{minipage}[c]{0.74\textwidth}
\centering
\includegraphics[height=4.05in]{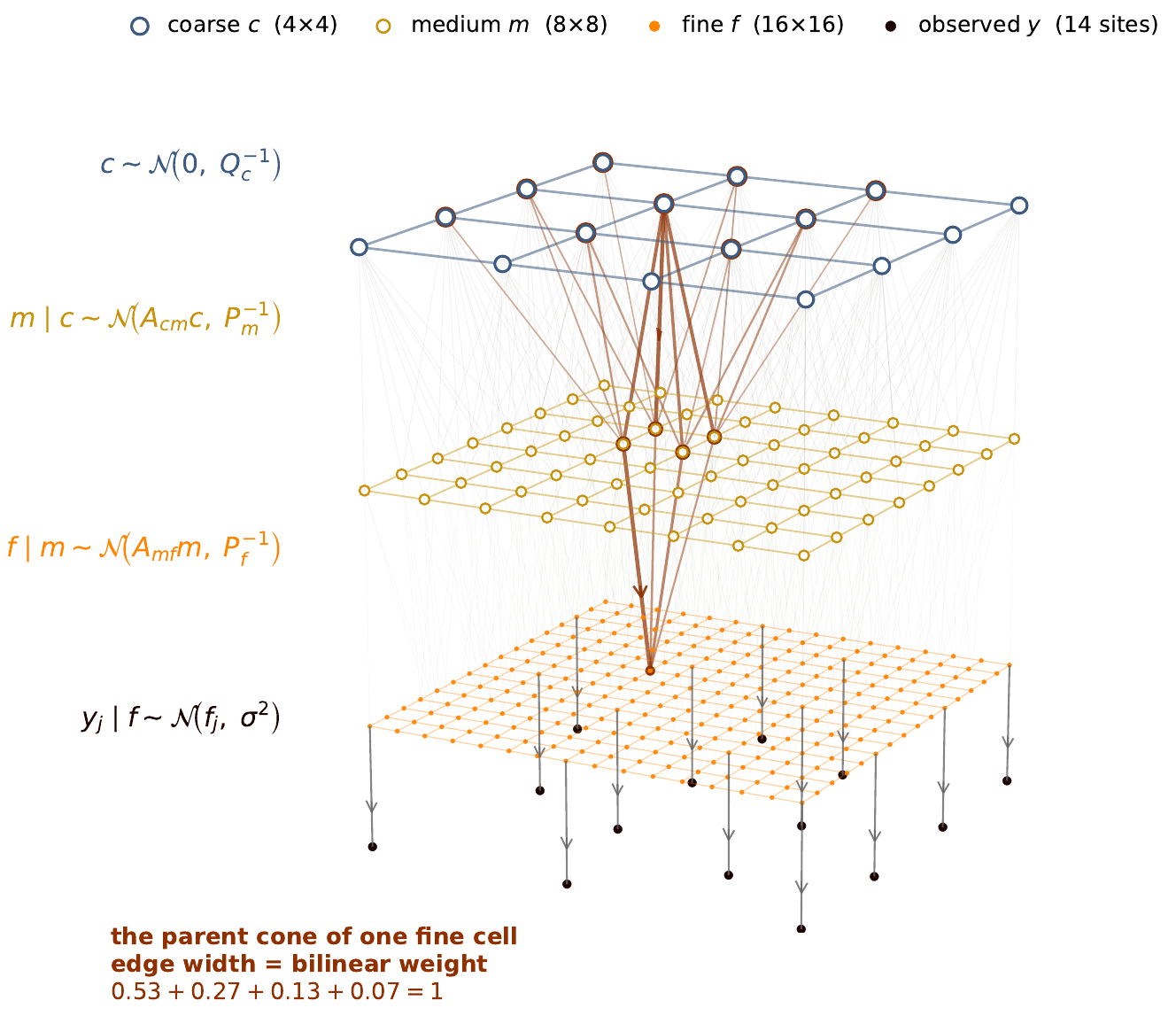}
\end{minipage}
\caption{\textbf{A directed Gaussian stochastic circuit.} Left: one draw from Eq.~\eqref{eq:gp_model}. The coarse field is upsampled and detailed into the medium field, the medium field into the fine field, and the fine field is measured at $14$ noisy sites. Each arrow is one factor of the circuit. Right: the same circuit as a directed factor graph. Every cell of every layer is a variable: open circles are latents, filled dark circles the $14$ observed sites. In-plane lattice edges carry the Gaussian Markov random field precisions $Q_c, P_m, P_f$. Edges between planes carry the factors of Eq.~\eqref{eq:gp_model}, one kernel per gap. The highlighted cone is the full ancestry of one fine cell: its four bilinear parents in $m$ (edge width $=$ interpolation weight, summing to $1$) and their parents in $c$. The demonstration clamps sets of planes of this graph and thermalizes the rest.}
\label{fig:gp_structure}
\end{figure*}

The demonstration is a sequential Bayesian experimental design task~\cite{chaloner1995}. We want to determine an unknown field, a real-valued random vector indexed by the cells of a two-dimensional grid. The field's values are drawn from a known Gaussian prior and accessible only through noisy pointwise measurements, one cell at a time. The goal is to reconstruct the field from as few measurements as possible, so the choice of where to measure next matters as much as the inference itself. At every step the posterior over the field given the measurements taken so far is computed, and the next measurement is taken at the cell of largest posterior variance, the acquisition rule known as uncertainty sampling~\cite{mackay1992}. The prior compiles to a thermodynamic hypergraphical model (THM) and each conditioning task, one per step of the loop, is just a different clamping pattern of its spins. The prior over the field is hierarchical, generated from coarse to fine. A coarse field $c$ on a $4\times4$ grid is drawn first, it is upsampled to an $8\times8$ grid and correlated Gaussian detail is added on top to give the medium field $m$, and the medium field is upsampled and detailed in the same way to give the fine field $f$ on the $16\times16$ grid, which is the field the measurements probe. Written as a directed factor graph, the model has one Gaussian factor per layer,
\begin{equation}\label{eq:gp_model}
\begin{aligned}
  c &\sim \mathcal{N}(0,\,Q_c^{-1}),\\
  m\mid c &\sim \mathcal{N}(A_{cm}\,c,\;P_m^{-1}),\\
  f\mid m &\sim \mathcal{N}(A_{mf}\,m,\;P_f^{-1}),
\end{aligned}
\end{equation}
where each line of Eq.~\eqref{eq:gp_model} is one factor of the DFG, the first the marginal law of the coarse field and the other two the conditional law of a layer given its parent. Each Gaussian is parametrized by its \emph{precision}, the inverse of its covariance: a zero-mean Gaussian with precision $Q$ has density $p(x)\propto\exp(-\tfrac{1}{2}x^TQx)$, so the entries of the precision are the couplings of a quadratic energy: $Q_{ij}=0$ exactly when $x_i$ and $x_j$ are conditionally independent given all the other variables. In the case considered here, we denote the precisions of each factor with $Q_c, P_m, P_f$. Flattening the grids gives $c\in\mathbb{R}^{16}$, $m\in\mathbb{R}^{64}$ and $f\in\mathbb{R}^{256}$. The upsampling is then carried out by the bilinear interpolation operators $A_{cm}=W_{4\to8}\otimes W_{4\to8}$ and $A_{mf}=W_{8\to16}\otimes W_{8\to16}$, where $W_{L\to2L}$ is the one-dimensional linear-interpolation matrix from a grid of side $L$ to one of side $2L$. In this way, the mean of every cell is a convex combination of at most four parent cells in the layer above, with weights summing to one (the highlighted cone of Fig.~\ref{fig:gp_structure}). For our experiment we fix,
\begin{equation}\label{eq:gp_precisions}
\begin{aligned}
  Q_c &= 0.8\,(\Delta_4+0.50^2 I),\\
  P_m &= 2.6\,(\Delta_8+1.4^2 I),\\
  P_f &= 6.5\,(\Delta_{16}+1.8^2 I),
\end{aligned}
\end{equation}
with $\Delta_L$ the graph Laplacian on the $L\times L$ grid; the mass term $\kappa^2 I$ makes each precision positive definite, so every factor in Eq.~\eqref{eq:gp_model} is a normalizable Gaussian. A precision of this form, $\Delta_L+\kappa^2 I$, is the Gaussian-Markov discretization of a Mat\'ern field in its stochastic-partial-differential-equation representation~\cite{lindgren2011}, so the compiled THM encodes a genuine Gaussian-process prior rather than an ad-hoc lattice model. Figure~\ref{fig:gp_structure} shows one draw from the prior and the corresponding DFG.

The target conditional of each layer is itself an EBM with a quadratic energy, so its kernel's couplings and fields are read off the interpolation operators $A$ and the precisions $Q_c, P_m, P_f$ of Eq.~\eqref{eq:gp_model} and the per-factor compilation error $\varepsilon_\ell$ of Sec.~\ref{sec:compile} is exactly zero. Because factor energies add, the negative logarithm of the joint density is, up to a constant,
\begin{equation}\label{eq:gp_energy}
\begin{aligned}
  E_0(z) ={}& \tfrac{1}{2}\,c^T Q_c\, c
  + \tfrac{1}{2}\,(m - A_{cm}c)^T P_m\,(m - A_{cm}c)\\
  &+ \tfrac{1}{2}\,(f - A_{mf}m)^T P_f\,(f - A_{mf}m),
\end{aligned}
\end{equation}
with $z=(c,m,f)$. Expanding the squares and collecting the terms quadratic in $z$ gives the block precision
\begin{widetext}
\begin{equation}\label{eq:gp_blocks}
\Lambda_0=
\begin{pmatrix}
Q_c+A_{cm}^TP_mA_{cm} & -A_{cm}^TP_m & 0\\[2pt]
-P_mA_{cm} & P_m+A_{mf}^TP_fA_{mf} & -A_{mf}^TP_f\\[2pt]
0 & -P_fA_{mf} & P_f
\end{pmatrix}.
\end{equation}
\end{widetext}
so $E_0(z)=\tfrac{1}{2}z^T\Lambda_0 z$ and the joint is a zero-mean Gaussian whose precision is $\Lambda_0$. On the hardware each real value is held by a signed bundle, a fixed-point register of binary spins with power-of-two weights, and since the encoding is linear the couplings and fields of the pairwise Ising THM follow from Eq.~\eqref{eq:gp_blocks} directly. After compilation to a THM, one can run the process forward, by clamping the coarse plane, or run it backward, by clamping the measured cells, with the same energy programming (Appendix~\ref{app:gp_circuit}, Fig.~\ref{fig:gp_circuit}). Perturb-and-MAP~\cite{papandreou2010} also draws Gaussian posterior fields, but offers no backward inverse of this kind.

A noisy fine-field measurement modifies the programmed THM. Each measurement is a noisy read of one fine cell, $y_j\mid f \sim \mathcal{N}(f_j,\,\sigma^2)$, shown in Fig.~\ref{fig:gp_structure} as the observation factor. With $H_z$ the binary matrix selecting the observed cells, the negative log-likelihood of the measurements adds a quadratic penalty to the energy,
\begin{equation}\label{eq:gp_clamp}
  E_{\mathrm{obs}}(z) = E_0(z) + \frac{1}{2\sigma^2}\,\lVert H_z z - y\rVert^2.
\end{equation}
All layers are linear and Gaussian, so the clamped energy is again quadratic in $z$ and the posterior over the full hierarchy is Gaussian. Collecting the quadratic and linear terms of Eq.~\eqref{eq:gp_clamp} gives
\begin{equation}\label{eq:gp_posterior}
\begin{aligned}
  \Lambda_{\mathrm{post}} &= \Lambda_0 + \sigma^{-2} H_z^T H_z,\\
  \mu_{\mathrm{post}} &= \sigma^{-2}\,\Lambda_{\mathrm{post}}^{-1} H_z^T y.
\end{aligned}
\end{equation}
Each measurement enters as one bias shift and one diagonal coupling change of the underlying THM. This closed-form conditional of the assumed model is the reference posterior for the accuracy comparisons below.
We set $\sigma=0.30\,\mathrm{median}_j\sqrt{(\Lambda_0^{-1})_{f_jf_j}}=0.2313$, large enough that measured cells keep visible uncertainty and the $14$ sites leave fine-field regions unconstrained.

The design loop turns that single-clamp update into a sequence of inferences. Each step thermalizes the clamped THM by chromatic block-Gibbs sampling ($120$ warm-up and $300$ measured sweeps across $12$ Gibbs chains), and then takes the next measurement at the fine cell of largest posterior variance. The acquisition signal, the per-cell posterior variance, is read from the same samples as the posterior mean, so uncertainty sampling~\cite{mackay1992} costs only the step's one relaxation, while the classical implementation of the same rule needs the diagonal of $\Lambda_{\mathrm{post}}^{-1}$, obtained from a sparse-Cholesky selected inversion, at every step~\cite{rue2005}.

\begin{figure}[tbp]
\centering
\includegraphics[width=\columnwidth]{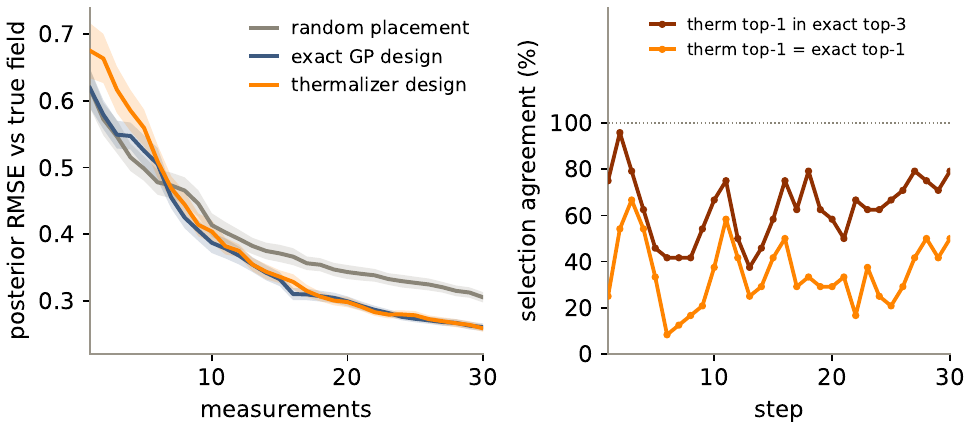}
\caption{\textbf{Design-loop performance of the compiled program.} Left: posterior RMSE against the true fine field versus the number of measurements, for the compiled THM, exact-posterior, and random-placement design loops, mean $\pm$ standard error over 24 fields. Right: how often the compiled THM's chosen site is the exact loop's top pick, and within its top three, per step.}
\label{fig:gp_active}
\end{figure}

The test is whether a variance map read from samples steers the loop to the same cells as the exact diagonal. We run 30-measurement loops on 24 hidden fields, comparing the compiled THM against the same acquisition rule evaluated on the exact posterior~\cite{rue2005} and against random placement of the measurements (Fig.~\ref{fig:gp_active}). The RMSE against the true field falls to $0.259\pm0.005$ for the compiled THM and $0.260\pm0.006$ for the exact loop, versus $0.305\pm0.008$ for random placement. Agreement on individual picks is weaker: the compiled THM selects the exact loop's top cell 35\% of the time and a top-three cell 63\% of the time. However, on a near-flat variance map the argmax is a tie decided by Monte Carlo noise, and the matched RMSE curves show that those ties cost nothing. Appendix~\ref{app:gp_loop} compares the sampled posterior against the exact one directly and gives the full error budget.

The compiled THM minor-embeds on the Z1 lattice (Fig.~\ref{fig:gp_embedding}); chain statistics and the energy projection are given in Appendix~\ref{app:gp_energy}. The eight-bit embedding runs to roughly $14{,}000$ p-bits, so a $250{,}000$-p-bit Z1 array reaches a fine grid near $32\times32$ to $48\times48$.

\begin{figure}[tbp]
\centering
\includegraphics[width=\columnwidth]{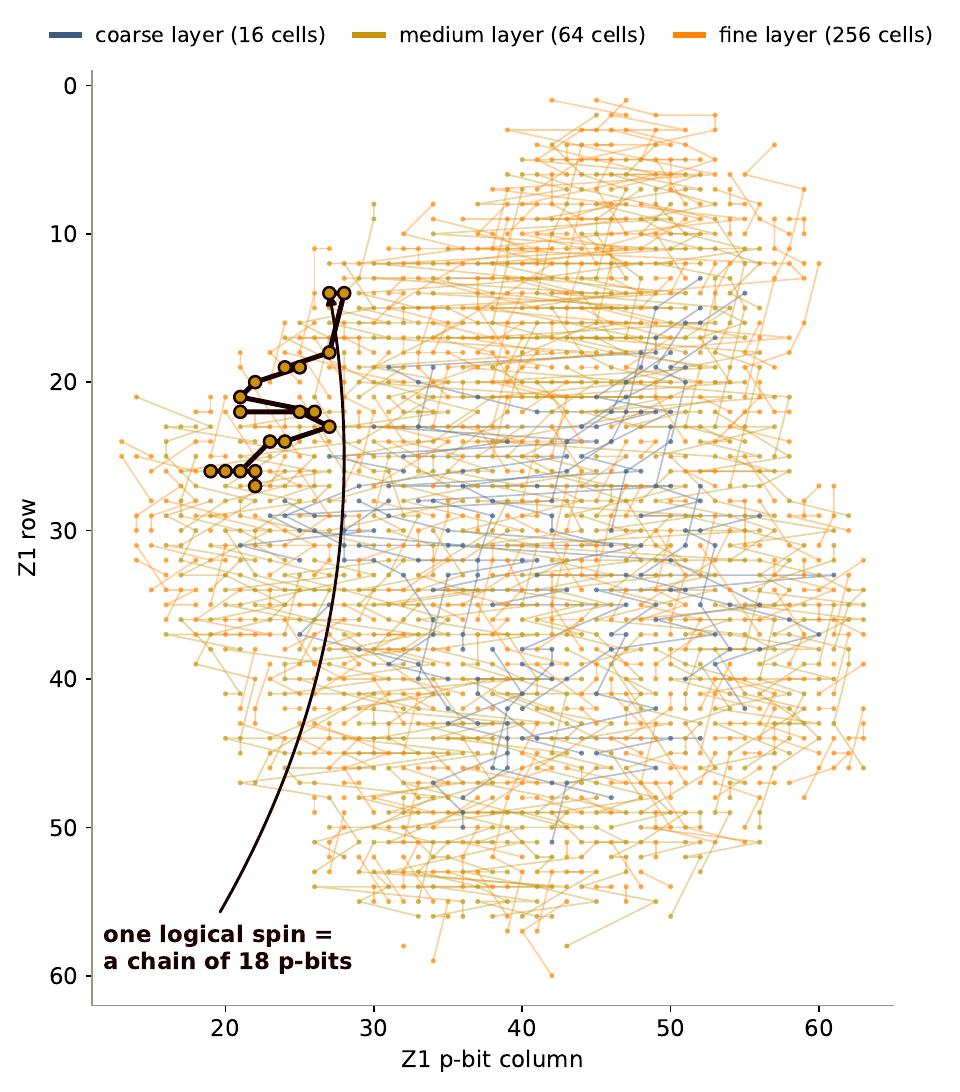}
\caption{\textbf{The composed program on the chip.} The three-layer
logical graph minor-embedded on the Z1 p-bit lattice, chains colored by
layer. One highlighted chain carries a single logical spin on $18$ p-bits.}
\label{fig:gp_embedding}
\end{figure}


\section{Discussion and conclusions}
\label{sec:discuss}

This manuscript introduced variational compilation as a way to turn directed stochastic processes into kernels that can be run on thermodynamic hardware. The starting point is a target process represented as a directed factor graph, or parametrized stochastic circuit~(\S\ref{sec:dfg})~\cite{frey2003unifying,lockwood2026torx}, whose factors are local conditional distributions. Each factor is replaced by a thermodynamic kernel, an energy-based object whose conditional law is sampled by clamping one set of variables and thermalizing the rest~(\S\ref{sec:compile}). This kernel is the compiler intermediate representation between directed probabilistic computation and the undirected equilibrium distributions realized by the substrate. Changing the clamping pattern exposes different uses of the same energy, including forward execution, backward or posterior sampling, and conditioning on observations~(Eqs.~\eqref{eq:forward} and \eqref{eq:backward}, Appendix~\ref{sec:backward-compilation}). The demonstrations show what kinds of structure fit this template: random walks~(\S\ref{sec:demos_random_walk}), interacting particle systems~(\S\ref{subsec:Eco}), a market simulator whose target kernel is known only through trajectory data~(\S\ref{sec:demos_market}), compiled Gibbs samplers for non-native energies~(\S\ref{sec:demos_meta}), and Gaussian graphical models~(\S\ref{sec:demos_gaussian}). They also show how input-distribution choice~(\S\ref{sec:input_distributions}), context matching, and trajectory-level post-training~(\S\ref{sec:reinforce}) enter once individually compiled kernels are composed into a program. Processes with comparable local conditional or graphical structure can be decomposed and compiled in this style.

The demonstrations should be read as evidence for the compilation framework and its error-mitigation tools, not as an explicit end-to-end demonstration of hardware energy efficiency. Every compiled program remains approximate, because finite coupling ranges, sparse topology, optimization residuals, and finite thermalization all contribute residual error. The analysis in this paper gives ways to account for these errors, from per-kernel residuals to trajectory-level and readout-level quantities~(Eq.~\eqref{eq:mitigation_chain}). Context matching, trajectory-level post-training, and hardware-in-the-loop refinement~\cite{kaiser2022insitu} can reduce the errors seen by the deployed program, but they do not remove the need to validate each application against its own task-level observable. Path-space KL is a useful conservative diagnostic for composed stochastic programs, while many downstream tasks are ultimately judged by readouts. For this reason, simulator results, calibrated hardware-model projections~\cite{jelincic2025dtm}, and physical hardware measurements should remain distinct when interpreting the evidence.

An important future direction is to find useful decompositions of other stochastic processes into local conditionals that can be compiled and run in this framework. A good decomposition must balance two opposing aspects: atomic kernels are easier to compile, reuse, and validate, while larger chain or block kernels can absorb more trajectory-level error but are harder to train and may require stronger mixing regularization~\cite{jelincic2025dtm}. Choosing this kernel size should eventually depend on the target process, the substrate connectivity, the coupling caps, the finite thermalization horizon, and the length of the composed trajectory.

A possibility open for future exploration relates to the discovery of novel stochastic algorithms. Instead of trying to decompose existing stochastic processes, one can start from atomic kernels implementing simple local conditionals and compose them to obtain a more complicated stochastic process. This is the approach that the \texttt{torx} framework \cite{lockwood2026torx} is spearheading. The precedent for this approach is quantum computing, where it has been applied successfully in the discovery of quantum algorithms. This has led to some of the most impactful algorithmic advances in the field \cite{kitaev1995quantum,shor1997,grover1996}.

Another direction to explore is connectivity-aware compilation, where the placement of input, hidden, and output spins on the actual hardware graph~\cite{choi2008minor,choi2011} is optimized together with the kernel parameters so that topology residuals can be measured and reduced rather than treated after the fact. The same study should help identify which kernels deserve pre-compiled libraries~(\S\ref{sec:libraries}) and which generic input distributions transfer well before context matching adapts them to a particular program. There is also room for training methods between per-kernel matching and leaf-only REINFORCE, aimed at reducing task-level error without relying only on high-variance end-to-end objectives. 

Larger hardware validation should separate compilation residuals from execution error, finite thermalization, calibration drift, and readout overhead. We hope \texttt{thermalizers} \cite{thermalizer} provides a practical place to explore these questions: express a structured stochastic process in \texttt{torx}~\cite{lockwood2026torx}, compile its local conditionals to a thermodynamic hypergraphical model with \texttt{thermalizers}, and sample them via \texttt{thrml}~\cite{thrml} to obtain thermodynamically computed results.

\begin{acknowledgments}
We acknowledge Arthur Parzygnat for helpful discussions around Markov chains. The authors thank Geremia Massarelli for careful review and comments on the manuscript. D.S. thanks Rida Hamadani for helpful discussions.
\end{acknowledgments}

\bibliography{biblio}

\vspace{1cm}

\appendix
\onecolumngrid

\section{Glossary}
Table~\ref{tab:glossary} collects the recurring objects and acronyms for reference.
\begin{table*}[h!]
    \caption{Glossary of the main objects, acronyms, and methods used in this manuscript.}
    \label{tab:glossary}
    \footnotesize
    \begin{tabular}{ll}
        \toprule
        \textbf{Term} & \textbf{Definition} \\
        \midrule
        EBM & \DefCell{Energy-based model: a Boltzmann distribution \(p_\vp(s)\propto e^{-E_\vp(s)}\).} \\
        THM & \DefCell{Thermodynamic hypergraphical model: an EBM factorized over a hypergraph.} \\
        DFG & \DefCell{Directed factor graph: a DAG of Markov kernels representing a stochastic program.} \\
        PSC & \DefCell{Parametrized stochastic circuit: a DFG organized into layers of gates~\cite{lockwood2026torx}.} \\
        Thermodynamic kernel & \DefCell{The compilation primitive: the forward conditional \(\Pf(y \mid x; \vp)\) of a THM, Eq.~\eqref{eq:forward}, expressed through the affinity \(\tk_\vp(x,y)=\sum_w e^{-E_\vp(x,w,y)}\).} \\
        Forward/backward cond. & \DefCell{The two normalizations of \(\tk_\vp\): \(\Pf(y\mid x)\) (clamp \(x\), Eq.~\eqref{eq:forward}) and \(\Pb(x\mid y)\) (clamp \(y\), Eq.~\eqref{eq:backward}).} \\
        Block Gibbs & \DefCell{Resampling a block of spins from its exact conditional given the rest; a chromatic schedule updates conditionally independent blocks in parallel.} \\
        CD & \DefCell{Contrastive divergence: the two-phase form \eqref{eq:cd_identity} of the compilation gradient, estimated by Gibbs sampling.} \\
        MET & \DefCell{Mixing expressivity tradeoff: sharper kernels thermalize slower, biasing sample-based gradients~\cite{jelincic2025dtm}.} \\
        Variational compilation & \DefCell{Fitting each DFG factor to a thermodynamic kernel by minimizing the trajectory-level KL, Eq.~\eqref{eq:compile_loss}, which decomposes into per-factor terms.} \\
        Input distribution \(\mu_\ell\) & \DefCell{The distribution of the clamped inputs under which kernel \(\ell\) is trained: the target marginal \(q_{\ell-1}\) (default), the model marginal \(\tilde q_{\ell-1}\), or a generic choice such as uniform (libraries), \S\ref{sec:input_distributions}.} \\
        Context matching & \DefCell{Re-optimizing a compiled kernel under a new input distribution: the target's marginal (\emph{target-context}) or the compiled circuit's own marginal (\emph{model-context}), \S\ref{sec:input_distributions}.} \\
        ATK & \DefCell{Analytic thermodynamic kernel: an enumerated core plus an analytically summed periphery, giving an exactly computable and differentiable conditional (Appendix~\ref{app:atk}).} \\
        Meta-EBM & \DefCell{A variationally compiled Gibbs sampling circuit: samples a non-native target EBM by compiling the block conditionals of a (chromatic) Gibbs schedule for it (\S\ref{sec:demos_meta}).} \\
        TCM & \DefCell{Thermodynamic consistency model: a chain of BM kernels in which every step maps its input directly to an approximation of clean data (Appendix~\ref{app:tcm}); its conditional variant is the kernel compiled in the market-simulator demonstration (\S\ref{sec:demos_market}).} \\
        p-bit & \DefCell{Probabilistic bit: a binary stochastic hardware unit.} \\
        Z1 & \DefCell{Reference hardware architecture~\cite{jelincic2025dtm}: capped pairwise Ising on a fixed planar lattice.} \\
        \bottomrule
    \end{tabular}
\end{table*}

\section{Example energy breakdown of Z1 hardware operation}\label{sec:z1-energy}

In this section, we present an energy breakdown of the Gibbs-update, read, and write operations at the level of each pBIT node, in the Z1 hardware of \S\ref{sec:hardware}. This analysis updates the high-level estimates reported in \cite{jelincic2025dtm} by incorporating the details of the Z1 hardware architecture and SPICE-based~\cite{nagel1973spice} energy estimates for its individual circuit modules. Note that the pBIT design has been further improved from \cite{jelincic2025dtm} to now support faster decorrealtion times across transistor variability supporting 50MHz Gibbs update speed.

\begin{table}[ht]
    \centering
    \begin{tabular}{lcc}
        \hline\hline
        Contribution &  &\\
        \hline\hline
        \multicolumn{3}{l}{\textit{(i) $\Esamp/KN$: sampling, per pBIT node per Gibbs cycle at $50$\,MHz} } \\
        \hline
        \textbf{Total = $\Erng$ + $\Eclk$ + $\Emac$ +  $\Enb$ + $\Eanalog$ + $\Esram$}                             & $\mathbf{\:=7.09 \; fJ}$ &  \\
        \hline\hline
        \noalign{\vskip 2ex}
        \multicolumn{3}{l}{\textit{(ii) $\Eread$ per pBIT node} \quad } \\
        \hline
        \textbf{Total = $E_{\textrm{route,buff}} +E_{\textrm{I/O}}$ }          & \multicolumn{1}{c}{$\mathbf{= 1.692 \; pJ}$} \\
        \hline\hline

        \noalign{\vskip 2ex}
        \multicolumn{3}{l}{\textit{(iii) $\Ewrite$ per pBIT node}} \\
        \hline
        \textbf{Total= $E_{\textrm{route,buff}} +E_{\textrm{I/O}}$ =}                & \multicolumn{1}{c}{$\mathbf{=153.6 \;pJ}$} \\
        \hline\hline
    \end{tabular}
    \caption{\textbf{Z1 average energy breakdown for a single pBIT node.} (i) Energy to implement a Gibbs update, i.e. compute the bias (MAC), generate a pBIT sample and communicate to its neighbors. (ii) Reading a pBIT state off-chip and (iii) Flashing (writing) the coupling weights, biases and/or clamp state. Both the read and write operations happen using a serial interface, whereas the Gibbs update clock can be set according to the autocorrelation times of the pBIT (at max. $50\,$MHz).}
    \label{tab:budget}
\end{table}

Below we describe the different contributors to the average energy consumption:
\begin{itemize}
    \item $\Erng$ : The pBIT signal chain. 
    \item $\Eclk$ : Broadcasting the clock signal from its source to the pBIT.
    \item $\Emac$ : The analog multiply-accumulate (MAC) circuitry used to compute the bias for the pBIT.
    \item $\Enb$ : Broadcasting the updated state to the 16 coupled neighbors.
    
    \item $\Eanalog$ : The analog supporting circuitry, such as DACs, temperature sensor circuits, voltage regulators etc. This shared by all the pBITs in the hardware.
    
    \item $\Esram$ : Each pBIT node has accompanied SRAM cells to store the coupling weights, biases and other necessary bits for orchestrating the different operations of the hardware.
    
    \item $\Eread$ : Reading one pbit state to the edge of the chip boundary to be read by a digital host, excluding the host's contributions.
    
    \item $\Ewrite$ : Writing all the SRAM bits local to the pBIT node from the digital host, excluding the host's contributions.
\end{itemize}

In the current Z1 hardware, read and write operations proceed across each core before the data are serialized and transmitted through the I/O interface. This introduces additional overhead because some of the non-algorithmic pBIT nodes must also be accessed. These limitations are expected to be reduced through architectural optimizations and additional features in the next Z1 revision.

To separate the intrinsic sampling cost from implementation-specific I/O overheads, we consider an idealized Z1 interface with node level access for read and write, while retaining the measured Z1 sampling and read/write energies per node. For the same $N=4900$, $N_{\mathrm{data}}=834$, and $K=250$ Denoising Thermodynamic Model (DTM) workload with $T$ diffusion layers, considered in Ref.~\cite{jelincic2025dtm}, this gives $15.6T$\,nJ for inference ($50$\,MHz Gibbs cycling frequency), compared to $1.6T$\,nJ coarse estimate. Thus, the refined estimate is higher by a factor of 9.8, while remaining within one order of magnitude of the original high-level projection.\par

Since, Z1 combines in-memory analog MACs with thermal-noise-driven pBIT sampling at ultra-low voltages, a Gibbs update consumes roughly three orders of magnitude less energy than a read or write operation. Its energy advantage over an equivalent digital implementation therefore grows for larger models and longer Gibbs chains, where many local updates are performed between infrequent I/O operations.

\section{Supplementary details on Markov chains and compositions of kernels}\label{sec:kernels-chains-appendix}

This appendix proves the three results that compose into the key display \eqref{eq:mitigation_chain}: the chain rule for the trajectory KL, the data-processing inequality for readouts, and the change of input distribution.

\paragraph{Conditional KL is a joint KL at a fixed input distribution.}
Let $P$ and $\tilde P$ be kernels from $X$ to $Y$, and let $\mu \in \cP(X)$ be an input distribution. Define joint laws on $X \times Y$ by
\begin{gather*}
	\Gamma_{P}(x,y) = \mu(x)P(y \mid x), \qquad \Gamma_{\tilde P}(x,y) = \mu(x)\tilde P(y \mid x).
\end{gather*}
If $P(\cdot \mid x) \ll \tilde P(\cdot \mid x)$ for every $x$ with $\mu(x) > 0$, then
\begin{equation}\label{conditional-joint-eq}
	\KL(\Gamma_{P} \| \Gamma_{\tilde P}) = \EE_{x \sim \mu} \, \KL(P(\cdot \mid x) \| \tilde P(\cdot \mid x)).
\end{equation}
Indeed,
\[
	\begin{aligned}
		\KL(\Gamma_{P} \| \Gamma_{\tilde P})
		& = \sum_{x,y}\mu(x)P(y \mid x)\log \frac{\mu(x)P(y \mid x)}{\mu(x)\tilde P(y \mid x)} \\
		& = \sum_x\mu(x)\sum_y P(y \mid x)\log \frac{P(y \mid x)}{\tilde P(y \mid x)}.
	\end{aligned}
\]
Thus the per-kernel compilation objective $\EE_{x \sim \mu_\ell}[J_\ell(x; \vp_\ell)]$, for any choice of input distribution $\mu_\ell$, is an ordinary KL divergence on a joint experiment with fixed input marginal.

\paragraph{Readouts and data processing.}
Let $R \colon X \times Y \to Z$ be a readout. The law it induces from $\Gamma_P$ is
\begin{equation*}
    (R_\#\Gamma_P)(z)
    \coloneqq
    \sum_{(x,y) \, : \, R(x,y)=z} \mu(x)\, P(y \mid x),
\end{equation*}
and likewise for $R_\#\Gamma_{\tilde P}$. The data-processing inequality~\cite{cover2006} gives
\begin{equation}\label{readout-kl-eq}
	\KL(R_\#\Gamma_P \| R_\#\Gamma_{\tilde P}) \leqslant \KL(\Gamma_P \| \Gamma_{\tilde P}).
\end{equation}

\paragraph{Chain rule for trajectory laws.}
Now pass to trajectories. Let $P$ and $\Pmod_\vp$ be the target and model trajectory laws of Eqs.~\eqref{eq:target_traj} and~\eqref{eq:model_traj} (in particular, sharing the same initial law $q_0$) and let $q_{\ell-1}$ be the marginal of $P$ on wire $\ell - 1$. Expanding the logarithm of the product ratio,
\[
	\log \frac{P(z_{0:L})}{\Pmod_\vp(z_{0:L})} = \sum_{\ell = 1}^L \log \frac{P_\ell(z_\ell \mid z_{\ell - 1})}{\Pf_\ell(z_\ell \mid z_{\ell - 1}; \vp_\ell)},
\]
taking the expectation under $P$ term by term, and noting that the $\ell$-th term depends only on $(z_{\ell-1}, z_\ell)$, whose law under $P$ is $q_{\ell-1}(z_{\ell-1})\, P_\ell(z_\ell \mid z_{\ell-1})$, Eq.~\eqref{conditional-joint-eq} yields the chain rule
\begin{equation}\label{path-kl-chain-eq}
        \KL(P \| \Pmod_\vp) = \sum_{\ell = 1}^L \EE_{x \sim q_{\ell - 1}}\big[ J_\ell(x; \vp_\ell) \big] = \sum_{\ell = 1}^L \varepsilon_\ell(\vp_\ell),
\end{equation}
with $J_\ell$ and $\varepsilon_\ell$ as defined around \eqref{eq:J_def}. For any readout $R$ on trajectories, the data-processing inequality then gives
\[
	\KL(R_\#P \| R_\#\Pmod_\vp) \leqslant \sum_{\ell = 1}^L \varepsilon_\ell(\vp_\ell);
\]
the endpoint estimate is the case $R(z_{0:L}) = z_L$.

\paragraph{Change of input distribution.}
Training may replace the target input marginal $q_{\ell - 1}$ with a different distribution $\mu_\ell$. If $q_{\ell - 1} \ll \mu_\ell$ and $q_{\ell-1}(x)/\mu_\ell(x) \leqslant c_\ell$ for all $x$, then
\[
	\varepsilon_\ell(\vp_\ell)
    = \EE_{x \sim \mu_\ell}\Big[ \tfrac{q_{\ell-1}(x)}{\mu_\ell(x)}\, J_\ell(x; \vp_\ell) \Big]
    \leqslant c_\ell\, \EE_{x \sim \mu_\ell}\big[ J_\ell(x; \vp_\ell) \big].
\]
Combining this bound with \eqref{path-kl-chain-eq} and \eqref{readout-kl-eq} produces the key display \eqref{eq:mitigation_chain}.

\section{Variational compilation: gradient derivation and full case analysis}
\label{app:vc_details}

This appendix derives the two-phase identity \eqref{eq:cd_identity} behind the compilation gradient and then works out the concrete loss and gradient estimator in each cell of the case analysis of \S\ref{sec:compile}, in which the target is given as (a) an explicit conditional, (b) a conditional sampler, or (c) trajectory data, and the model kernel is either (1) tractable or (2) intractable.

\subsection{Derivation of the two-phase identity}
\label{app:vc_grad}

Fix one kernel and drop the index $\ell$. Write $E_\vp$ for its energy, $p_\vp(x, w, y) \propto e^{-E_\vp(x, w, y)}$ for the THM distribution it lives on, and $\Phi_\vp = -\nabla_\vp E_\vp$ for the sufficient statistics. From the definitions \eqref{eq:kernel_def} and \eqref{eq:forward},
\[
    \log \Pf(y \mid x; \vp) = \log \tk_\vp(x, y) - \log \Zf(x; \vp),
\]
and the two terms differentiate into expectations of $\Phi_\vp$ under two conditionals of $p_\vp$:
\begin{align}
    \nabla_\vp \log \tk_\vp(x, y)
    &= \frac{\sum_w \big({-}\nabla_\vp E_\vp(x, w, y)\big)\, e^{-E_\vp(x, w, y)}}{\tk_\vp(x, y)}
    = \EE_{w \sim p_\vp(\cdot \mid x, y)}\big[ \Phi_\vp(x, w, y) \big],
    \label{eq:app_grad_psi}\\[1mm]
    \nabla_\vp \log \Zf(x; \vp)
    &= \frac{\sum_{w, y'} \big({-}\nabla_\vp E_\vp(x, w, y')\big)\, e^{-E_\vp(x, w, y')}}{\Zf(x; \vp)}
    = \EE_{(w, y') \sim p_\vp(\cdot,\, \cdot \mid x)}\big[ \Phi_\vp(x, w, y') \big].
    \label{eq:app_grad_Z}
\end{align}
Equation~\eqref{eq:app_grad_psi} clamps both $x$ and $y$ and thermalizes the hidden spins (the positive phase); Eq.~\eqref{eq:app_grad_Z} clamps only $x$ and thermalizes hidden and output spins together (the negative phase). Their difference is the per-sample identity \eqref{eq:cd_identity} of the main text, and averaging it over target pairs, $x \sim q_{\ell-1}$ (or $x \sim \mu_\ell$ under a different input distribution) and $y \sim P_\ell(\cdot \mid x)$, gives the two-phase form of the gradient \eqref{eq:compile_grad}.

\subsection{The four estimators}
\label{app:vc_cases}

As noted in \S\ref{sec:compile}, the expectations of \eqref{eq:compile_grad} concern only the target and the integrand only the model, so the two case analyses combine freely. Cases (b) conditional sampler and (c) trajectory data supply the same object, input--output pairs from the joint $\mu_\ell(x) P_\ell(y \mid x)$, and share their estimators, so the grid has four distinct cells (Table~\ref{tab:vc_cases}).

\begin{table}[t]
    \caption{The compilation loss and gradient in each cell of the case analysis. Rows distinguish how the target kernel is specified, columns whether the model kernel's conditional is tractable.}
    \label{tab:vc_cases}
    \footnotesize
    \newcommand{\CaseCell}[1]{\parbox[t]{0.32\textwidth}{\raggedright #1}}
    \begin{tabular}{lll}
        \toprule
         & \textbf{(1) tractable model} & \textbf{(2) intractable model} \\
        \midrule
        \textbf{(a) explicit target} &
        \CaseCell{exact loss and gradient. Closed-form matching when the kernel can express the target (App.~\ref{sec:pnot})} &
        \CaseCell{target sums exact. Phases estimated by Gibbs sampling} \\
        \addlinespace
        \textbf{(b/c) sampled target} &
        \CaseCell{exact autodiff on Monte Carlo pairs ($=$ conditional maximum likelihood)} &
        \CaseCell{phases estimated by Gibbs sampling on Monte Carlo pairs ($=$ conditional CD)} \\
        \bottomrule
    \end{tabular}
\end{table}

\paragraph*{(a, 1): everything exact.}
Per input $x$, the loss is the finite sum
\[
    J_\ell(x; \vp) = \sum_y P_\ell(y \mid x)\, \log \frac{P_\ell(y \mid x)}{\Pf_\ell(y \mid x; \vp)},
\]
with $\Pf_\ell$ evaluated through \eqref{eq:forward} (in practice a \texttt{logsumexp} over the enumerated states), and its gradient
\[
    \nabla_\vp J_\ell(x; \vp) = -\sum_y P_\ell(y \mid x)\, \nabla_\vp \log \Pf_\ell(y \mid x; \vp)
\]
is computed by automatic differentiation. The training loss averages $J_\ell(\cdot\,; \vp)$ over the chosen input distribution, enumerating the inputs when they are few and sampling them otherwise. The optimization is deterministic. The only residuals are the hypothesis-class limitation and the optimizer's tolerance. When the kernel can reproduce the target's log-odds exactly, the minimizer is available in closed form by matching them. Appendix~\ref{sec:pnot} works out the probabilistic-NOT gate as an example. The Trotter gates of the random-walk demonstration (\S\ref{sec:demos_random_walk}) are also an example of this case.

\paragraph*{(a, 2): exact outer sums, sampled phases.}
The sum over the target's outputs $y$ is exact, so the positive phase becomes the exactly weighted mixture
\[
    \sum_y P_\ell(y \mid x)\; \EE_{w \sim p_\vp(\cdot \mid x, y)}\big[ \Phi_\vp(x, w, y) \big],
\]
in which each conditional hidden-spin expectation is one Gibbs run with $(x, y)$ clamped. The negative phase is one Gibbs run with $x$ clamped. If one positive-phase run per output state is too many, draw $y \sim P_\ell(\cdot \mid x)$ from the enumerated conditional instead and fall back to the (b/c, 2) estimator below.

\paragraph*{(b/c, 1): conditional maximum likelihood.}
Given pairs $(x_m, y_m)$, $m = 1, \ldots, M$, from the joint $\mu_\ell(x) P_\ell(y \mid x)$,
\[
    \widehat{\nabla} \;=\; -\frac{1}{M} \sum_{m=1}^{M} \nabla_\vp \log \Pf_\ell(y_m \mid x_m; \vp),
\]
with each per-pair gradient computed exactly by automatic differentiation. Monte Carlo enters only through the pairs themselves: this is classical conditional maximum-likelihood training, free of any sampling bias on the model side.

\paragraph*{(b/c, 2): conditional contrastive divergence.}
For each pair $(x_m, y_m)$, run one positive phase (clamp $x_m$ and $y_m$; thermalize $w$) and one negative phase (clamp $x_m$; thermalize $(w, y')$), average the measured statistics $\Phi_\vp$ within each phase across the minibatch, and subtract. Both phases are run at a finite horizon of $K$ Gibbs sweeps, which biases the estimator when the clamped kernel does not mix within $K$; the discussion of the MET and its mitigation in \S\ref{sec:compile} applies.

In every cell with target case (a) or (b), evaluating the gradient produces samples from the target kernel's output distribution as a by-product. In case (a) the conditional has just been enumerated, so drawing from it costs one categorical sample, and in case (b) such draws are the estimator's own input. These by-product samples are what makes target-input rollouts free during training, and Appendix~\ref{app:input_distributions} builds its bookkeeping on this observation.

\section{Closed-form compilation example}
\label{sec:pnot}
An example of closed-form compilation for a small non-trivial conditional is the probabilistic-NOT gate. With the parameterization used below, the gate preserves its input spin with probability $\sigma(\theta)$ and flips it with probability $\sigma(-\theta)$:
\begin{equation}
  P_\mathrm{target}(y \mid x; \theta) \;=\; \begin{cases} \sigma(\theta) & y = x \\ \sigma(-\theta) & y = -x. \end{cases}
\end{equation}

The $n_\mathrm{in} = n_\mathrm{out} = 1$, $n_h = 0$ thermodynamic kernel matching this conditional has energy $-E(x, y) = J_{xy}\,xy + h_y\,y$, with two free parameters and $x, y \in \{-1,+1\}$. Its forward conditional assigns $y = +1$ with probability
\begin{align}
  \Pf(y{=}{+}1 \mid x{=}{-}1) &= \sigma\big(2(h_y - J_{xy})\big), \\
  \Pf(y{=}{+}1 \mid x{=}{+}1) &= \sigma\big(2(h_y + J_{xy})\big).
\end{align}
Matching to the target (preserve $y=x$ with probability $\sigma(\theta)$) gives the two scalar equations $2(J_{xy} + h_y) = \theta$ and $2(J_{xy} - h_y) = \theta$, whose solution is
\begin{equation}\label{eq:pnot_solution}
  h_y \;=\; 0, \qquad J_{xy} \;=\; \theta/2.
\end{equation}
The compiled gate uses two p-bits and one coupling, with the bias vanishing. The solution is exact for finite $\theta$ within the uncapped parameterization. Exact identity and exact NOT are obtained only in the limits $\theta\to +\infty$ and $\theta\to -\infty$, respectively, under this convention; however finite coupling strengths can only approximate those deterministic endpoints without realizing exact zeros.

\section{Analytic thermodynamic kernels}
\label{app:atk}

The tractable-model case (1) of \S\ref{sec:compile} extends beyond kernels with a handful of spins. A thermodynamic kernel's conditional is intractable only through the sums over hidden spins in \eqref{eq:kernel_def} and over all possible outputs in the partition function of \eqref{eq:forward}. A sum over spins becomes cheap in two situations, when the spins are few enough to enumerate, or when they can be summed out analytically, which is possible in closed form when they are conditionally independent of the other summed spins. An \emph{analytic thermodynamic kernel} is a kernel whose spins are organized so that both mechanisms apply. A small core of spins, which may be coupled among themselves arbitrarily, is enumerated, while the remaining hidden spins form a periphery that is summed out analytically. This requires the periphery spins to be mutually uncoupled with all their neighbors in the core or among the inputs. The conditional of such a kernel is a closed-form differentiable function of the parameters, so it can be trained exactly, without Gibbs sampling, CD, or partition-function estimation.

\subsection{Construction}
\label{app:atk_construction}

Partition the kernel's hidden spins as $w = (u, v)$. The spins of the kernel then fall into three sets. The \emph{core} $C$ consists of the output spins $y$ together with the hidden spins $v$ and is enumerated, so $2^{|C|}$ must be affordable. In exchange, core spins may be coupled among themselves arbitrarily, so the kernel retains hidden--hidden couplings. The \emph{periphery} consists of the remaining hidden spins $u$, indexed by $a$, which are summed out analytically. Finally, the \emph{inputs} $x$ are clamped.

Two requirements make the analytic sum possible, (R1) that $2^{|C|}$ is small enough to enumerate, and (R2) that the periphery is an \emph{independent set}, meaning that no coupling connects two periphery spins and that every neighbor of a periphery spin lies in $C$ or among the inputs. Under (R2), on the pairwise substrate \eqref{eq:ising}, no energy term contains a product of two periphery spins, so the energy is affine in each $u_a$,
\begin{equation}\label{eq:atk_energy}
    E_\vp(x, u, v, y)
    \;=\; E^{C}_\vp(x, v, y) \;-\; \sum_a \eta_a(x, v, y)\, u_a,
\end{equation}
where $E^C_\vp$ collects all terms not involving the periphery (an unrestricted THM energy on the core and inputs) and the local field
\begin{equation}\label{eq:atk_field}
    \eta_a(x, v, y) \;=\; h_a + \sum_{i \in C} J_{ia}\, s_i + \sum_{k \in \mathrm{in}} J_{ka}\, x_k
\end{equation}
collects everything multiplying $u_a$, with $s_i$ ranging over the core spins $(v, y)$.

\begin{figure*}[t]
    \centering
    \begin{tabular}{ccc}
        \includegraphics[width=0.28\textwidth]{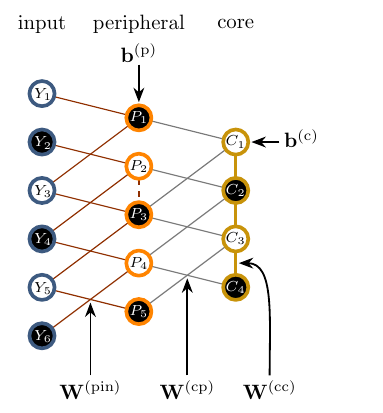} &
        \includegraphics[width=0.36\textwidth]{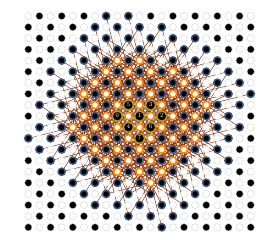} &
        \includegraphics[width=0.36\textwidth]{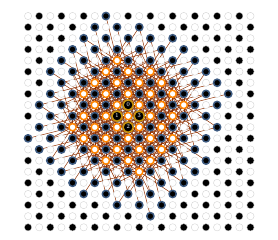} \\
        (a) Schematic, drawn in layers. &
        (b) An 8-bit ATK on the Z1 lattice. &
        (c) A 4-bit ATK on the Z1 lattice. \\
        &
        Compiled to Fig.~\ref{fig:local-updates}\subref{fig:local-reactions} in \S\ref{subsec:Eco} &
        Compiled to Fig.~\ref{fig:local-updates}\subref{fig:local-migration} in \S\ref{subsec:Eco}
    \end{tabular}

    \caption{\textbf{(a)} An ATK as three groups of spins: clamped inputs (blue), periphery (orange, summed out analytically), core (green, enumerated). The core carries internal couplings; the periphery is an independent set, with its one periphery--periphery lattice edge disabled (dashed red). \textbf{(b,c)} ATKs on the Z1 lattice: the user selects the core spins (green); the periphery ring (orange) and the clamped input boundary (blue) follow from the lattice geometry, and the disabled periphery--periphery couplings keep the periphery an independent set. A four-spin core supports tens of periphery spins and over a hundred boundary inputs while the conditional remains computable in closed form. }
    \label{fig:atk}
\end{figure*}

On a fixed lattice such as Z1, an ATK is carved out as follows (Fig.~\ref{fig:atk}). Pick any small set of spins as the core $C$, take the periphery to be the core's one-step neighborhood, and take the inputs to be the neighbors of the periphery outside $C \cup \{\text{periphery}\}$, which form the outer boundary of the region. Because the periphery is the entire neighborhood of the core, the free spins ($C$ and the periphery) have their whole Markov blanket inside the clamped boundary, so the kernel is self-contained and the core communicates with the rest of the chip only through the periphery. When the core spans both colors of the lattice's chromatic partition, the periphery may contain lattice-adjacent pairs, and each such periphery--periphery coupling is disabled (pinned to zero) to satisfy (R2). Moving a spin from the periphery into the core buys it arbitrary connectivity at the price of doubling the enumeration. This is the single design knob of an ATK, which trades classical training cost against expressivity.

\subsection{The exact conditional}
\label{app:atk_conditional}

Because \eqref{eq:atk_energy} is affine in each $u_a$ and contains no $u_a u_{a'}$ term, the periphery sum in the affinity \eqref{eq:kernel_def} factorizes spin by spin:
\begin{align}
    \tk_\vp(x, y)
    &= \sum_{v} e^{-E^C_\vp(x, v, y)} \prod_a \sum_{u_a \in \{-1,+1\}} e^{\eta_a(x, v, y)\, u_a} \notag\\
    &= \sum_{v} e^{-E^C_\vp(x, v, y)} \prod_a 2\cosh\big(\eta_a(x, v, y)\big) \notag\\
    &= \sum_{v} \exp\Big[ {-E^C_\vp(x, v, y)} + \sum_a \log\!\big(2\cosh\eta_a(x, v, y)\big) \Big],
    \label{eq:atk_affinity}
\end{align}
using $\sum_{u_a \in \{-1,+1\}} e^{\eta_a u_a} = 2\cosh(\eta_a)$. (With $\{0,1\}$ spins the sum would instead be $1 + e^{\eta_a}$, contributing a softplus $\softplus(\eta) \coloneqq \log(1 + e^{\eta})$ per spin.) Each summed-out periphery spin thus contributes one $\log(2\cosh)$ term to an effective \emph{core free energy}
\begin{equation}\label{eq:atk_freeenergy}
    E^{\mathrm{core}}_\vp(v, y \mid x)
    \;\coloneqq\;
    E^C_\vp(x, v, y) \;-\; \sum_a \log\!\big(2\cosh\eta_a(x, v, y)\big),
\end{equation}
in terms of which $\tk_\vp(x, y) = \sum_v e^{-E^{\mathrm{core}}_\vp(v, y \mid x)}$ and, by \eqref{eq:forward},
\begin{equation}\label{eq:atk_conditional}
    \Pf(y \mid x; \vp)
    \;=\;
    \frac{\sum_{v} e^{-E^{\mathrm{core}}_\vp(v, y \mid x)}}
         {\sum_{v', y'} e^{-E^{\mathrm{core}}_\vp(v', y' \mid x)}},
\end{equation}
a ratio of two \texttt{logsumexp}s over the $2^{|C|}$ core states alone: the periphery, however large, is gone from the state space. Evaluating $E^{\mathrm{core}}_\vp$ at one core state costs $O(n_u + \mathrm{nnz})$, with $n_u$ the number of periphery spins (one $\log(2\cosh)$ each) and $\mathrm{nnz}$ the number of couplings feeding the local fields, so the full normalized conditional costs
\begin{equation}\label{eq:atk_cost}
    O\big( 2^{|C|} \cdot (n_u + \mathrm{nnz}) \big)
\end{equation}
per input: exponential in the small core, linear in the periphery. The log-cosh sum in \eqref{eq:atk_freeenergy} is the standard free energy of a bipolar restricted Boltzmann machine~\cite{hinton2002,mnih2011crbm}, here supplied by the summed-out periphery. The ATK adds to it an arbitrary energy $E^C_\vp$ on the enumerated core, which a flat hidden layer cannot express.

\section{Input distributions: algorithms and analysis}
\label{app:input_distributions}

This appendix supplies the two results promised in \S\ref{sec:input_distributions}. The first part shows how target inputs are obtained at no additional sampling cost when calculating the gradients, including the bookkeeping that lets us train all kernels in parallel. In the second part, we analyze the difference between training under target inputs and under model inputs, and prove that target inputs ensure per-kernel residuals never amplify in propagation, while model inputs control the depth-independent error floor of contracting programs.

\subsection{Training schedules under target inputs}
\label{app:inputs_free}

Target-input training draws the clamped inputs of kernel $\ell$ from $q_{\ell-1}$, the marginal its wire carries under the target law. In target case (c) trajectory data, the training pairs come from recorded target trajectories, so their inputs follow $q_{\ell-1}$ automatically. In cases (a) explicit conditional and (b) conditional sampler, producing one input for kernel $\ell$ in principle requires a rollout of the target program through kernels $1, \ldots, \ell-1$, which would make the per-kernel training cost grow with depth. However, as noted at the end of Appendix~\ref{app:vc_details}, computing kernel $k$'s gradient at input $x$ already materializes a sample from $P_k(\cdot \mid x)$. This means that, while kernel $k$ trains, it also produces the target outputs that its children will consume as inputs, at no additional sampling cost.

\paragraph{Sequential schedule.}
In the simplest schedule, the kernels are trained one at a time, in topological order. While kernel $k$ trains, one target-output sample is stored per consumed input, tagged with the identity of the rollout it extends, which we call the rollout's \emph{provenance}. By the time kernel $\ell$ starts, the stored outputs of its parents form a pool of target rollouts $z_{0:\ell-1} \sim P$, from which its inputs are drawn directly. In this way, training performs no target sampling beyond that required by the gradients.

\paragraph{Parallel schedule: bags with provenances.}
The same bookkeeping allows all kernels to train simultaneously, and Algorithm~\ref{alg:bags} states one training iteration of the resulting scheme. The shared state is a family of \emph{bags} $B_0, \ldots, B_L$, one per wire, where $B_\ell[p]$ holds the target's output at wire $\ell$ on the rollout with provenance $p$ and $B_0$ holds global inputs $z_0 \sim q_0$. A single \emph{gatekeeper} process mints all new provenances into $B_0$, up to a cap $p_{\max}$ that bounds the memory, each kernel's training process is the only writer of its own bag, and a published entry is immutable. Every bag therefore has a single writer and the scheme requires no locks.

\begin{algorithm}[t]
\caption{One training iteration of kernel $\ell$ under the parallel target-input schedule (batch size $B$). For a source kernel, $\mathrm{pa}(\ell) = \{0\}$.}
\label{alg:bags}
$\mathrm{ready} \gets \{ p : B_j[p] \text{ published for all } j \in \mathrm{pa}(\ell) \}$\;
$\mathrm{batch} \gets$ the $B$ lowest provenances in $\mathrm{ready}$ that $\ell$ has not yet consumed\;
\If{$|\mathrm{batch}| < B$ {\normalfont\textbf{and}} $\ell$ is a source}{
    ask the gatekeeper to mint up to $B - |\mathrm{batch}|$ fresh global inputs into $B_0$ (respecting $p_{\max}$); add their provenances to $\mathrm{batch}$\;
}
\If{$|\mathrm{batch}| < B$}{
    pad $\mathrm{batch}$ with provenances $\ell$ has already consumed, chosen uniformly at random (\emph{reuse}); if none exist, pad with generic inputs carrying no provenance (\emph{cold start})\;
}
gather the inputs $\big( B_j[p] \big)_{j \in \mathrm{pa}(\ell)}$ for each $p$ in $\mathrm{batch}$\;
compute the gradient \eqref{eq:compile_grad} on the batch, obtaining one target-output sample per input as a by-product\;
\For{each provenance $p$ consumed for the first time}{
    publish $B_\ell[p] \gets$ the target-output sample obtained at $p$ (publish once; never overwrite)\;
}
\end{algorithm}

The scheme produces coherent rollouts. For a fixed provenance $p$, the collection $\{B_\ell[p]\}_\ell$ is a single trajectory drawn from $P$, which follows by induction along the topological order, since $B_0[p]$ is drawn from $q_0$ by the gatekeeper, kernel $\ell$ writes $B_\ell[p]$ only by sampling its target conditional at its parents' entries under the same $p$, and the immutability of published entries guarantees that a child never consumes a value revised at a later time.

Reuse does not bias the input distribution. The inputs of a reused provenance remain exact draws from the target rollout law, because the choice of which provenances a kernel consumes, lowest-first for fresh ones and uniform among consumed ones for reuse, is made without inspecting the stored values. Each training input therefore stays distributed as $q_{\ell-1}$, and the only cost of reuse is a correlation between the gradient estimates of different iterations, which reduces the effective sample size until fresh rollouts become available. The preference for the lowest fresh provenances makes all kernels advance through the provenances in the same order, so that several parents feeding a common child publish along a common prefix rather than starving the child with mismatched random picks. The only biased phase is the cold start, since generic-input padding occurs before the pipeline fills, no bag entry is ever published from a generic input, and the bias is thus confined to the early gradient steps without reaching the stored rollouts.

Finally, the rate at which fresh provenances become ready for kernel $\ell$ is set by the slowest of its ancestors, and a starved kernel keeps training on reused rollouts in the meantime. The memory is bounded by the provenance cap, since once $p_{\max}$ rollouts exist the gatekeeper stops minting and all kernels continue on the existing pool, which by then constitutes a large i.i.d.\ sample of target trajectories.

\subsection{Target inputs versus model inputs}
\label{app:inputs_model}

If every kernel could match its target exactly, the input distribution would be irrelevant, since any $\mu_\ell$ with full support finds the same optimum. In reality the residual $J_\ell(\cdot\,; \vp_\ell)$ of Eq.~\eqref{eq:J_def} is not identically zero (Fig.~\ref{fig:expressivity_hierarchy}), and the input distribution determines where the kernel spends its limited accuracy. In this subsection we compare the two context-matching choices of \S\ref{sec:input_distributions} and determine which one controls the deployed marginals $\tilde q_\ell$, the quantities on which the deployed circuit is evaluated.

We work in total variation. For two distributions $\alpha, \beta$ on the same finite set, $\| \alpha - \beta \|_\TV = \tfrac12 \sum_z |\alpha(z) - \beta(z)|$, which is the largest difference in probability that the two distributions assign to any event. The per-input error in this metric,

\[
    J^{\TV}_\ell(x; \vp_\ell) \;\coloneqq\; \big\| P_\ell(\cdot \mid x) - \Pf_\ell(\cdot \mid x; \vp_\ell) \big\|_\TV,
\]
is controlled by the KL residual through Pinsker's inequality~\cite{pinsker1964information}, $J^{\TV}_\ell \leq \sqrt{J_\ell / 2}$. The quantity we track is the deployed-marginal error

\[
    \tilde\delta_\ell \;\coloneqq\; \big\| \tilde q_\ell - q_\ell \big\|_\TV,
\]
and we write $P\alpha$ for the output distribution of a kernel $P$ applied to the input distribution $\alpha$, $(P\alpha)(y) = \sum_x P(y \mid x)\, \alpha(x)$.

Since $q_\ell = P_\ell\, q_{\ell-1}$ and $\tilde q_\ell = \Pf_\ell\, \tilde q_{\ell-1}$, adding and subtracting $P_\ell\, \tilde q_{\ell-1}$ splits the error at wire $\ell$ into an inherited part and a freshly injected part:
\begin{equation}\label{eq:deployed_split}
    \tilde q_\ell - q_\ell
    \;=\; \underbrace{P_\ell \big( \tilde q_{\ell-1} - q_{\ell-1} \big)}_{\text{inherited}}
    \;+\; \underbrace{\big( \Pf_\ell - P_\ell \big)\, \tilde q_{\ell-1}}_{\text{injected at } \ell}.
\end{equation}
Note that the inherited part is propagated by the target kernel. To bound it we need to know by how much a kernel can stretch the distance between two input distributions. Write
\begin{equation}\label{eq:dobrushin}
  \rho(P) \;\coloneqq\; \sup_{\alpha \neq \beta}
  \frac{\|P\alpha - P\beta\|_\TV}{\|\alpha - \beta\|_\TV}
\end{equation}
for the largest such stretch factor, the supremum running over pairs of distinct input distributions. Every Markov kernel has $\rho(P) \le 1$ (kernels never expand total variation) and $\rho(P) < 1$ means the kernel strictly contracts. This quantity is known as the Dobrushin ergodicity coefficient~\cite{dobrushin1956clt,levin2017markov}. Taking total-variation norms in \eqref{eq:deployed_split} gives the recursion
\begin{equation}\label{eq:deployed_recursion}
    \tilde\delta_\ell \;\leq\; \rho(P_\ell)\, \tilde\delta_{\ell-1} \;+\; \iota_\ell,
    \quad
    \iota_\ell \coloneqq \big\| \big( \Pf_\ell - P_\ell \big)\, \tilde q_{\ell-1} \big\|_\TV.
\end{equation}
The injected term is bounded by the per-input error averaged under the model input marginal. By the triangle inequality,
\begin{equation}
    \iota_\ell
    = \frac{1}{2} \sum_{y} \Big| \sum_{x} \tilde q_{\ell-1}(x) \big( \Pf_\ell - P_\ell \big)(y \mid x) \Big|
    \;\leq\; \sum_{x} \tilde q_{\ell-1}(x)\, \frac{1}{2} \sum_{y} \big| \big( \Pf_\ell - P_\ell \big)(y \mid x) \big|
    \;=\; \EE_{x \sim \tilde q_{\ell-1}}\big[ J^{\TV}_\ell(x; \vp_\ell) \big],
    \label{eq:iota_bound}
\end{equation}
where $(\Pf_\ell - P_\ell)(y \mid x)$ abbreviates the difference of the two conditionals, and by Pinsker's inequality and Jensen's inequality the right-hand side is at most $\big( \tfrac12 \EE_{x \sim \tilde q_{\ell-1}}[J_\ell(x; \vp_\ell)] \big)^{1/2}$. The error that kernel $\ell$ injects into the deployed marginals is therefore its error on the inputs the deployed circuit feeds it, which are distributed as $\tilde q_{\ell-1}$, and up to the square root the bound \eqref{eq:iota_bound} is the model-context objective of \S\ref{sec:input_distributions}.

The two weightings are connected only through a change of measure,
\begin{equation}\label{eq:context_com}
    \EE_{x \sim \tilde q_{\ell-1}}\big[ J_\ell(x; \vp_\ell) \big]
    \;\leq\;
    \Big( \max_x \tfrac{\tilde q_{\ell-1}(x)}{q_{\ell-1}(x)} \Big)\,
    \EE_{x \sim q_{\ell-1}}\big[ J_\ell(x; \vp_\ell) \big],
\end{equation}
where the constant on the right is inflated by upstream residuals and becomes infinite as soon as the deployed circuit visits an input the target never produces. Generically, suppose the compiled ancestors leak probability onto a set $A$ of inputs that the target rarely produces, $q_{\ell-1}(A) \approx 0$ but $\tilde q_{\ell-1}(A) > 0$. The conservation leakage of the random-walk demonstration (\S\ref{sec:demos_random_walk}) is of this type, with $A$ the set of multi-particle configurations. Training under target inputs puts almost no weight on $A$, so a capacity-limited optimizer sacrifices accuracy on $A$ in exchange for accuracy on the target's support. The deployed circuit, however, visits $A$, where the training leaves the behavior of the kernel uncontrolled, so $\iota_\ell$ can be large even while $\varepsilon_\ell$ is small and the drift propagates to later wires. Training under model inputs weights $A$ in proportion to how often the deployed circuit visits it.

However, model inputs are not the only choice that bounds the deployed error. Target inputs also bound it, and the two guarantees differ in a precise way. Unrolling \eqref{eq:deployed_recursion} from a matched start $\tilde\delta_0 = 0$ gives
\begin{equation}\label{eq:deployed_unrolled}
    \tilde\delta_L \;\leq\; \sum_{\ell=1}^{L} \rho(P_L) \cdots \rho(P_{\ell+1})\; \iota_\ell.
\end{equation}
Two regimes follow.
\begin{itemize}[leftmargin=*]
    \item \emph{Without contraction} ($\rho \leq 1$ only), \eqref{eq:deployed_unrolled} is the linear budget $\tilde\delta_L \leq \sum_\ell \iota_\ell$. Target inputs deliver a budget of exactly the same form. Telescoping through the hybrid laws $H_k \coloneqq \Pf_L \cdots \Pf_{k+1}\, P_k \cdots P_1\, q_0$ (target kernels up to wire $k$, compiled kernels after, so that $H_L = q_L$ and $H_0 = \tilde q_L$) gives the exact identity
    \begin{equation}\label{eq:telescope_target}
        \tilde q_L - q_L
        \;=\; \sum_{\ell=1}^{L} \Pf_L \cdots \Pf_{\ell+1} \big( \Pf_\ell - P_\ell \big)\, q_{\ell-1},
    \end{equation}
    and hence, using only that the compiled kernels are non-expansive,
    \begin{equation}\label{eq:budget_target}
        \tilde\delta_L \;\leq\; \sum_{\ell=1}^{L} \EE_{x \sim q_{\ell-1}}\big[ J^{\TV}_\ell(x; \vp_\ell) \big].
    \end{equation}
    Note what switched between \eqref{eq:deployed_recursion} and \eqref{eq:telescope_target}: the injected errors are now weighted by the \emph{target} marginals, but they propagate through the \emph{compiled} kernels. At the level of worst-case linear budgets, the two input distributions are therefore interchangeable: per-kernel residuals never amplify in propagation, whichever inputs they were trained under, because Markov kernels cannot expand total variation.
    \item \emph{With contracting targets}, $\rho(P_\ell) \leq \rho_0 < 1$, the bound \eqref{eq:deployed_unrolled} saturates:
    \begin{equation}\label{eq:deployed_floor}
        \tilde\delta_L \;\leq\; \sum_{\ell=1}^{L} \rho_0^{\,L - \ell}\, \iota_\ell
        \;\leq\; \frac{\max_\ell \iota_\ell}{1 - \rho_0},
    \end{equation}
    a depth-independent floor whose value is the model-weighted per-kernel error of \eqref{eq:iota_bound}, i.e.,\ the model-context objective. The target-weighted identity \eqref{eq:telescope_target} does not reproduce this floor, because there the propagation runs through the compiled kernels, whose behavior on drifted inputs is left uncontrolled by training under target inputs. A compiled kernel that misbehaves outside the target's support can fail to contract there, in which case only the linear budget remains available.
\end{itemize}

{\it Remark}. The above analysis also goes through verbatim if we replace $\iota_\ell$ with $\eta_\ell$, the per-layer TV error defined in Eq.~\eqref{eq:error-defn}. Repeating the derivations above, we obtain

\begin{align}
\widetilde{\delta}_L \leq \frac{\max_\ell \eta_\ell}{1-\rho_0} \label{eq:eta_floor}
\end{align}

\subsection{When to switch to model inputs}
Two considerations determine when the switch to model inputs should happen. First, the two objectives coincide to first order when the residual is bounded. If $\sup_x J_\ell(x; \vp_\ell)$ is finite, then
\[
    \big| \EE_{\tilde q_{\ell-1}}[J_\ell] - \EE_{q_{\ell-1}}[J_\ell] \big|
    \;\leq\; 2\, \tilde\delta_{\ell-1} \sup_x J_\ell(x; \vp_\ell),
\]
so model inputs are a second-order refinement when residuals are small, and the two objectives separate only in the regime where $\sup_x J_\ell$ grows large on drifted inputs, the failure mode of \eqref{eq:context_com} identified above. Second, the marginal $\tilde q_{\ell-1}$ depends on the upstream parameters, so the simultaneous re-optimization of all kernels is a fixed-point iteration rather than descent on a fixed objective, and early in training it spends capacity on input distributions that later parameters will not produce. For both reasons, training should start under target inputs and switch to model inputs late, which is the schedule recommended in \S\ref{sec:input_distributions}.

The switch also carries a feasibility requirement. Model-context matching queries the target conditional at model-generated inputs $x \sim \tilde q_{\ell-1}$, so it requires target case (a) or (b) and is unavailable under case (c). The inputs themselves are obtained by rolling out the compiled circuit, on the hardware if available, in which case the input stream also carries the upstream execution error, such as finite-horizon readout and calibration drift, and the re-optimization absorbs it into the trained parameters.

\section{Compiling backward conditionals via input reweighting and Gibbs refinement}
\label{sec:backward-compilation}

The variational compilation of Sec.~\ref{sec:compile} fits a
forward conditional $\Pf(y \mid x;\vp)$ to a target kernel. The compiled backward
$\Pb(x \mid y)$ is then accessible on the same programming by switching
the clamping pattern, but it equals a specific Bayes inverse determined by the
compilation rather than a target backward of independent design. This Appendix explores the possibilities available in ensuring that the backward also matches a specified target. We present two distinct methods that can be used. The first is \emph{input reweighting} (Eq.~\eqref{eq:gauge-def} below), which
modifies the implicit input prior of a single kernel without touching its forward.
The second is Gibbs refinement on the composed circuit, which corrects an
inter-factor bias that no per-kernel reweighting can remove. (Throughout this Appendix, ``prior'' is meant literally: the backward direction inverts the forward by Bayes' rule, the input is the unknown, and the implicit input prior is the distribution that inversion is taken against.)

Recall that the kernel admits two normalizations from one energy programming,
\begin{align}
\Pf(y \mid x) \;=\; \frac{\psi(x,y)}{Z^F(x)},
\qquad
\Pb(x \mid y) \;=\; \frac{\psi(x,y)}{Z^B(y)},
\label{eq:two-conditionals-recap}
\end{align}
with $Z^F(x) = \sum_{y'} \psi(x,y')$ and $Z^B(y) = \sum_{x'} \psi(x',y)$.
Multiplying and dividing $\Pb$ by $Z^F(x)$ rewrites it as Bayes' rule,
\begin{align}
\Pb(x \mid y)
\;=\;
\frac{Z^F(x) \, \Pf(y \mid x)}
     {\sum_{x'} Z^F(x') \, \Pf(y \mid x')},
\label{eq:bayes-rewrite}
\end{align}
with the identifications $p(y \mid x) \leftrightarrow \Pf(y \mid x)$
and $p(x) \leftrightarrow Z^F(x)$: the forward is the likelihood and the
forward partition function is the implicit prior. Whatever prior the compiler
happens to produce is the prior against which the hardware backward inverts.

For a generic compilation this implicit prior $p_\psi(x) \propto Z^F(x)$ is
not the prior the application demands. In a Trotterized denoising chain the
per-step $Z^F$ is uniform up to $O(\delta t)$ corrections, which is the regime
where the free backward is unbiased. Outside that regime the implicit prior
drifts away from uniform, and the drift is what we must correct.

\subsection{Installing a prior by input reweighting}

As recorded in \S\ref{sec:compile}, compiling the forward conditional leaves the backward undetermined. The freedom at play is \emph{input reweighting}, the energy-shift freedom introduced in \S\ref{sec:ebm} (Eq.~\eqref{eq:gauge_energy}), here written on the affinity: for any real-valued function $\lambda$ of the input alone, define the reweighted affinity
\begin{align}
\psi_\lambda(x,y) \;:=\; e^{\lambda(x)} \, \psi(x,y).
\label{eq:gauge-def}
\end{align}
The forward
partition function rescales pointwise,
\begin{align}
Z^F_\lambda(x) \;=\; \sum_{y'} \psi_\lambda(x,y') \;=\; e^{\lambda(x)} Z^F(x),
\end{align}
and the forward conditional is invariant,
\begin{align}
\Pf_\lambda(y \mid x)
\;=\; \frac{e^{\lambda(x)}\psi(x,y)}{e^{\lambda(x)} Z^F(x)}
\;=\; \Pf(y \mid x).
\end{align}
Conversely, any two affinities with the same forward conditional differ by exactly such a factor: if $\psi_1$ and $\psi_2$ share a forward, then $\psi_2(x,y) = [\Zf_2(x)/\Zf_1(x)]\,\psi_1(x,y)$, a ratio depending on $x$ alone. Forward compilation therefore determines the affinity only up to input reweighting.
However, the backward partition function is not invariant under the reweighting: it acquires an
$x$-dependent factor under the sum,
\begin{align}
Z^B_\lambda(y) \;=\; \sum_{x'} e^{\lambda(x')} \psi(x',y),
\end{align}
and the compiled backward becomes
\begin{align}
\Pb_\lambda(x \mid y)
&\;=\; \frac{e^{\lambda(x)} \psi(x,y)}{\sum_{x'} e^{\lambda(x')} \psi(x',y)} \\
&\;=\; \frac{e^{\lambda(x)} Z^F(x) \, \Pf(y \mid x)}
          {\sum_{x'} e^{\lambda(x')} Z^F(x') \, \Pf(y \mid x')}.
\label{eq:gauge-backward}
\end{align}
Comparing with (\ref{eq:bayes-rewrite}), the implicit prior under the
reweighting $\lambda$ is
\begin{align}
p_{\psi_\lambda}(x) \;\propto\; e^{\lambda(x)} \, Z^F(x).
\label{eq:implicit-prior-lambda}
\end{align}
Therefore, we have the freedom to choose a function of $n_{\rm in}$ bits to modify the prior while leaving the forward untouched.

Suppose the application
specifies a desired prior $p^\star(x)$ and we want the hardware backward to
equal the Bayes posterior of $\Pf$ against $p^\star$,
\begin{align}
\Pb_\lambda(x \mid y)
\;\stackrel{!}{=}\;
\frac{p^\star(x) \, \Pf(y \mid x)}{\sum_{x'} p^\star(x') \Pf(y \mid x')}.
\end{align}
Matching (\ref{eq:gauge-backward}) to this expression requires
$e^{\lambda(x)} Z^F(x) \propto p^\star(x)$, hence
\begin{align}
\lambda(x) \;=\; \log p^\star(x) \;-\; \log Z^F(x) \;+\; c,
\label{eq:lambda-choice}
\end{align}
where $c$ is an arbitrary additive constant that does not affect sampling.
The expression has two pieces with distinct interpretations. The term
$\log p^\star(x)$ is given by the problem and imposes the target prior. The
term $-\log Z^F(x)$ depends on the compiled kernel and undoes the implicit
prior produced during the compilation of the forward. The two pieces are independently meaningful:
pure uniformization $\lambda = -\log Z^F$ produces the maximum-likelihood
inverse $\Pb \propto \Pf(y \mid x)$, while pure prior
forcing $\lambda \approx \log p^\star$ suffices when the compilation
already lands in a near-Trotter regime with $Z^F$ approximately constant.

The reweighting function enters the energy as an additive shift on the input block,
\begin{align}
-E_\lambda(x,w,y) \;=\; -E(x,w,y) + \lambda(x).
\label{eq:lambda-into-energy}
\end{align}
During forward execution $x$ is clamped, so $\lambda(x)$ is a constant under
sampling and does not change the readout; during backward execution $x$ is
sampled and $\lambda(x)$ reweights the marginal exactly as
(\ref{eq:gauge-backward}) requires.

The correction $\lambda(x)$ is a function of the $n_{\rm in}$ input bits, which
can always be decomposed as a sum of products of those bits:
\begin{equation}
\lambda(x) \;=\; S_0 \;+\; \sum_i S_i \, x_i \;+\; \sum_{i<j} S_{ij} \, x_i x_j \;+\; (\text{higher-order products}),
\label{eq:lambda-expansion}
\end{equation}
with coefficients fixed by the $2^{n_{\rm in}}$ values $\{\lambda(x)\}$. The constant ($|S_0| = 0$) is absorbed into the
overall energy reference and is sampling-irrelevant. Linear terms ($|S_1| = 1$)
shift the bias on input spin $i$ by $S_{\{i\}}$. Pairwise terms
($|S_2| = 2$) populate couplings $J_{ij}$ between input spins. These input-block parameters, i.e.,\ biases and couplings supported entirely on the input spins, are constants under forward sampling, so compilation may as well fix them to zero; we call that choice the \emph{canonical form} of a compiled kernel, and it pins $\lambda \equiv 0$. The reweighting correction reactivates them while keeping the energy pairwise on the same spin graph.

Higher-order terms $|S| \ge 3$ are not representable by a pairwise energy on
the input spins alone. One option is to compile
against a reweighting-penalized objective: add
\begin{align}
\mathcal{L}_{\mathrm{rew}}(\vp) \;=\; \eta \sum_{|\chi| \ge 3} S_\chi(\vp)^2
\end{align}
to the compile loss, which favors reweighting functions $\lambda(x)$ that can be expressed with at most pairwise Ising coupling terms.

\subsection{Gibbs refinement}
\label{app:gibbs}

A single-kernel reweighting picks one Bayes inverse against one prior on one input
set. For a DFG with $L$ factors and clamped terminal state
$z_L$, the natural backward target is the full posterior over the
rest of the trajectory,
\begin{align}
p\!\left(z_{0:L-1} \,\big|\, z_L\right)
\;\propto\;
\prod_{\ell=1}^{L} \psi_\ell\!\left(z_{\ell-1}, z_\ell\right),
\label{eq:true-posterior}
\end{align}
treating the data-side endpoint $z_0$ as part of the sampled trajectory. The free backward of Sec.~\ref{sec:compile}, run factor-by-factor from the clamped
end, samples a strictly different distribution. Starting at $z_L$, draw
$z_{L-1} \sim \Pb_L$, clamp, draw $z_{L-2} \sim \Pb_{L-1}$,
clamp, and so on. The joint distribution of the resulting trajectory is
\begin{align}
q^B\!\left(z_{0:L-1} \,\big|\, z_L\right)
&\;=\;
\prod_{\ell=1}^{L} \Pb_\ell\!\left(z_{\ell-1} \,\big|\, z_\ell\right) \\
&\;=\;
\frac{\prod_{\ell} \psi_\ell\!\left(z_{\ell-1}, z_\ell\right)}
     {\prod_{\ell} Z^B_\ell\!\left(z_\ell\right)}.
\label{eq:free-backward}
\end{align}
The numerator agrees with the true posterior (\ref{eq:true-posterior}). The
denominator does not: $p$ has a single trajectory-independent normalizer (omitted from Eq.~\eqref{eq:true-posterior}), while
$q^B$ accumulates one state-dependent factor $Z^B_\ell(z_\ell)$ per factor.

The standard tool for sampling a joint specified by conditionals is Gibbs sampling, and the joint Ising structure of a composed circuit makes the requisite conditionals natively available on hardware. Clamp every $z_k$ with $k \ne \ell$. The
only factors in (\ref{eq:true-posterior}) that depend on $z_\ell$ are
$\psi_\ell(z_{\ell-1}, z_\ell)$ and $\psi_{\ell+1}(z_\ell, z_{\ell+1})$,
all others are constants in $z_\ell$ and cancel out.
Hence
\begin{equation}
p\!\left(z_\ell \,\big|\, z_{\ell-1}, z_{\ell+1}\right) =
\frac{\psi_\ell\!\left(z_{\ell-1}, z_\ell\right) \, \psi_{\ell+1}\!\left(z_\ell, z_{\ell+1}\right)}
     {\sum_{z_\ell'} \psi_\ell\!\left(z_{\ell-1}, z_\ell'\right) \, \psi_{\ell+1}\!\left(z_\ell', z_{\ell+1}\right)}.
\label{eq:gibbs-conditional}
\end{equation}
Two adjacent kernels appear, both neighbors of $z_\ell$ are present, and
no $Z^B_\ell$ factor survives.

Expand the two kernels back into their Ising
definitions,
\begin{equation}
\psi_\ell \, \psi_{\ell+1}
\;=\;
\sum_{w_\ell,\, w_{\ell+1}}
\exp\!\Big[\!-\!E_\ell\!\left(z_{\ell-1},w_\ell,z_\ell\right)
            -E_{\ell+1}\!\left(z_\ell,w_{\ell+1},z_{\ell+1}\right) \Big].
\end{equation}
This is the Boltzmann factor of a pairwise Ising energy over the two-factor
block of spins $(w_\ell, z_\ell, w_{\ell+1})$ with $z_{\ell-1}$ and
$z_{\ell+1}$ clamped at their current values. On hardware this corresponds to clamping the two outer slices, thermalizing the spins between them and reading out
$z_\ell$. After marginalizing the hidden block, the readout distribution is
\begin{equation}
\frac{\sum_{w,w'} e^{-E_\ell - E_{\ell+1}}}
     {\sum_{z_\ell'}\sum_{w,w'} e^{-E_\ell - E_{\ell+1}}}
= \frac{\psi_\ell(z_{\ell-1}, z_\ell) \, \psi_{\ell+1}(z_\ell, z_{\ell+1})}
       {\sum_{z_\ell'}\psi_\ell(z_{\ell-1}, z_\ell') \, \psi_{\ell+1}(z_\ell', z_{\ell+1})}
\label{eq:hardware-equals-gibbs}
\end{equation}
where the hidden sums factor because $w_\ell$ and $w_{\ell+1}$ couple to
disjoint state spins, and collapse termwise back into the kernel definitions.
The right-hand side is exactly the Gibbs conditional
(\ref{eq:gibbs-conditional}). The state-dependent partition functions
$Z^B_\ell$ never appear in (\ref{eq:hardware-equals-gibbs}) because the two
factors are normalized as a joint block, not one at a time.

Therefore, the DFG backward sampler is now a two-stage procedure.
\begin{enumerate}
\item \emph{Initialize.} A single factorized backward pass through
$\Pb_L, \Pb_{L-1}, \ldots, \Pb_1$ seeds a full trajectory
$(\hat z_0, \ldots, \hat z_{L-1})$ from $q^B$. Biased relative to $p$ but adequate as a starting point.

\item \emph{Sweep.} For $\ell$ ranging over the intermediate slices: clamp
$\hat z_{\ell-1}$ and $\hat z_{\ell+1}$, thermalize the two-factor Ising
block as in (\ref{eq:hardware-equals-gibbs}), read the new $\hat z_\ell$,
overwrite. One sweep is one pass over every intermediate $\ell$. Repeat for
$K$ sweeps.
\end{enumerate}
The procedure converges to $p(z_{0:L-1} \mid z_L)$ regardless of where it starts, because Gibbs sampling converges to the joint distribution defined by its conditionals and (\ref{eq:hardware-equals-gibbs}) is exactly the Gibbs conditional of (\ref{eq:true-posterior}). Depending on the mixing time of each two-factor block, the number of sweeps required to converge to the true posterior of the DFG may vary. For fast-mixing kernels, this means that executing the sampling from the posterior has similar runtime cost as forward sampling.

The two-factor block update described above is a conceptually clean Gibbs sampler whose stationary distribution is the joint posterior. It is not the only valid choice. The joint posterior (\ref{eq:true-posterior}) is the Boltzmann distribution of a pairwise Ising on the full spacetime ribbon, including all the hidden blocks and all the intermediate state slices, so any Gibbs sampler on that joint Ising (block sweeps, single-spin sweeps, chromatic decompositions, mixed schedules) targets the same stationary distribution.

\section{Trajectory-level REINFORCE: derivation and estimator}
\label{app:reinforce}

In this appendix we derive the energy-based REINFORCE gradient of \S\ref{sec:reinforce}, give an unbiased estimator for the chain, and expand on three choices which we left implicit in the main text. First, the reason for the factor gradient to treat hidden and output spins on the same footing. Second, why the energy-gradient mean subtracted in~\eqref{eq:conditional_reinforce_grad} is part of the score rather than a free baseline. Third, why we subtract it instead of a baseline formed from $F$. The appendix closes with the derivation of the effective reward \eqref{eq:f_eff} for functionals of a readout law.

\subsection{The energy-based REINFORCE gradient}
\label{app:reinforce_deriv}

Consider a single EBM with energy $E_\vp$ and law $p_\vp(y) = e^{-E_\vp(y)}/Z(\vp)$, $Z(\vp) = \sum_y e^{-E_\vp(y)}$, and the objective $\mathcal{L}(\vp) = \EE[F(y)]$ with $F$ independent of $\vp$. Throughout this subsection and the next, every expectation $\EE[\cdot]$ is over $y\sim p_\vp$, with $y'$ an independent copy from the same law, and we drop the subscript. The score-function identity~\cite{williams1992,glynn1990} reads
\[
    \nabla_\vp \mathcal{L}(\vp) = \EE\big[ F(y)\, \nabla_\vp \log p_\vp(y) \big].
\]
For a Boltzmann law the score splits into an energy term and the log-partition gradient,
\[
    \nabla_\vp \log p_\vp(y) = -\nabla_\vp E_\vp(y) - \nabla_\vp \log Z(\vp),
\]
where $\nabla_\vp \log Z(\vp) = \EE[ -\nabla_\vp E_\vp(y') ]$, which is the standard fact that the gradient of the log-partition function is the model average of the energy gradient. Substituting and collecting terms gives three equivalent forms,
\begin{align}
    \nabla_\vp \mathcal{L}
    &= \EE\big[ {-}\nabla_\vp E_\vp(y)\, F(y) \big]
       - \EE\big[ {-}\nabla_\vp E_\vp(y) \big]\, \EE\big[ F(y) \big]
       \label{eq:app_rg_cov} \\
    &= \EE\Big[ F(y) \big( {-}\nabla_\vp E_\vp(y) - \EE[ {-}\nabla_\vp E_\vp(y') ] \big) \Big]
       \label{eq:app_rg_energy_baseline} \\
    &= \EE\Big[ {-}\nabla_\vp E_\vp(y) \big( F(y) - \EE[ F(y') ] \big) \Big].
       \label{eq:app_rg_F_baseline}
\end{align}
The first line is a covariance between the energy gradient and the reward, while the second and third move the subtracted mean onto one factor or the other. Equation~\eqref{eq:app_rg_energy_baseline} is the single-kernel form of~\eqref{eq:conditional_reinforce_grad} and is the one used in the main text.

\subsection{The chain gradient and hidden spins}
\label{app:reinforce_chain}

The main-text gradient \eqref{eq:conditional_reinforce_grad} treats each factor's hidden spins on the same footing as its outputs. To state this precisely, define the \emph{augmented} model law over wires and hidden spins,
\begin{equation}\label{eq:augmented_law}
    \Pmod^+_\vp(z_{0:L}, w_{1:L})
    \;=\; q_0(z_0) \prod_{\ell=1}^{L} p_{\vp_\ell}(w_\ell, z_\ell \mid z_{\ell-1}),
    \qquad
    p_{\vp_\ell}(w, y \mid x) \coloneqq \frac{e^{-E_{\vp_\ell}(x, w, y)}}{\Zf(x; \vp_\ell)},
\end{equation}
whose marginal over the wires is the model trajectory law $\Pmod_\vp$ of \eqref{eq:model_traj}. Since $F$ does not read the hidden spins,
\[
    \LRF(\vp) = \EE_{z_{0:L} \sim \Pmod_\vp}\big[ F(z_{1:L}) \big]
    = \EE_{(z_{0:L}, w_{1:L}) \sim \Pmod^+_\vp}\big[ F(z_{1:L}) \big],
\]
so we may differentiate the augmented law instead. Its logarithm splits as $\sum_\ell \log p_{\vp_\ell}(w_\ell, z_\ell \mid z_{\ell-1})$ plus $\vp$-independent terms, so only factor $\ell$'s term carries $\vp_\ell$, and the score-function identity gives
\[
    \nabla_{\vp_\ell} \LRF
    = \EE_{\Pmod^+_\vp}\Big[ F(z_{1:L})\, \nabla_{\vp_\ell} \log p_{\vp_\ell}(w_\ell, z_\ell \mid z_{\ell-1}) \Big].
\]
The conditional $p_{\vp_\ell}(w, y \mid x)$ is itself a Boltzmann law with partition function $\Zf(x; \vp_\ell)$, so exactly as in \eqref{eq:app_rg_energy_baseline},
\[
    \nabla_\vp \log p_\vp(w, y \mid x)
    = \Phi_\vp(x, w, y)
    - \EE_{(w', y') \sim p_\vp(\cdot,\, \cdot \mid x)}\big[ \Phi_\vp(x, w', y') \big],
\]
which, substituted above, yields \eqref{eq:conditional_reinforce_grad}.

One may ask why the hidden spins are not marginalized here, as compilation does. Nothing in the derivation requires it, and marginalizing would only complicate the gradient. A spin is ``hidden'' in this derivation exactly when $F$ ignores it, so hidden and output spins can be sampled and treated together, jointly playing the role of the single-kernel $y$ above, and $F$ is simply chosen not to read the hidden ones (and, if desired, some of the outputs too). Marginalizing $w$ would replace the energy $E_\vp$ in the score by the free energy $-\log \tk_\vp$, whose gradient requires the additional clamped-output expectation \eqref{eq:app_grad_psi}, with no benefit in return. This is also why the two training stages differ in the thermalizations they need. Compilation works at the level of the marginalized conditional $\Pf$ and needs both clamping patterns, while REINFORCE works at the level of the joint conditional $p_\vp(w, y \mid x)$ and needs only input-clamped ones.

\subsection{Why the energy-gradient mean is not an optional baseline}
\label{app:reinforce_baseline}

In ordinary REINFORCE one may subtract from $F$ any constant $b$ independent of the sampled $y$ without changing the gradient, because the score has mean zero, $\EE[\nabla_\vp \log p_\vp(y)] = 0$, so
\begin{equation}\label{eq:app_baseline_freedom}
    \EE\big[ (F(y) - b)\, \nabla_\vp \log p_\vp(y) \big]
    = \EE\big[ F(y)\, \nabla_\vp \log p_\vp(y) \big].
\end{equation}
This is what justifies the usual variance-reducing baselines. For an EBM, however, the energy gradient $-\nabla_\vp E_\vp(y)$ is not the score, since it differs from it by the model average $\EE[-\nabla_\vp E_\vp(y')]$, which is generally nonzero. Pairing the bare energy gradient with an arbitrary $(F(y) - b)$ therefore does not reproduce the gradient. For $b \neq \EE[F(y')]$,
\[
    \EE\big[ {-}\nabla_\vp E_\vp(y)\,(F(y) - b) \big] \;\neq\; \nabla_\vp \mathcal{L}(\vp),
\]
the two sides differing by $\big(\EE[F(y')] - b\big)\,\EE[-\nabla_\vp E_\vp(y)]$. Among baselines paired with the bare energy gradient, only the choice $b = \EE[F(y')]$ recovers the gradient, which is then the form~\eqref{eq:app_rg_F_baseline}. The mean $\EE[-\nabla_\vp E_\vp(y')]$ subtracted in~\eqref{eq:app_rg_energy_baseline} is thus not a free baseline but the $\nabla_\vp \log Z$ part of the score itself, and dropping it biases the gradient. Once the mean is restored, the full score $\nabla_\vp \log p_\vp(y) = -\nabla_\vp E_\vp(y) - \EE[-\nabla_\vp E_\vp(y')]$, which integrates to zero, is recovered, the baseline freedom of~\eqref{eq:app_baseline_freedom} returns, and an arbitrary constant may again be subtracted from $F$.

\subsection{Estimating the chain gradient}
\label{app:reinforce_estimator}

Equation~\eqref{eq:conditional_reinforce_grad} is estimated by ancestral sampling with one auxiliary draw per factor. We sample a single \emph{main} trajectory forward, $(w_\ell, z_\ell) \sim p_{\vp_\ell}(\cdot,\, \cdot \mid z_{\ell-1})$ for $\ell = 1, \ldots, L$, recording each factor's hidden spins alongside its output, and only the outputs $z_\ell$ are propagated forward. At each factor we additionally draw one \emph{reference} sample $(w_\ell', z_\ell') \sim p_{\vp_\ell}(\cdot,\, \cdot \mid z_{\ell-1})$ from the same parent $z_{\ell-1}$, independently of the main draw. The reference forms the factor-$\ell$ subtraction and is never propagated forward (Fig.~\ref{fig:ancestry-tree}). The single-sample estimator
\begin{equation}\label{eq:reinforce_estimator}
    \widehat{g}_\ell \coloneqq F(z_{1:L})\,\big( \Phi_{\vp_\ell}(z_{\ell-1}, w_\ell, z_\ell)
    - \Phi_{\vp_\ell}(z_{\ell-1}, w_\ell', z_\ell') \big)
\end{equation}
is then unbiased for $\nabla_{\vp_\ell}\LRF$, since the reference is drawn from the same conditional whose mean appears in~\eqref{eq:conditional_reinforce_grad} and is independent of the main rollout given $z_{\ell-1}$. Averaging $\widehat{g}_\ell$ over a minibatch of main trajectories reduces its variance, as does drawing several references per main sample~\cite{greensmith2004}. Both $\Phi_{\vp_\ell}$ evaluations in \eqref{eq:reinforce_estimator} are analytic, since they are spin products read off known configurations. Only obtaining the configurations requires sampling, at a total cost of $2L$ input-clamped thermalizations per rollout.

\begin{figure}[t]
\centering
\begin{tikzpicture}[>=Stealth, thick]
  \tikzset{
    ebm/.style={circle, draw, thick, minimum size=9mm, inner sep=1pt, align=center},
    samp/.style={inner sep=1pt, font=\normalsize}
  }
  \def\Ltree{4}
  \def\dxQ{1.05cm}  
  \def\dxY{0.85cm}  
  \def\dyT{1.05cm}  
  \node[ebm]  (q1) {$\Pf_1$};
  \node[samp, right=\dxY of q1] (y1) {$z_1$};
  \node[samp, below=\dyT of y1]  (yp1) {$z_1'$};
  \draw[->] (q1) -- (y1);
  \draw[->] (q1) -- (yp1);
  \foreach \l in {2,...,\Ltree} {
    \pgfmathtruncatemacro{\lmone}{\l-1}
    \node[ebm,  right=\dxQ of y\lmone] (q\l) {$\Pf_{\l}$};
    \node[samp, right=\dxY of q\l]     (y\l) {$z_{\l}$};
    \node[samp, below=\dyT of y\l]      (yp\l) {$z_{\l}'$};
    \draw[->] (y\lmone) -- (q\l);
    \draw[->] (q\l) -- (y\l);
    \draw[->] (q\l) -- (yp\l);
  }
\end{tikzpicture}
\caption{Ancestral sampling for the chain estimator~\eqref{eq:reinforce_estimator}. The main trajectory $z_1 \to z_2 \to \cdots$ runs along the top. At each factor the kernel emits, besides the main sample $(w_\ell, z_\ell)$, an independent reference $(w_\ell', z_\ell')$ from the same parent (downward arrow). Only the main outputs are propagated forward, and each reference enters only through the factor-$\ell$ subtraction.}
\label{fig:ancestry-tree}
\end{figure}

\subsection{Energy-gradient baseline versus \texorpdfstring{$F$}{F}-baseline}
\label{app:reinforce_fbaseline}

The covariance~\eqref{eq:app_rg_cov} can be rewritten so that the subtracted mean sits on the $F$ side instead, as in~\eqref{eq:app_rg_F_baseline}. Carried over to the chain, this $F$-baseline form is
\[
    \nabla_{\vp_\ell}\LRF = \EE\big[ \Phi_{\vp_\ell}(z_{\ell-1}, w_\ell, z_\ell)\,( F(z_{1:L}) - \bar F_\ell ) \big],
\]
where $\bar F_\ell = \EE[ F(z_{1:\ell-1}, z_\ell', \ldots, z_L') \mid z_{\ell-1} ]$ is the expected loss over a fresh reference rollout that branches at factor $\ell$, obtained by drawing $z_\ell' \sim \Pf_\ell(\cdot \mid z_{\ell-1}; \vp_\ell)$, propagating it forward through factors $\ell+1, \ldots, L$ to a reference endpoint, and evaluating $F$ there. Its single-sample estimator subtracts $F$ at a perturbed terminal rather than the local energy-gradient difference of~\eqref{eq:reinforce_estimator}.

The two forms are equal in expectation but differ in cost. The $F$-baseline needs, at every factor $\ell$, a reference rollout through all downstream factors $\ell+1, \ldots, L$, which over a chain of length $L$ amounts to $\sum_{\ell=1}^{L}(L - \ell + 1) = L(L+1)/2$ conditional samples per main trajectory, i.e.,\ $O(L^2)$. The energy-gradient subtraction of~\eqref{eq:reinforce_estimator} needs only the single extra draw at each factor, never propagated, for $O(L)$ samples in total. Since each conditional sample is itself a Gibbs run on a kernel, the constant in front is large and the $F$-baseline becomes impractical already at moderate depth. We also expect the energy-gradient form to have lower variance. Its per-factor signal $\Phi_{\vp_\ell}(z_{\ell-1}, w_\ell, z_\ell) - \Phi_{\vp_\ell}(z_{\ell-1}, w_\ell', z_\ell')$ is a difference of two energy gradients at the same parent that differ only in the factor-$\ell$ draw, whereas the $F$-baseline signal is a difference of two terminal-loss evaluations from rollouts whose downstream draws are independent. This variance comparison is heuristic, as we have not measured it directly.

\subsection{Linearized functionals of a readout law}
\label{app:reinforce_feff}

The second route for $F$ in \S\ref{sec:reinforce} optimizes a differentiable functional of a readout law. For the squared error $\mathcal{D}(\vp) = \lVert m_\vp - t \rVert^2$ with $m_\vp = \EE_{z_{0:L} \sim \Pmod_\vp}[f(z_L)]$, the chain rule and the score-function identity give
\begin{align*}
    \nabla_\vp \mathcal{D}
    &= 2\, (m_\vp - t)^\top \nabla_\vp m_\vp \\
    &= 2\, (m_\vp - t)^\top\, \EE\big[ f(z_L)\, \nabla_\vp \log \Pmod_\vp(z_{0:L}) \big] \\
    &= \EE\big[ F_{\mathrm{eff}}(z_L)\, \nabla_\vp \log \Pmod_\vp(z_{0:L}) \big],
\end{align*}
with $F_{\mathrm{eff}}$ as in \eqref{eq:f_eff}. This is exactly the REINFORCE gradient of the objective \eqref{eq:rf_loss} with the reward $F_{\mathrm{eff}}$. The mean $m_\vp$ enters $F_{\mathrm{eff}}$ only as a coefficient, which is estimated by the minibatch mean and treated as a constant under differentiation (a stop-gradient). The factor of two in $F_{\mathrm{eff}}$ comes from the product rule, so holding the coefficient fixed is exact rather than an approximation. Applying the per-factor decomposition \eqref{eq:conditional_reinforce_grad} with $F \to F_{\mathrm{eff}}$ completes the recipe used by the random-walk benchmark, Eq.~\eqref{eq:rw_leaf_loss}.

\subsection{Scalar-reward versus target-based objectives}
\label{app:objectives}

The REINFORCE gradient consumes a scalar reward, using neither a per-sample target nor a gradient of $F$ in sample space, which is what makes it applicable to discrete EBM samples that cannot be differentiated through. Both routes for $F$ in \S\ref{sec:reinforce} produce such a scalar, a discriminator score or the linearization of a marginal functional. A different family of generative objectives is instead \emph{target-based}. Score-based diffusion and denoising models~\cite{ho2020ddpm,song2021sde,karras2022edm} supply, for each sample, a target direction in output space and train by regression toward it. For an EBM this target-based signal is exactly what contrastive divergence already uses, a positive phase clamped to target samples. Within the present framework the diffusion-style objective is therefore realized not as a post-training reward but as the per-factor compilation of \S\ref{sec:compile}, since a denoising chain compiled step by step is a diffusion model in EBM form, which is precisely the DTM/TCM construction~\cite{jelincic2025dtm}. REINFORCE is complementary to it, supplying the global scalar-reward signal that per-step regression cannot provide, at the cost of the higher variance discussed in \S\ref{sec:reinforce}. This is why we treat the diffusion route as part of compilation rather than as an alternative post-training objective.

\section{Meta-EBM construction and methods}
\label{app:meta_ebm}

This appendix collects the formal construction behind the meta-EBM demonstration of \S\ref{sec:demos_meta}. We define the block-Gibbs kernels and schedules compiled by a meta-EBM, show why their restrictions are native, give the execution semantics and error law of the composed sampler, derive the kernel logit \eqref{eq:kernel_logit}, describe the numerical protocol of the $d=12$ demonstration, and extend the construction to continuous targets.

\subsection{Block-Gibbs kernels and schedules}
\label{app:meta_blockgibbs}

Let $\pi(x) \propto e^{-E(x)}$ be a logical target on $x \in \{-1,+1\}^d$, where ``logical'' means specified at the algorithm level with no commitment to a hardware realization. Given a block $A \subseteq \{1, \dots, d\}$ with complement $A^c$, the block-Gibbs kernel resamples $x_A$ from its exact conditional while holding $x_{A^c}$ fixed,
\begin{equation}\label{eq:blockgibbs}
  K_A(x' \mid x)
  \;=\;
  \mathbbm{1}\!\left[x'_{A^c} = x_{A^c}\right]\,
  \pi\!\left(x'_A \mid x_{A^c}\right),
\end{equation}
with the conditional given in \eqref{eq:meta_conditional}. Each kernel is in detailed balance with the target,
\begin{equation}\label{eq:detailed_balance}
  \pi(x)\, K_A(x' \mid x) \;=\; \pi(x')\, K_A(x \mid x') ,
\end{equation}
which holds by inspection, since both sides vanish unless $x$ and $x'$ agree on $A^c$, and otherwise both equal $e^{-E(x)}\, e^{-E(x')}$ over the common normalization $Z =\sum_{x''_A} e^{-E(x''_A,\, x_{A^c})}$. Every $K_A$ therefore leaves $\pi$ invariant.

A schedule fixes an ordered cover $A_1, \dots, A_{n_{\mathrm{blk}}}$ of the sites and applies the kernels in a fixed cyclic order (systematic scan), by drawing the next block at random (random scan), or, in the hardware-natural choice, over the color classes of a proper coloring of the interaction hypergraph (chromatic scan, \S\ref{sec:ebm})~\cite{gonzalez2011parallel}. Same-color sites share no hyperedge, so each site's Markov blanket lies in the complement of its color class. The color-class conditional therefore factorizes into single-site conditionals that can be updated in parallel, and one sweep is one pass over the color classes. Single-site Gibbs is the all-singletons schedule $A_j = \{j\}$. A fixed cycle of $\pi$-invariant kernels is $\pi$-invariant (and a random mixture of them likewise), and under mild irreducibility the composed chain is ergodic, so all these schedules share the same stationary law.

Block Gibbs is a useful compilation target because the conditional $\pi(x'_A \mid x_{A^c})$ is itself a normalized EBM over the block variables, with $x_{A^c}$ entering only as a conditioning input. Its energy is the restriction of the target energy to the block,
\begin{equation}\label{eq:restriction}
  E_A\!\left(x_A \mid x_{A^c}\right) \;=\; E\!\left(x_A,\, x_{A^c}\right),
\end{equation}
read as a function of $x_A$ with $x_{A^c}$ held as a parameter. This is the object that is rendered native by the compilation framework of \S\ref{sec:compile}, with $x_{A^c}$ playing the role of the clamped kernel input and the resampled block $x_A$ the role of the output. The restriction can be realized natively even when the full target cannot, because conditioning on $x_{A^c}$ lowers the order of every interaction that crosses the block's boundary. For single-site blocks on the three-body target of \S\ref{sec:demos_meta}, the restriction is the single-spin energy $-\theta_n(x_{\setminus n})\,x_n$, whose half log-odds is exactly \eqref{eq:logit}.

The composed sampler applies the compiled kernels $K_{A_1}, \dots, K_{A_{n_{\mathrm{blk}}}}$ one at a time, each application a fresh clamped thermalization with its hidden spins re-equilibrated, so the realized kernels $\tilde K_A$ are mutually independent approximations of the exact conditionals. This generality comes at a cost in the error accounting. The realized conditionals are in general the full conditionals of no common energy, so no common invariant measure exists, and the stationary bias of the composed sampler is governed by the perturbation bound of Appendix~\ref{app:inputs_model}, of order $\bar\varepsilon/(1-\rho_0)$ by \eqref{eq:deployed_floor}, the per-step floor amplified by the relaxation time. The residual $\bar\varepsilon$ and contraction $\rho$ are given concrete definitions in \S\ref{app:meta_numerics}.

\subsection{Derivation of the kernel logit}
\label{app:meta_kernel}

The single-site kernel of \S\ref{sec:demos_meta} is a pairwise Ising EBM on the clamped blanket inputs $x$, hidden spins $w \in \{-1,+1\}^{n_h}$, and one output $y$. With couplings entering the energy with a minus sign, as throughout the paper, its energy is
\begin{equation*}
  E(x, w, y) \;=\; -\bigl(J_{xy}^\top x + h_y\bigr)\, y
  \;-\; \sum_{a=1}^{n_h} \bigl(\alpha_a^\top x\bigr)\, w_a
  \;-\; \sum_{a=1}^{n_h} \beta_a\, w_a\, y ,
\end{equation*}
up to terms in $x$ alone, which cancel from the conditional. Here $\alpha_a$ collects the input couplings of hidden spin $a$ and $\beta_a$ is its coupling to the output. The hidden spins do not couple to one another, so the sum over them factorizes,
\begin{equation*}
  P(y \mid x) \;\propto\; e^{(J_{xy}^\top x + h_y)\, y}
  \prod_{a=1}^{n_h} 2\cosh\!\bigl(\alpha_a^\top x + \beta_a\, y\bigr),
\end{equation*}
each factor coming from $\sum_{w_a = \pm 1} e^{(\alpha_a^\top x + \beta_a y)\, w_a} = 2\cosh(\alpha_a^\top x + \beta_a y)$. The forward half log-odds follows by differencing at $y = \pm 1$,
\begin{equation*}
  \theta_y(x) \;=\; J_{xy}^\top x + h_y
  + \tfrac{1}{2} \sum_{a=1}^{n_h}
  \bigl[\log 2\cosh\bigl(\alpha_a^\top x + \beta_a\bigr)
  - \log 2\cosh\bigl(\alpha_a^\top x - \beta_a\bigr)\bigr].
\end{equation*}
The identity
\begin{equation}\label{eq:logcosh}
  \log\!\big(2\cosh c\big) \;=\; c \;+\; \softplus(-2c),
  \qquad \softplus(z) = \log\!\big(1 + e^z\big),
\end{equation}
converts each log-cosh difference into a softplus difference plus a linear part. The linear parts contribute a constant $\beta_a$ per hidden spin, absorbed into the output bias, and each hidden spin leaves exactly one softplus feature,
\begin{equation*}
  \theta_y(x) \;=\; J_{xy}^\top x + h_y + \sum_{a} \beta_a
  + \sum_{a=1}^{n_h}
  \tfrac{1}{2}\Bigl[
    \softplus \bigl(-2(\alpha_a^\top x + \beta_a)\bigr)
    -
    \softplus \bigl(-2(\alpha_a^\top x - \beta_a)\bigr)
  \Bigr].
\end{equation*}
This is the logistic-regression form quoted in the main text, an affine part plus one additive softplus feature per hidden spin. Equation \eqref{eq:kernel_logit} is recovered by the relabeling $\beta_a \to -\beta_a$, with the constant $\sum_a \beta_a$ absorbed into $h_y$. The relabeling leaves the compiled family unchanged, since the $\beta_a$ are free variational parameters.

Placement on the Z1 lattice adds an obstruction that is independent of the coupling cap. Laid out on the chessboard, the output and the hidden spins sit on opposite colors, and each blanket input takes the color that lets it reach its dominant partner. A spin that is simultaneously a pairwise neighbor and a three-body partner of site $n$ needs both an affine edge to the output and a bilinear edge to a hidden spin, and on a bipartite graph it cannot keep both. The lattice topology therefore contributes an irreducible per-site residual even before the cap is imposed, so topology and cap act as two independent sources of compilation error.

\subsection{Numerical protocol for the \texorpdfstring{$d=12$}{d=12} demonstration}
\label{app:meta_numerics}

All quantities entering Fig.~\ref{fig:results_banner} are computed exactly. The per-step residual $\bar\varepsilon$ is the worst-case total-variation distance between a compiled per-site kernel and the exact conditional, with the worst case running over all 12 per-site kernels and all blanket inputs. This is the uniform bound $\varepsilon = \max_\ell \eta_\ell$, with $\eta_\ell$ defined in Eq.~\eqref{eq:error-defn}, computed by exact enumeration. The contraction $\rho_0$ is the Dobrushin coefficient \eqref{eq:dobrushin} of the ideal systematic sweep. It involves only the target, which is why the main text reports it as fixed by the target rather than the substrate. One macro-step is one systematic sweep of the 12 sites. The marginal error at sweep $t$ is the exact total-variation distance $\tilde{\delta}_t=\|\tilde{q}_t-q_t\|_\TV$ between the laws of the compiled and ideal chains, started from a common initial distribution so that $\tilde\delta_0 = 0$ and computed by full enumeration of the $2^{12}$ states, and its large-$t$ plateau is the stationary bias with respect to the target $\pL$.

The cap sweep runs over $J_{\max} \in \{0.3,\, 0.5,\, 0.75,\, 1,\, 1.5,\, 2,\, 3,\, 6,\, 10\}$. At each cap we compile the per-site kernels by the constructive recipe of \S\ref{app:meta_kernel} with every coupling and field clipped to $|J| \le J_{\max}$, chain them, and run to stationarity. Each cap contributes one point to Fig.~\ref{fig:results_banner}b, with the predicted floor $\bar\varepsilon/(1-\rho_0)$ on the horizontal axis and the measured plateau on the vertical axis. The dashed line marks the bound $y = x$ and the dotted line the empirical $0.6\times$ trend. At the worst cap, $J_{\max} = 0.3$, the per-step residual is $\bar\varepsilon = 0.5052$, which reproduces the predicted floor $0.70215 = \bar\varepsilon/(1-\rho_0)$ up to the rounding of $\rho_0 = 0.28$.

All numbers quoted here and in \S\ref{sec:demos_meta} derive from the same exact enumeration.

\subsection{Continuous targets by linear encoding}
\label{app:meta_continuous}

The construction extends beyond discrete targets. A continuous logical target $\pi(u) \propto e^{-E(u)}$ on $u \in \mathbb{R}^d$ is brought into the pbit setting by a \emph{linear encoding}, in which each continuous coordinate is represented by a block of $b$ pbits and recovered as a fixed affine combination of them, $u = u_0 + R\,x$, with a fixed decoding matrix $R$ and $x \in \{-1,+1\}^n$ a vector of $n = d\,b$ pbits. A natural choice is a fixed-point expansion, $u_i = u_i^0 + \delta \sum_{k=0}^{b-1} 2^{k}\, x_{ik}$, which lays a uniform grid of $2^{b}$ levels per coordinate at resolution $\delta$. Finer resolution costs only more pbits per coordinate. Substituting $u = u_0 + Rx$ turns the continuous energy into a pbit energy $\hat E(x) := E(u_0 + Rx)$, to which the block-Gibbs compilation above (\S\ref{app:meta_blockgibbs}) applies verbatim. Because the encoding is linear it preserves the polynomial order of the energy, so a quadratic (Gaussian) target maps to a pairwise pbit energy and a degree-$p$ interaction in $u$ maps to a degree-$p$ interaction in $x$. Conditioning therefore still lowers the order of every interaction, and every block conditional remains native. The encoded energy $\hat E$ may itself be dense and high-order, since a pairwise term $u_i u_j$ already expands into the full $b \times b$ block of products $x_{ik}\, x_{jl}$. This is no obstacle, because the substrate never programs $\hat E$ directly, only the block conditionals it induces, and conditioning collapses each block conditional to a native kernel.

\section{Compiling Gaussian stochastic circuits: implementation and cost}
\label{app:gp_circuit}

\begin{figure}[tbp]
\centering
\includegraphics[width=0.6\columnwidth]{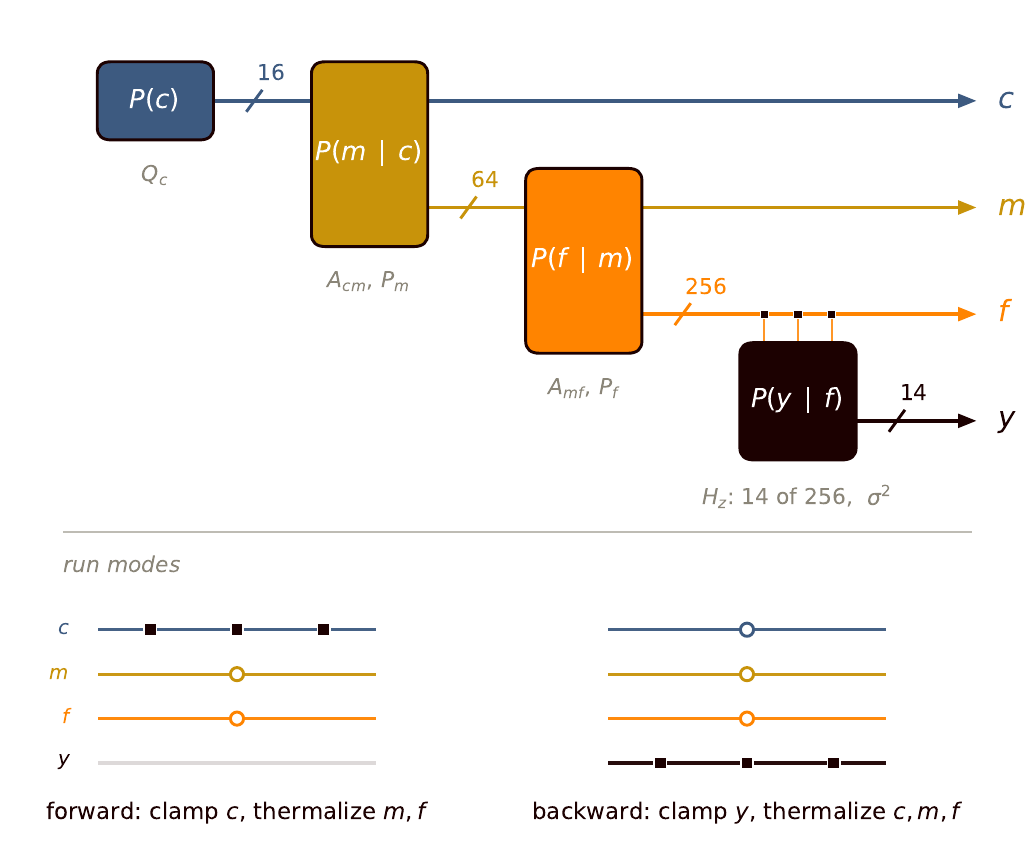}
\caption{\textbf{The program as a stochastic circuit.} Wires are the registers $c$, $m$, $f$ and the measurement record $y$, with bus widths $16$, $64$, $256$, and $14$. Each gate is one factor of Eq.~\eqref{eq:gp_model}, with its parameters listed beneath it. The measurement gate does not consume $f$. It reads the $14$ measured cells through the taps of $H_z$ and carries the clamp of Eq.~\eqref{eq:gp_clamp}. The two run modes of the same circuit appear below, where filled squares mark clamped wires and open circles mark thermalized ones. In the forward mode the measurement wire carries no clamp.}
\label{fig:gp_circuit}
\end{figure}

This appendix collects the implementation, error budget, energy projection, and embedding-scaling detail for the directed Gaussian demonstration of Sec.~\ref{sec:demos_gaussian}. Figure~\ref{fig:gp_circuit} shows the program as a stochastic circuit and marks the clamped and thermalized wires of each run mode.

\subsection{Posterior faithfulness and error budget}
\label{app:gp_loop}
\label{app:gp_budget}

\begin{figure}[tbp]
\centering
\includegraphics[width=0.5\columnwidth]{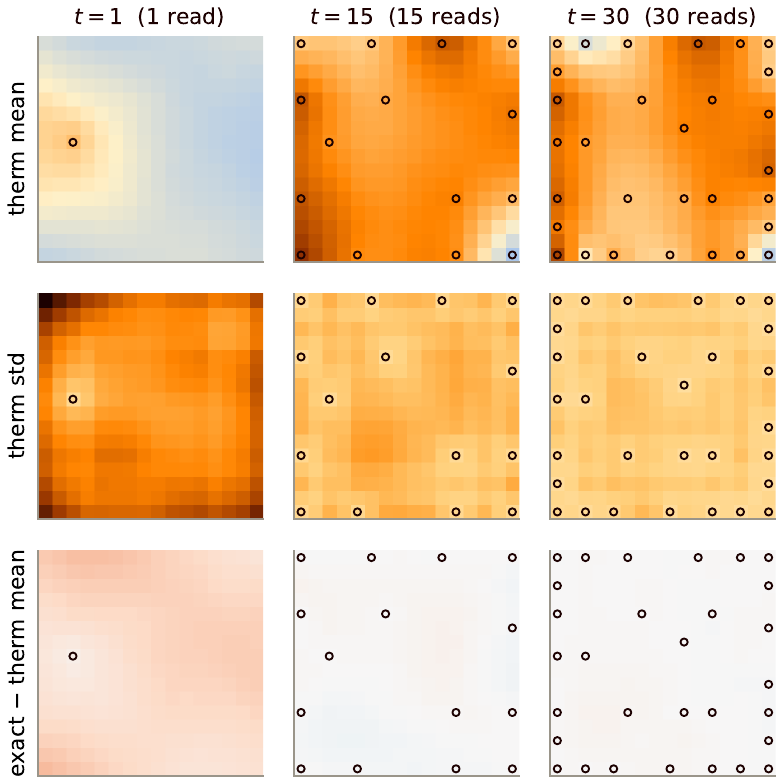}
\caption{\textbf{Sampled posterior across the design loop.} Fine-field posterior of the compiled THM at the first, middle, and last step. The top row shows the posterior mean, the middle row the per-cell standard deviation, and the bottom row the exact posterior mean minus the sampled mean on the same measurements, on the color scale of the mean. Circles mark measured sites. The difference is visible only at the first step, where the broad posterior mixes slowly, and washes out once the field is constrained.}
\label{fig:gp_loop}
\end{figure}

Matched RMSE curves do not by themselves rule out a miscalibrated posterior, so we also compare the sampled posterior against the exact posterior directly. On the compiled THM's own measurements the exact mean reaches RMSE $0.258$, matching the sampled $0.259\pm0.005$, and the sampled variance estimate agrees with the exact one to a median 4.3\% once the field is constrained (Fig.~\ref{fig:gp_loop}). The one large disagreement occurs at the first step, where the variance error reaches 41\%. With a single measurement the posterior is broad and the chain mixes slowly, so the variance map is near-uniform and the argmax is decided by noise. Warm starts do not bias the variance estimate. A cold-started chain matches the loop's variance error across the 24 fields, with a paired difference of $+0.5\pm0.4$ percentage points.

Any disagreement that remains between the sampled and exact posteriors comes from the compilation pipeline. Analytic compilation leaves zero KL residual, so the error budget of~\eqref{eq:mitigation_chain} reduces to bundle discretization and Monte Carlo error. Discretization shifts the posterior mean by 53.2\%, 2.1\%, and 1.8\% of the field spread at $4$, $6$, and $8$ bits, tracking the round-off scale $\Delta/\sqrt{12}$ (0.21, 0.050, 0.0125 in prior-std units). The eight-bit value is already at the Monte Carlo floor. The floor falls as $1/\sqrt{N}$, from a 15.0\% variance error at 600 samples to 4.7\% at 9{,}600, with a lag-one autocorrelation of 0.53 to 0.76 setting the effective sample size. The only term left is the embedding, which we defer to the on-substrate study.

\subsection{Embedding and energy cost}
\label{app:gp_energy}

At one bit per cell the compiled THM occupies $1774$ p-bits of the Z1 lattice, with the $336$ logical spins and $3140$ couplings on chains of length up to $18$ (Fig.~\ref{fig:gp_embedding}). The eight-bit bundles reuse the same coupling pattern. The embedding cost stated in Sec.~\ref{sec:demos_gaussian} includes a layout margin. Because the couplings are local, layout can push chain lengths toward the fine layer's $2.9$-p-bit average rather than the medium layer's $13.3$. Chain length matters for the energy projection, since the loop is sampling-bound and the substrate's energy advantage applies to the sampling workload.

The energy and time numbers below are cost-model projections on the Z1 side and converted wall-clock measurements on the GPU side, so neither is a measured silicon power trace. On the device model of \S\ref{sec:hardware} ($7.09~\mathrm{fJ}$ per p-bit per sweep, $20~\mathrm{ns}$ sweeps, $25~\mathrm{\mu s}$ readout; Table~\ref{tab:budget}, $50$\,MHz column), one design-loop step ($12$ chains, $120+300$ sweeps, eight-bit embedding) projects to $0.50~\mathrm{\mu J}$ and $7.5~\mathrm{ms}$, with readout dominating the time. The projection charges energy to the sweeps alone and carries no readout or idle term. Over the $30$-step loop this totals $15~\mathrm{\mu J}$. The same per-step update measured on an H100 (dense Cholesky and inverse, median of $500$ reps, launch-latency-bound at this size) takes $488~\mathrm{\mu s}$. At board power this corresponds to $195~\mathrm{mJ}$, a factor of $3.9\times10^{5}$ in energy. All results in this appendix were obtained on a block-Gibbs simulator of the compiled THM. We leave the on-substrate run to the hardware-deployment study.

\section{Thermodynamic consistency models}
\label{app:tcm}

A Thermodynamic Consistency Model (TCM) is a chain of Boltzmann-machine kernels built around a synthetic noise process, in which the chain starts from pure noise and every step maps its input directly to an approximation of clean data, refining the output of the step before it. TCMs are a variant of the Denoising Thermodynamic Models (DTMs) of Ref.~\cite{jelincic2025dtm}, and they take their name from consistency models~\cite{song2023consistency}, with which they share the property that every step targets clean data. A conditional TCM (\S\ref{app:tcm_conditional}) is the kernel compiled in the market-simulator demonstration of \S\ref{sec:demos_market}. The present appendix describes the construction, while a systematic study of TCMs on image benchmarks will be reported elsewhere. We first recall why a single expressive BM is hard to train and sample from (\S\ref{app:tcm_met}), then describe the noise process (\S\ref{app:tcm_forward}), the model (\S\ref{app:tcm_steps}, \S\ref{app:tcm_conditional}), and its training (\S\ref{app:tcm_training}) and inference (\S\ref{app:tcm_inference}) procedures.

\subsection{The mixing--expressivity tradeoff}
\label{app:tcm_met}

Suppose a single BM (\S\ref{sec:hardware}) is asked to model a complex data distribution: some of its spins are designated \emph{visible} and carry the data dimensions, the rest are hidden, and training shapes the energy so that the marginal of the Gibbs distribution \eqref{eq:bm_stat_dist} on the visible spins matches the data. If the data distribution has several well-separated modes, the trained energy must place a deep valley at each mode, and the valleys are necessarily separated by barriers. Gibbs sampling crosses a barrier only through a sequence of individually unlikely spin flips, so the crossing probability decays exponentially with the barrier height, and the mixing time, the number of sweeps after which the chain forgets its initialization, grows exponentially with the scale of the couplings.

Let us consider a two-spin example. Take $E(s_1, s_2) = -J s_1 s_2$ with $J > 0$, a model of the bimodal ``data'' $\{(+1,+1),\, (-1,-1)\}$. Both modes have energy $-J$ and the two intermediate states have energy $+J$, so moving between the modes requires passing through a state $2J$ above them, which the sampler does with probability $\propto e^{-2J}$, and the mixing time scales as $e^{2J}$. Modeling the data faithfully requires large $J$, while sampling within any fixed budget requires small $J$. This tension is the mixing--expressivity tradeoff (MET)~\cite{jelincic2025dtm} already encountered in \S\ref{sec:compile}. The tradeoff enters twice, biasing the truncated CD phases of Eq.~\eqref{eq:cd_identity} during training and inflating the number of sweeps the kernel needs per sample at deployment.

Clamped spins do not enter this tradeoff. A spin held at a fixed value contributes a constant term to the local field of each of its neighbors: it shifts the distribution the free spins relax to, but it creates no barrier among them. (Clamping $s_2 = +1$ in the example above reduces $s_1$ to a single Bernoulli draw for any value of $J$.) The inputs of a thermodynamic kernel, and any conditioning information, may therefore be coupled arbitrarily strongly to the free spins at no cost in mixing time. Both the denoising couplings and the conditioning introduced below exploit this property.

\subsection{The forward noise process}
\label{app:tcm_forward}

Denoising models circumvent the MET by splitting generation across several BMs, none of which has to represent the sharp data distribution on its own. The starting point is a forward process that destroys the data gradually. Let $z(t) \in \{-1,1\}^D$ with $z(0) \sim Q_0$, the clean-data distribution, and let every bit flip independently at rate $\lambda > 0$, so that
\begin{equation}\label{eq:tcm_flip}
  \Prob\big( z^{(i)}(t) \neq z^{(i)}(0) \big) \;=\; \tfrac{1}{2}\big(1 - e^{-\lambda t}\big).
\end{equation}
As $t \to \infty$ the state becomes uniform on $\{-1,1\}^D$ and the process retains no information about $z(0)$. Because the bits flip independently, the time-$t$ kernel factorizes per bit and can be written in an exponential form that slots directly into a BM energy,
\begin{equation}\label{eq:tcm_forward_kernel}
  Q^F_t(z' \mid z) \;\propto\; \exp\big\{ J(t)\, z' \cdot z \big\},
  \qquad
  J(t) \;=\; \arctanh\big(e^{-\lambda t}\big),
\end{equation}
which follows from \eqref{eq:tcm_flip}, since the odds of a bit being preserved rather than flipped are $e^{2 J(t)} = (1 + e^{-\lambda t})/(1 - e^{-\lambda t})$. The \emph{coupling weight} $J(t)$ diverges as $t \to 0$ (the kernel copies its input) and vanishes as $t \to \infty$ (the kernel ignores it).

Now fix noise levels $0 = t_0 < t_1 < \cdots < t_N = \infty$. A DTM~\cite{jelincic2025dtm} places one BM kernel per increment: step $n$ is trained to sample the reverse conditional $Q\big(z(t_n) \mid z(t_{n+1})\big)$, undoing one increment of noise, and generation runs the steps from $n = N$ down to $1$. If the increments are small, each reverse conditional is nearly unimodal (it mostly has to decide which few bits to unflip), so moderate weights suffice and each step mixes quickly.

\subsection{Direct-to-data steps}
\label{app:tcm_steps}

A TCM keeps this construction but changes the target of every step: step $n$ maps its input at noise level $t_n$ \emph{directly to clean data}, rather than one noise level down. Write $x$ for a state noised to time $t_n$ and $y$ for the clean state it came from. Bayes' rule against the forward kernel \eqref{eq:tcm_forward_kernel} gives the exact reverse conditional
\begin{equation}\label{eq:tcm_bayes}
  Q(y \mid x) \;\propto\; Q_0(y)\, \exp\big\{ J(t_n)\, y \cdot x \big\}.
\end{equation}
The right-hand side has two parts, a pairwise coupling term fixed entirely by the forward process and the clean-data log-probability $\log Q_0(y)$, which is the only unknown object.

Each TCM step approximates \eqref{eq:tcm_bayes} by a thermodynamic kernel in exactly the sense of \S\ref{sec:compile}: input spins carry $x$ and stay clamped, output spins carry $y$, hidden spins $w$ enlarge the hypothesis class, and the energy is
\begin{equation}\label{eq:tcm_energy}
  E_n(x, w, y;\, \vp_n) \;=\; -J(t_n)\, x \cdot y \;+\; E_{\vp_n}(w, y),
\end{equation}
a frozen one-to-one coupling of strength $J(t_n)$ between each input spin and the matching output spin, plus a trainable pairwise energy $E_{\vp_n}$ of the hardware's form \eqref{eq:bm_energy} on the output and hidden spins. A comparison of the kernel conditional $\Pf(y \mid x;\, \vp_n)$ of Eq.~\eqref{eq:forward} with \eqref{eq:tcm_bayes} shows that the trainable part, with its hidden spins summed out, has to approximate $-\log Q_0(y)$ up to a constant. At the noisiest level $J(t_N) = J(\infty) = 0$, so the input is ignored and the step reduces to an unconditional BM.

Targeting clean data from every level looks more ambitious than the DTM's incremental steps: at high noise, \eqref{eq:tcm_bayes} is close to the full multimodal data distribution, which is precisely the kind of distribution that the MET puts out of reach of a fast-mixing BM. However, no step is expected to represent it on its own. Each step is trained under an explicit mixing budget (\S\ref{app:tcm_training}) to be the best approximation available at that budget, and its output is a coarse clean-data sample, which the next step, less noisy and therefore more strongly coupled to its input, takes in and improves. In this way, the chain acts as an iterative refiner. A one-step TCM is already a complete, if crude, generative model, and every additional step only has to repair the residual errors of its predecessor.

On sparse hardware a further degree of freedom becomes available. Each step assigns its input and output spins to freshly drawn random sites of the hardware graph (\S\ref{sec:hardware}). On a local topology, two output spins at graph distance $d$ can only be correlated through paths of intermediate spins, and along a uniform chain of couplings $J$ the induced correlation decays as $\tanh(J)^d$, so that holding a fixed correlation across a growing distance requires growing couplings and, by the MET, slower mixing. Under a fixed mixing budget, each placement makes some pairwise correlations cheap to represent and leaves others out of reach. Drawing a different placement for every step allows different steps to specialize in different parts of the data's correlation structure, with the refinement chain combining their contributions.

\subsection{Conditional TCMs}
\label{app:tcm_conditional}

A TCM becomes a conditional model by giving every step a second block of always-clamped input spins carrying a context $c$ (in the market demonstration, the recent history of the panel). The context spins hold $c$ fixed during both CD phases and at inference, and the hardware-graph edges between them and the free spins are trainable, so the context enters the trainable energy $E_{\vp_n}(w, y, c)$ as a learned, input-dependent field on the hidden and output spins. Since clamped spins do not enter the MET (\S\ref{app:tcm_met}), this added expressivity comes at no cost in mixing time. Every step then targets the clean conditional: the forward process noises $y$ alone and leaves $c$ untouched, so \eqref{eq:tcm_bayes} holds with $Q_0(y)$ replaced by $Q_0(y \mid c)$. In particular, the noisiest step becomes a single conditional BM that samples an approximation of $Q_0(y \mid c)$ directly.

\subsection{Training}
\label{app:tcm_training}

Step $n$ is trained to minimize the expected divergence from the true reverse conditional,
\begin{equation}\label{eq:tcm_loss}
  \min_{\vp_n}\;
  \EE_{y \sim Q_0}\,
  \EE_{x \sim Q^F_{t_n}(\cdot \mid y)}
  \Big[ \KL\big( Q(\cdot \mid x) \,\big\|\, \Pf(\cdot \mid x;\, \vp_n) \big) \Big].
\end{equation}
Up to a $\vp_n$-independent constant, \eqref{eq:tcm_loss} is the conditional maximum-likelihood objective \eqref{eq:compile_mle} on input--output pairs $(x, y)$ obtained by noising data, so per-step TCM training is variational compilation under target inputs. In the classification of \S\ref{sec:compile} the target is sampled and the model is intractable, and the gradient is therefore the conditional-CD estimator \eqref{eq:cd_identity}: a positive phase clamping $x$ and $y$ and thermalizing $w$, and a negative phase clamping $x$ and thermalizing $(w, y)$, both truncated at $K$ sweeps. Because every step's training inputs come from the forward process rather than from other steps, the steps decouple and can be trained in parallel.

The truncation at $K$ sweeps is only sound while the step mixes within $K$ sweeps, and the MET implies that training pushes the step away from this regime. The \emph{correlation penalty} (CP)~\cite{jelincic2025dtm} counteracts this pressure directly. Let $m_i = \EE_{p_{\vp}}[s_i]$ be the model means of the free spins (at clamped inputs), and let $p_{\mathrm{MF}}(s) = \prod_i \tfrac{1}{2}(1 + s_i m_i)$ be the product distribution with those means. The penalty is the divergence of the model from its own product approximation,
\begin{equation}\label{eq:tcm_cp}
  \mathrm{CP}(\vp) \;=\; \KL\big( p_{\mathrm{MF}} \,\big\|\, p_{\vp} \big),
\end{equation}
and minimizing it pulls the free spins toward independence, flattening the energy barriers that slow mixing. Its gradient is
\begin{equation}\label{eq:tcm_cp_grad}
  \nabla_{\vp}\, \mathrm{CP}(\vp)
  \;=\;
  \EE_{p_{\vp}}\big[ \Phi_{\vp}(s) \big] \;-\; \Phi_{\vp}(m),
\end{equation}
where $\Phi_\vp = -\nabla_\vp E_\vp$ is the vector of sufficient statistics as in Eq.~\eqref{eq:cd_identity}, $\Phi_{\vp}(m)$ evaluates the spin products at the mean vector $m$, and $m$ is treated as a constant under the gradient. The first term coincides with the negative-phase statistics of CD, and $m$ is estimated from the same samples, so the penalty comes at a negligible additional cost. The training loss of step $n$ is the CD loss plus $\gamma_n \cdot \mathrm{CP}$, with a per-step coefficient $\gamma_n \geq 0$.

The right coefficient cannot be fixed in advance, since it depends on the data, the noise level, and the stage of training, and it is therefore set by feedback. Mixing is monitored through the autocorrelation of the negative-phase Gibbs chain at lag $K$, defined as the correlation between the free-spin configurations $K$ sweeps apart, averaged over spins and over clamped contexts. If the chain mixes within $K$ sweeps this autocorrelation is near zero, while a large value means that the CD phases are truncated too early and the gradient estimate is unreliable. The \emph{adaptive correlation penalty} (ACP)~\cite{jelincic2025dtm} fixes this issue: whenever the measured autocorrelation sits above a threshold $\varepsilon_{\mathrm{ACP}}$ and is not improving, $\gamma_n$ is increased multiplicatively, and whenever it sits below the threshold, $\gamma_n$ is decreased, down to a small floor. Training with ACP amounts to approximate constrained optimization, minimizing \eqref{eq:tcm_loss} subject to the step mixing within its sweep budget. The controller is insensitive to its own hyperparameters over broad ranges~\cite{jelincic2025dtm}.

\subsection{Inference}
\label{app:tcm_inference}

Generation runs the chain from the noisiest step to the cleanest. The state $x$ is initialized uniformly at random on $\{-1,1\}^D$, the marginal of the forward process at $t_N = \infty$. For $n = N, \ldots, 1$, step $n$ receives $x$ (and the context $c$, in the conditional case) on its clamped input spins, thermalizes the free spins, and returns the output $\hat y_n$, which for $n > 1$ becomes the input of the next step, $x \leftarrow \hat y_n$. The final output $\hat y_1$ is the generated sample. Each step's Gibbs run needs a number of sweeps on the order of the training horizon $K$, since ACP has kept its mixing time within that budget throughout training.

Between steps, the output could be partially re-noised toward the next step's training noise level, as consistency models do between successive applications~\cite{song2023consistency}. For BM chains, experiments outside the scope of this paper (part of the image-benchmark study mentioned above) find that feeding the output forward verbatim works better, since each step's output is an imperfect reconstruction whose errors already play the role of noise for the next step, and injecting more noise on top of them only degrades that step's input.

\section{Details of the market-simulator demonstration}
\label{app:market}

This appendix gives a self-contained account of the market-simulator demonstration of \S\ref{sec:demos_market}: the data and the modeling task (\S\ref{app:market_task}), the discrete state representation (\S\ref{app:market_disc}), the target kernel and the compiled program (\S\ref{app:market_kernel}), the evaluation protocol (\S\ref{app:market_eval}), the architecture of the compiled kernel (\S\ref{app:market_arch}), the two training stages (\S\ref{app:market_vc}, \S\ref{app:market_rg}), and additional results (\S\ref{app:market_addl}). This demonstration represents an example where the target is a real system known only through recorded trajectories (target case (c)), and the compiled kernel, a conditional thermodynamic consistency model (Appendix~\ref{app:tcm}) trained by conditional contrastive divergence, is intractable (model case (2)). 

\subsection{The modeling task}
\label{app:market_task}

The data is a panel of $N = 14$ daily financial time series chosen to cover the major asset classes: US and international stock indices, US government bonds (Treasuries), corporate credit, exchange rates, gold, oil, copper, and a stock-market volatility index (Table~\ref{tab:market_panel}). Each series contributes one number per trading day, the day's \emph{move}, computed as the change of the logarithm of the price (the log-return) or, for the volatility index, of the log of the index level. The panel spans April 2007 to July 2026, that is 4{,}802 trading days, including the 2008 financial crisis, the crises of 2011 and 2015, the COVID crash of 2020, and the 2022 interest-rate shock. All series are taken from Yahoo Finance. Wherever an asset class is represented by an exchange-traded fund rather than an index, we make the deliberate choice to use the price of the fund as an actually traded quantity, so the series is free of the hindsight biases of reconstructed index histories. The panel is aligned to US trading days, and days on which any series is missing are dropped.

\begin{table}[t]
\caption{The market panel: 14 daily series across the major asset classes. All series enter as daily log-returns of the (dividend-adjusted) closing price, except the VIX, which enters as the daily change of the log index level. Data from Yahoo Finance, April 2007 to July 2026.}
\label{tab:market_panel}
\footnotesize
\begin{tabular}{lll}
    \toprule
    \textbf{Asset class} & \textbf{Series} & \textbf{Tracks} \\
    \midrule
    US equity & SPY, QQQ, IWM & S\&P 500, Nasdaq-100, Russell 2000 \\
    Intl.\ equity & EFA, EEM & developed ex-US, emerging markets \\
    Treasuries & TLT, IEF & $\geq 20$-year, 7--10-year US bonds \\
    Credit & HYG & high-yield corporate bonds \\
    FX & EURUSD, USDJPY & euro--dollar, dollar--yen rates \\
    Commodities & GLD, USO, HG & gold, WTI oil, copper futures \\
    Volatility & VIX & S\&P 500 implied volatility \\
    \bottomrule
\end{tabular}
\end{table}

A large body of evidence indicates that markets are close to informationally efficient: the direction of tomorrow's movement is nearly unpredictable from public history, since any predictable pattern would be traded away~\cite{fama1970efficient}. For this reason we do not attempt to build a predictor; instead we focus on distributional structure, statistical regularities of the joint law of the moves which are stable across decades and markets, known in empirical finance as the stylized facts of asset returns~\cite{cont2001empirical}. In this demonstration we target the following:
\begin{enumerate}[label=(\roman*), leftmargin=*]
    \item \emph{No return autocorrelation.} The sign of today's move carries almost no information about the sign of tomorrow's, which is the efficiency statement above.
    \item \emph{Volatility clustering.} The size of the moves, unlike their direction, is predictable: calm days cluster into calm stretches and turbulent days into turbulent stretches, with the autocorrelation of the move magnitudes decaying slowly over weeks~\cite{mandelbrot1963variation,cont2001empirical}. This is the structure that the classical ARCH/GARCH model families were built to capture~\cite{engle1982arch,bollerslev1986garch}.
    \item \emph{Cross-asset correlation.} Same-day moves are strongly dependent: stock indices move together, and stocks and government bonds tend to move in opposite directions.
    \item \emph{Tail co-movement.} Extreme moves arrive together far more often than independence would allow, i.e.,\ markets crash jointly~\cite{longin2001extreme}.
    \item \emph{Flight to quality.} The dependence structure itself shifts with market conditions: on days when stocks fall, money tends to move into government bonds, so that the stock--bond relationship on down days differs from the one on up days~\cite{baur2009flights}.
\end{enumerate}

A generative model of this structure is a market simulator, a source of synthetic multi-year market trajectories which are statistically realistic without replaying history. Synthetic data of this kind is used to stress-test portfolios, estimate risk, and backtest trading strategies beyond the single trajectory that actually occurred~\cite{assefa2020generating}. Learned market simulators are an active topic of research, with proposed generators based on restricted Boltzmann machines~\cite{kondratyev2020market}, generative adversarial networks~\cite{wiese2020quant}, and signature-based models~\cite{buehler2020data}. Furthermore, the task fits the framework of this paper naturally: the object to learn is a joint, multimodal conditional distribution (mostly-up days and mostly-down days are separate modes, coupled across all 14 series), which is the regime where a single Boltzmann machine runs into the mixing--expressivity tradeoff and the multi-step TCM chain becomes advantageous.

\subsection{Discretization}
\label{app:market_disc}

Per series, a day's move is mapped to one of eight signed levels, given by the sign of the move together with a magnitude bucket $m \in \{1, 2, 3, 4\}$ determined by the quartiles of the absolute moves of that series. The quartile boundaries are fit per series on the training split only, so that no information from held-out data enters the representation. Each level is encoded in 4 bits, one sign bit and a 3-bit thermometer code for the magnitude (bits $m \geq 2$, $m \geq 3$, $m \geq 4$), and a day of the panel is then a state $x \in \{-1, 1\}^{56}$. The thermometer code is chosen for robustness: every 4-bit pattern decodes to a valid level (sign bit plus the count of set magnitude bits), and Hamming distance approximates level distance, so that a single flipped spin moves a series by one magnitude bucket rather than to an arbitrary level.

It is important to note that, with quantile binning, the four magnitude buckets of every series are equally likely by construction (on the training data). The shape of the marginal distribution of each series, which for asset returns is heavy-tailed~\cite{cont2001empirical}, is therefore erased by the representation and is accordingly not among the evaluated statistics. All learnable structure, and all evaluation, concerns the dependence between series and across time.

\subsection{Target kernel and compiled program}
\label{app:market_kernel}

Write $x_t \in \{-1,1\}^{56}$ for the state of the panel on day $t$ and
\begin{equation*}
    c_t \;\coloneqq\; (x_{t-B}, \ldots, x_{t-1})
\end{equation*}
for the context of the $B$ preceding days. The target is the one-day transition kernel
\begin{equation}\label{eq:market_kernel}
    P\big(x_t \mid c_t\big),
\end{equation}
i.e.,\ the panel is modeled as a Markov chain of order $B$. This truncation is a modeling choice: real markets carry memory longer than any fixed $B$, and $B$ sets how much of it is seen by the kernel. We swept $B = 1, \ldots, 6$ and found it to be the single most influential hyperparameter, acting almost entirely through the volatility-clustering error of \S\ref{app:market_eval}, which roughly halves from $B = 1$ to $B = 5$ with little further gain at $B = 6$. The final model uses $B = 5$, giving a context of $280$ bits.

The kernel \eqref{eq:market_kernel} is compiled to a thermodynamic kernel $\Pf(x_t \mid c_t;\, \vp)$ realized as a conditional TCM (Appendix~\ref{app:tcm_conditional}), whose internal structure is described in \S\ref{app:market_arch}. The simulator is the rollout of this kernel as a directed factor graph: starting from a real $B$-day window $c_1$, we sample $x_1 \sim \Pf(\cdot \mid c_1; \vp)$, form $c_2$ by appending $x_1$ and dropping the oldest day, and repeat,
\begin{equation}\label{eq:market_rollout}
    \Pmod_\vp\big(x_{1:L} \mid c_1\big) \;=\; \prod_{t=1}^{L} \Pf\big(x_t \mid c_t;\, \vp\big).
\end{equation}
The same kernel, with the same parameters, is applied on every day, i.e.,\ the program is a weight-tied chain over process time (Fig.~\ref{fig:market_dfg}).

The training signal comes from the recorded panel. Every day of the training split whose $B$ predecessors also lie in the split yields one $(c, x)$ pair, for a total of about $3{,}300$ pairs. This is target case (c) trajectory data of \S\ref{sec:compile}. The contexts of the pairs are automatically distributed as the input marginal of the target, so that compilation trains under target inputs by default. However, the market cannot be re-run at inputs of our choosing, so querying the target at model-generated contexts, the ingredient of model-context matching (\S\ref{sec:input_distributions}), is not possible. On the model side, the conditional \eqref{eq:market_kernel} over $2^{56}$ states, realized with over a thousand hidden spins, cannot be normalized, which places the model in case (2) intractable kernel, trained with the conditional-CD estimator of Eq.~\eqref{eq:cd_identity}.

\subsection{Evaluation protocol}
\label{app:market_eval}

In this demonstration, the target kernel is not known explicitly here which means that there is no reference against which to compute a per-input total variation. The model is therefore judged at the readout level: long feedback rollouts of \eqref{eq:market_rollout} are compared against held-out real data on the stylized facts of \S\ref{app:market_task}. This is also the deployment condition, and the most demanding test available, since from day one the kernel consumes its own outputs as context and any drift left uncontrolled by the compilation compounds over the rollout.

\paragraph{Data split.}
Ten contiguous windows of 120 trading days each, placed at random along the timeline, are held out for evaluation ($1{,}200$ days). The remaining $3{,}402$ days form the training split. Each held-out window is flanked by a 10-day buffer excluded from both splits, so that no training context overlaps held-out days (a purged split with embargo, standard practice for serially dependent financial data~\cite{lopezdeprado2018advances}; the buffer exceeds $B$). We use random windows instead of the more obvious chronological split (train on the past, hold out the most recent years) because the dependence structure of financial data shifts over decades, most prominently with the 2022 interest-rate shock which reversed the usual negative stock--bond correlation, so that a chronological holdout measures regime novelty rather than model fit. Under the random-window design both splits share the same mix of market conditions. The chronological split is retained and reported separately as a regime-shift stress test.

\paragraph{Rollout geometry.}
All reported scores use 256 independent rollouts of $1{,}200$ days each, started from real held-out $B$-day windows. Autocorrelation-type statistics are computed on 120-day segments, both of the rollouts and of the real held-out windows, because such statistics depend on the length of the segment they are estimated on. Fixing one segment length, equal to the length of the holdout windows, keeps real and simulated values comparable. Every system, the compiled models as well as all baselines, is scored with identical geometry.

\paragraph{Scored statistics.}
Three error terms, each targeting one stylized fact, are computed between the simulated and the real held-out trajectories, on the signed levels:
\begin{itemize}[leftmargin=*]
    \item $\delta_{\mathrm{corr}}$, \emph{cross-asset correlation}: the root-mean-square difference of the off-diagonal entries of the $14 \times 14$ correlation matrix of same-day move signs.
    \item $\delta_{\mathrm{vol}}$, \emph{volatility clustering}: we form the daily panel-wide volatility series, the absolute level averaged over the 14 series on each day, and take the $\ell_2$ distance between its real and simulated autocorrelation functions at lags 1--20. Aggregating across the panel before the autocorrelation is the discriminating choice: independent per-series volatility clustering averages out across 14 series, so the panel-wide volatility is persistent only if volatility moves in common, a joint temporal feature which neither a static cross-sectional model nor independent per-series dynamics can produce.
    \item $\delta_{\mathrm{tail}}$, \emph{tail co-movement}: the summed absolute error, over $m \in \{4, 6, 8\}$, of the probability that at least $m$ of the 14 series sit in their extreme-down level on the same day.
\end{itemize}
Each term is normalized by the same error computed for the baseline that lacks the corresponding structure (defined below), and the three ratios are summed,
\begin{equation}\label{eq:market_composite}
    \mathcal{C} \;=\;
    \frac{\delta_{\mathrm{corr}}}{\delta_{\mathrm{corr}}^{\mathrm{MK}}}
    + \frac{\delta_{\mathrm{vol}}}{\delta_{\mathrm{vol}}^{\mathrm{IID}}}
    + \frac{\delta_{\mathrm{tail}}}{\delta_{\mathrm{tail}}^{\mathrm{MK}}}.
\end{equation}
A model with every ratio below one beats each baseline on the axis that baseline is built to fail. The composite error $\mathcal{C}$ is used for model selection and ranking throughout the demonstration. One further statistic is monitored as a sanity check but not scored: the maximum absolute sign autocorrelation over lags 1--20, which is $\approx 0.04$ on the real held-out data and must remain comparably small in the rollouts, since a simulator must not introduce the momentum ruled out by stylized fact (i). Statistics beyond the scored ones (the co-movement of extreme up-moves, the flight-to-quality conditional correlations, and the correlation between today's sign and future magnitudes) are tracked but not scored.

\paragraph{Baselines.}
Two training-free baselines probe the two sides of the problem, the cross-section and the dynamics. \emph{IID day-resampling} draws whole days (full 14-series cross-sections) with replacement from the training split: it reproduces the unconditional cross-sectional distribution exactly, including $\delta_{\mathrm{corr}}$ and most of $\delta_{\mathrm{tail}}$, but has no temporal structure at all, so it fails $\delta_{\mathrm{vol}}$ and supplies $\delta_{\mathrm{vol}}^{\mathrm{IID}}$. \emph{Independent per-series Markov chains} fit one 8-state first-order transition matrix per series by counting (with add-one smoothing): they capture per-series dynamics but render the series independent, so they fail the two cross-sectional axes and supply $\delta_{\mathrm{corr}}^{\mathrm{MK}}$ and $\delta_{\mathrm{tail}}^{\mathrm{MK}}$. The third comparison is the \emph{single conditional BM}, the noisiest step of the compiled TCM run alone, which by Appendix~\ref{app:tcm_conditional} is itself a complete conditional model of \eqref{eq:market_kernel}. This comparison is meant to isolate the advantage provided by the multi-step chain.

\subsection{Architecture of the compiled kernel}
\label{app:market_arch}

The kernel is a two-step conditional TCM. Each step is a BM on a $40 \times 40$ grid with the Z1 connection rules of \S\ref{subsec:graph_arch}, here with periodic boundaries and with no cap imposed on the coupling magnitudes: $1{,}600$ spins, of which 56 are the output spins carrying $x_t$, 280 are the always-clamped context spins carrying $c_t$, and the remaining $1{,}264$ are hidden. The denoising input occupies 56 further clamped spins coupled one-to-one to the output spins. The noise levels are $(t_1, t_2) = (0.7, \infty)$ at unit flip rate $\lambda = 1$. The noisiest step has coupling $J(\infty) = 0$: it ignores its denoising input and directly samples an approximation of $P(x_t \mid c_t)$, which is the single-conditional-BM baseline defined above. The cleaner step receives that sample through the coupling $J(0.7) \approx 0.54$ and refines it; during its training, the denoising input is the true next day with each bit flipped with probability $(1 - e^{-0.7})/2 \approx 0.25$. Sweeping the noise level of the cleaner step over $[0.1, 0.7]$ changed little, and a third step brought no further gain.

The placement of the visible spins on the grid turned out to be an influential design choice: across the layout families we explored, placement alone moved the composite error\eqref{eq:market_composite} by several tens of percent, and we describe only the final recipe. All output and context spins are confined to one color class of the bipartite lattice, so that no two visible spins are adjacent and every visible--visible correlation must route through hidden spins. Within that class, the bits of each asset are placed as a local cluster: the spins of the asset are scattered around a randomly drawn center with a Gaussian density of scale $\sigma$ (in grid units). On the noisiest step, the 4 output spins and the 20 context spins of each asset form two separate, wide clusters ($\sigma = 8$). On the cleaner step, the output spins of each asset are placed amid its own context spins in a single tight cluster ($\sigma = 2$), so that the refining step finds the recent history of each asset in the immediate neighborhood of the spins it must refine. As everywhere in a TCM, the placements are re-drawn independently for each step (Appendix~\ref{app:tcm_steps}).

\subsection{Training stage 1: variational compilation}
\label{app:market_vc}

Each step is trained independently by the conditional-CD procedure of Appendix~\ref{app:tcm_training}, using minibatches of $(c, x)$ pairs from the training split, with $x$ freshly noised to the level of the step at every visit, a Gibbs horizon of $K = 800$ sweeps per phase, and Adam at learning rates $5$--$8 \times 10^{-3}$, annealed roughly tenfold over $1$--$3 \times 10^{4}$ updates. Mixing is controlled by the adaptive correlation penalty~\cite{jelincic2025dtm} at autocorrelation thresholds of $0.03$--$0.05$, with one refinement specific to this demonstration: since the kernel is conditional, the penalty targets the covariance of the free spins given the clamped context, estimated across several independently initialized chains per context and averaged over contexts. Covariance induced by varying the context reflects the model correctly responding to its input and is not penalized; only within-context covariance reflects slow mixing.

Model checkpoints are selected on deployed behavior: every few epochs, the current parameters are scored with a short version of the rollout protocol (256 rollouts of 240 days), and the checkpoint with the best composite error is kept. This selection is needed because most runs reach their best rollout composite error in the first quarter of training and then degrade, even as one-step-ahead fit statistics stay flat, a pattern consistent with the bias of the CD estimator growing as training sharpens the kernel (a manifestation of the MET). Tighter ACP thresholds slowed the degradation but did not remove it, so we select the best checkpoint rather than the last one.

\subsection{Training stage 2: trajectory-level REINFORCE}
\label{app:market_rg}

Compilation controls the error of the kernel on real contexts. Deployed as a simulator, however, the kernel sees its own outputs: after a few simulated days every context is model-generated, and the input distribution drifts away from the data the kernel was fitted on. The correction designated by the framework for this situation, re-training under model inputs (\S\ref{sec:input_distributions}), is unavailable in target case (c) trajectory data, since it requires querying the target conditional at model-generated contexts and the market cannot be queried. Trajectory-level REINFORCE (\S\ref{sec:reinforce}) is the one stage of the framework that trains the program under the conditions of its deployment, and this demonstration uses it as a second training stage.

The trajectory is the rollout itself: a real 5-day training window seeds the chain \eqref{eq:market_rollout}, which generates 20 consecutive days, all through the same weight-tied kernel. One update processes a batch of 128 such rollouts with Adam at learning rate $10^{-4}$. Gradients follow the matched-reference estimator of Appendix~\ref{app:reinforce} with four reference samples per factor. All weights are trainable, including the inter-step couplings: the synthetic noise process of Appendix~\ref{app:tcm} plays no role in this stage, and the coupling values it prescribed are merely the initialization of ordinary parameters. Per-BM ACP runs throughout, with the autocorrelation now measured at contexts drawn from the model's own rollouts, since that is where the deployed kernel must mix. The target cannot be evaluated at model-generated contexts, but the mixing of the model can be measured there, so this half of the model-context idea survives in case (c) trajectory data.

The reward follows the linearized-functional route of \S\ref{sec:reinforce}, Eq.~\eqref{eq:f_eff}: the objective is a weighted squared error between the mean statistics of the generated 25-day windows (context plus generated days) and the same statistics averaged over all real 25-day training windows. The statistics vector contains the 91 pairwise same-day sign products, the autocovariance of the panel-wide volatility of the window at lags 1--10, the panel-averaged sign autocovariances at lags 1--3 (whose real targets are $\approx 0$), the frequencies of days with at least $m \in \{3, 4, 5, 6, 8\}$ series at an extreme level, and the mean absolute level; the weights are inverse standard deviations over the real windows. All entries are raw products or frequencies, so the linearization behind Eq.~\eqref{eq:f_eff} is exact. One update costs a few tens of thousands of clamped thermalizations (128 rollouts $\times$ 20 days $\times$ 2 steps, each with one main and four reference draws), and the selected model is reached after roughly $1.5 \times 10^4$ updates, with the same short-rollout checkpoint selection as in stage 1 (and the same tendency to degrade past the peak, which makes moderate learning rates the reliable choice).

These statistics were chosen to parallel the three scored axes of \S\ref{app:market_eval}, plus the no-momentum sanity axis. It is important to note that post-training therefore directly optimizes short-window, training-split relatives of the statistics scored by the evaluation. The evaluation then tests generalization along two axes not contained in the training signal, from 25-day windows to $1{,}200$-day feedback rollouts and from the training split to held-out data. We did not measure statistics beyond the scored ones, so the improvement shows that the learning transfers, but it does not establish that quantities absent from the reward would improve as well.

\subsection{Additional results}
\label{app:market_addl}

This section collects supporting results: the behavior of statistics not scored by the composite, the effect of the context length, the training dynamics of both stages, and a qualitative look at the simulated trajectories.

\begin{figure*}[t]
    \centering
    \includegraphics[width=\textwidth]{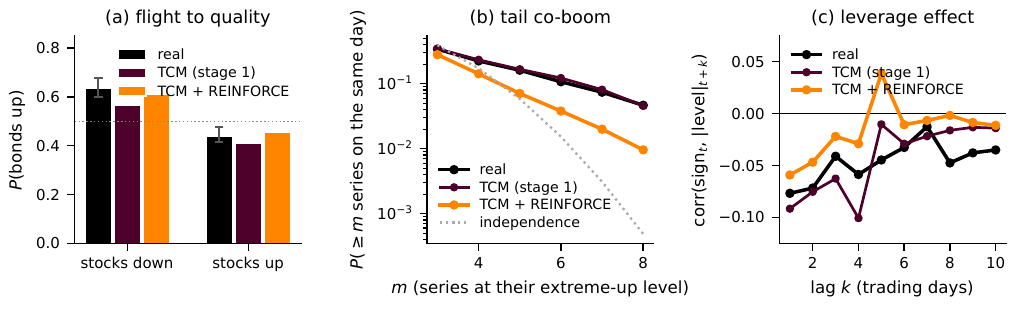}
    \caption{\textbf{Statistics outside the composite}, real vs.\ both training stages (holdout protocol of Fig.~\ref{fig:market_results}). \textbf{(a)}~Flight to quality: the probability that the long-maturity Treasury series (TLT) rises on a day when the equity index (SPY) falls vs.\ rises; whiskers are 95\% bootstrap intervals on the real values. \textbf{(b)}~Tail co-boom, the up-side mirror of Fig.~\ref{fig:market_results}c: the probability that at least $m$ series post an extreme up-move on the same day. Stage 1 tracks the real curve, while post-training loses much of the joint up-tail (see text). \textbf{(c)}~Leverage effect: correlation between the sign of today's move and the move magnitude $k$ days later, averaged over the five equity series; negative values mean that down days precede turbulent days.}
    \label{fig:market_unscored}
\end{figure*}

\paragraph{Statistics outside the composite.}
Figure~\ref{fig:market_unscored} tracks three statistics not scored by the composite \eqref{eq:market_composite}, before and after post-training. Two survive post-training intact. Flight to quality (panel a): on days when stocks fall, bonds rise with probability $0.64$ in the real data, versus $0.44$ on days when stocks rise, and both stages reproduce the asymmetry ($0.57$ vs.\ $0.41$ after stage 1, $0.61$ vs.\ $0.45$ after stage 2). The leverage effect (panel c): in real markets a down day tends to be followed by turbulent days, which appears as a negative correlation between the sign of a day's equity move and the size of the moves that follow, and both stages show it at roughly the real size. The third statistic is traded away. Tail co-boom (panel b) is the up-side mirror of the scored tail term, the probability that many series post an extreme up-move on the same day. Stage 1 matches it closely ($0.046$ against the real $0.047$ at $m = 8$), while the post-trained model largely loses it ($0.010$).

This behavior originates in the reward. Its co-extreme counts are two-sided, a series counts as extreme regardless of the sign of its move, and the scored tail term covers down-moves only, so that joint up-moves are pinned by neither. The optimizer met the two-sided counts mostly with down-moves and spent the capacity of the model on the statistics with the largest remaining error, above all the volatility ones. This is not a failure of the estimator: REINFORCE reduced exactly the loss it was given, and any property the loss does not measure is available to be traded for ones it does measure. The burden placed on the user is loss design, since a per-sample reward must contain every property the application needs preserved. A learned reward such as a discriminator score (\S\ref{sec:reinforce}) could in principle guard the whole distribution at once rather than a prescribed list of moments; we leave this route to future work. The momentum sanity check of \S\ref{app:market_eval} tells a milder version of the same story: the largest absolute sign autocorrelation is $0.044$ in the real held-out data, $0.031$ after stage 1, and $0.055$ after stage 2. The value remains small, but post-training, whose reward pins the sign autocovariances only at lags 1--3, is the stage that pushes it past the real value.

\begin{figure}[t]
\centering
\includegraphics[width=3.35in]{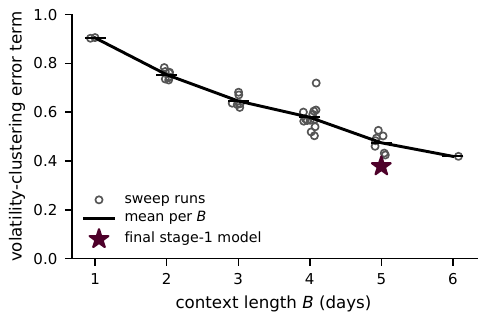}
\caption{\textbf{Context length $B$ vs.\ the volatility-clustering error term} for all sweep models (open circles; per-$B$ means in black; star: the final stage-1 model). The runs differ in grid size and training budget, so individual points are not directly comparable; the trend in $B$ dominates.}
\label{fig:market_b_ladder}
\end{figure}

\paragraph{Context length.}
Figure~\ref{fig:market_b_ladder} isolates the hyperparameter that mattered most, the context length $B$. It plots the volatility-clustering error term, the axis through which $B$ acts (\S\ref{app:market_kernel}), for every model of the hyperparameter sweeps, against the $B$ of that model. The term falls steadily as the kernel is shown more history, from $\approx 0.90$ at $B = 1$ to $\approx 0.47$ at $B = 5$, with little further gain at $B = 6$.

\begin{figure*}[t]
    \centering
    \includegraphics[width=\textwidth]{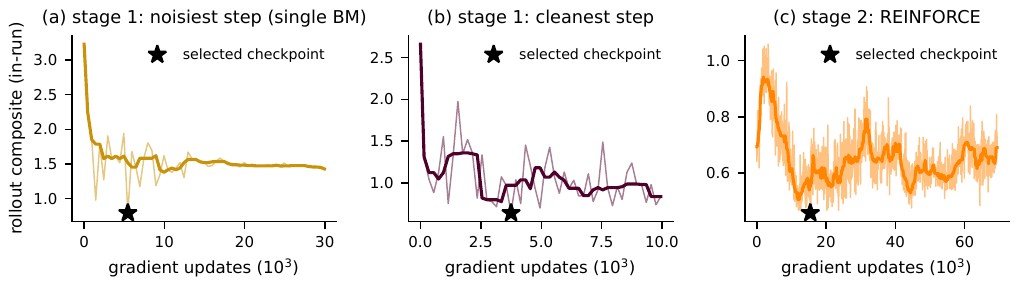}
    \caption{\textbf{Training curves of the three runs behind the final model:} the in-run rollout composite (256 rollouts of 240 days, single noisy evaluations at regular intervals during training; thin line raw, thick line running median) against gradient updates. Stars mark the selected checkpoints. In-run values are noisier and slightly offset from the definitive $1{,}200$-day numbers quoted in the text. \textbf{(a)}~The noisiest stage-1 step (the single conditional BM). \textbf{(b)}~The cleanest stage-1 step. \textbf{(c)}~Stage-2 REINFORCE, which shows a transient degradation before its gains.}
    \label{fig:market_curves}
\end{figure*}

\begin{figure}[ht!]
\centering
\includegraphics[width=3.35in]{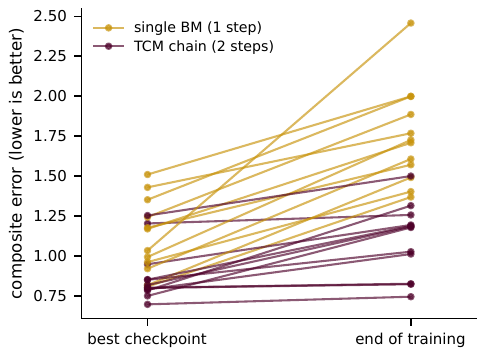}
\caption{\textbf{Best-checkpoint vs.\ end-of-training composite} for twelve matched single-BM/chain pairs trained from scratch for 300 epochs under otherwise identical settings. Both model types degrade when trained past their best point (the CD-bias mechanism of \S\ref{app:market_vc}), the single BMs much more steeply.}
\label{fig:market_checkpoint}
\end{figure}

\paragraph{Training dynamics and checkpoint selection.}
Figure~\ref{fig:market_curves} shows the rollout composite over the course of training for the three runs behind the final model, the two stage-1 steps and the stage-2 post-training. The shape described in \S\ref{app:market_vc} is visible in all three: quality improves quickly, peaks well before the training budget is exhausted, and then degrades or wanders, so that the selected checkpoint (star) predates the end of training in every case. Figure~\ref{fig:market_checkpoint} shows that the size of the loss past the peak depends on the type of model. On a homogeneous campaign, twelve single-BM/chain pairs from the layout exploration of \S\ref{app:market_arch} trained from scratch for 300 epochs under otherwise identical settings, everything degrades, but the median single BM goes from $1.10$ at its best checkpoint to $1.72$ at the final epoch, while the median two-step chain goes from $0.81$ to $1.18$. The chain both reaches better composites and holds them longer, which is the basis of the corresponding claim in \S\ref{sec:demos_market}.

\begin{figure*}[ht!]
    \centering
    \includegraphics[width=0.8\textwidth]{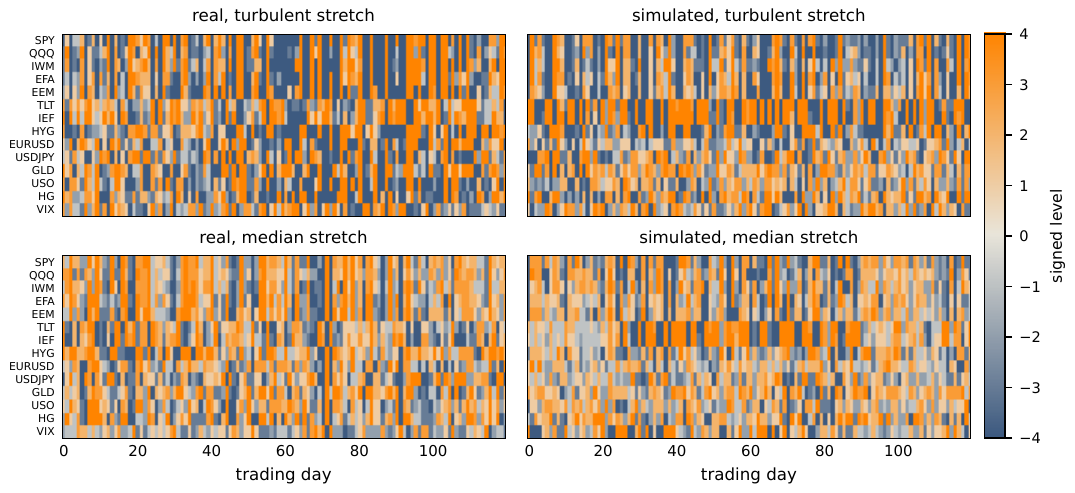}
    \caption{\textbf{Real vs.\ simulated 120-day stretches} of the 14-series panel, colored by signed level. The stretches are selected by a fixed rule applied identically to both sides: all 120-day stretches (of the held-out real windows and of the post-trained model's rollouts) are ranked by average absolute level, and the top-ranked (turbulent, top row) and median (bottom row) stretches are shown.}
    \label{fig:market_ribbons}
\end{figure*}

\paragraph{Qualitative samples.}
Figure~\ref{fig:market_ribbons} shows the data behind these statistics: 120-day stretches of the real held-out panel next to stretches simulated by the post-trained model, drawn as heatmaps of the signed levels. To avoid cherry-picking, the stretches are chosen by a fixed rule applied identically to real and simulated data: all 120-day stretches are ranked by their average absolute level, and the top-ranked (turbulent) one and the median one are shown. The simulated panel reproduces the visible features of the real one: days on which a single deep color spans most of the panel (joint extreme days), sustained turbulent and calm episodes, and Treasury rows (TLT, IEF) that repeatedly run against the equity block at the top.

\end{document}

%% file: figures/eco_local_updates.tex
\begingroup
\newsavebox{\inlineBGlyphBox}
\newsavebox{\inlineDGlyphBox}
\newsavebox{\inlineTGlyphBox}
\newlength{\inlineBGlyphHeight}
\newlength{\inlineDGlyphHeight}
\newlength{\inlineTGlyphHeight}
\newlength{\inlineEtaGlyphWidth}
\newlength{\inlineHalfEtaGlyphWidth}
\newlength{\inlineNGlyphWidth}
\newlength{\inlineHalfNGlyphWidth}
\newlength{\inlineSterileAnnotationShift}
\newlength{\inlineTimesGlyphWidth}
\newlength{\inlineHalfTimesGlyphWidth}
\newlength{\inlineFGlyphWidth}
\newlength{\inlinePeriodGlyphWidth}
\newlength{\inlineDownShift}
\newlength{\inlineFactorNudge}
\newlength{\inlineFactorLeftNudge}
\newlength{\inlineFactorRowReduction}
\newlength{\inlineFactorBottomHoldShift}
\newlength{\inlineSterileFactorXShift}
\newlength{\inlineSterileFactorYShift}
\newlength{\inlineFertileTitleShift}

\sbox{\inlineBGlyphBox}{{\scriptsize b}}
\sbox{\inlineDGlyphBox}{{\scriptsize d}}
\sbox{\inlineTGlyphBox}{{\footnotesize t}}
\setlength{\inlineBGlyphHeight}{\ht\inlineBGlyphBox}
\setlength{\inlineDGlyphHeight}{\ht\inlineDGlyphBox}
\setlength{\inlineTGlyphHeight}{\ht\inlineTGlyphBox}
\settowidth{\inlineEtaGlyphWidth}{{\scriptsize \(\eta\)}}
\setlength{\inlineHalfEtaGlyphWidth}{0.5\inlineEtaGlyphWidth}
\settowidth{\inlineNGlyphWidth}{{\footnotesize n}}
\setlength{\inlineHalfNGlyphWidth}{0.5\inlineNGlyphWidth}
\setlength{\inlineSterileAnnotationShift}{\inlineHalfNGlyphWidth}
\addtolength{\inlineSterileAnnotationShift}{\inlineNGlyphWidth}
\settowidth{\inlineTimesGlyphWidth}{{\scriptsize \(\times\)}}
\setlength{\inlineHalfTimesGlyphWidth}{0.5\inlineTimesGlyphWidth}
\settowidth{\inlineFGlyphWidth}{{\footnotesize f}}
\settowidth{\inlinePeriodGlyphWidth}{{\footnotesize .}}
\setlength{\inlineDownShift}{\inlineDGlyphHeight}
\addtolength{\inlineDownShift}{\inlineHalfTimesGlyphWidth}
\setlength{\inlineFactorNudge}{0.5pt}
\setlength{\inlineFactorLeftNudge}{0.5pt}
\setlength{\inlineFactorRowReduction}{2pt}
\setlength{\inlineFactorBottomHoldShift}{0.86pt}
\setlength{\inlineSterileFactorXShift}{\inlineTimesGlyphWidth}
\addtolength{\inlineSterileFactorXShift}{-\inlineFactorNudge}
\addtolength{\inlineSterileFactorXShift}{-\inlineFactorLeftNudge}
\addtolength{\inlineSterileFactorXShift}{-\inlineHalfTimesGlyphWidth}
\setlength{\inlineSterileFactorYShift}{\inlineDownShift}
\addtolength{\inlineSterileFactorYShift}{\inlineFactorNudge}
\addtolength{\inlineSterileFactorYShift}{\inlineFactorBottomHoldShift}
\settowidth{\inlineFertileTitleShift}{{\footnotesize fer}}

\newsavebox{\localUpdatesLeftColumnBox}
\newsavebox{\localUpdatesTorusBox}
\newlength{\localUpdatesColumnGap}
\newlength{\localUpdatesLeftColumnWidth}
\newlength{\localUpdatesRightColumnWidth}
\newlength{\localUpdatesLeftColumnHeight}
\newlength{\localUpdatesTorusNaturalWidth}
\newlength{\localUpdatesTorusNaturalHeight}
\newlength{\localUpdatesTorusLabelGap}
\newlength{\localUpdatesTorusLabelShift}
\def\localUpdatesTorusScale{1}
\def\localUpdatesTorusInverseScale{1}

\setlength{\localUpdatesTorusLabelGap}{0.8cm}
\setlength{\localUpdatesTorusLabelShift}{\localUpdatesTorusLabelGap}

\definecolor{eastcol}{HTML}{0072B2}
\definecolor{southcol}{HTML}{009E73}
\definecolor{westcol}{HTML}{D62728}
\definecolor{northcol}{HTML}{CC79A7}

\tikzset{
  site/.style={circle, draw, thick, minimum size=1.15cm, inner sep=1pt,
               font=\small, fill=white},
  mig/.style={-{Stealth[length=5pt]}, dashed, thick,
              shorten >=1.5pt, shorten <=1.5pt},
}

\def\S{2.5}
\def\stub{0.8}

\newcommand{\localUpdatesTorusPicture}[1]{%
\begin{tikzpicture}

  \node[
      overlay,
      anchor=east,
      font=\small,
      xshift=-\localUpdatesTorusLabelShift
  ] at (-0.575,-3.75) {\scalebox{#1}{(\thesubfigure)}};

  \foreach \i in {0,1,2,3}{
    \foreach \j in {0,1,2,3}{
      \node[site] (n\i\j) at ({\j*\S},{-\i*\S}) {$n_{(\i,\j)}$};
    }
  }

  \foreach \i in {0,1,2,3}{
    \foreach \c/\col in {0/eastcol, 1/westcol, 2/eastcol}{
      \pgfmathtruncatemacro{\d}{\c+1}
      \draw[mig,\col] (n\i\c) to[bend left=12] (n\i\d);
      \draw[mig,\col] (n\i\d) to[bend left=12] (n\i\c);
    }
  }

  \foreach \j in {0,1,2,3}{
    \foreach \r/\col in {0/southcol, 1/northcol, 2/southcol}{
      \pgfmathtruncatemacro{\d}{\r+1}
      \draw[mig,\col] (n\r\j) to[bend left=12] (n\d\j);
      \draw[mig,\col] (n\d\j) to[bend left=12] (n\r\j);
    }
  }

  \foreach \i in {0,1,2,3}{
    \draw[mig,westcol] ($(n\i3.east)+(0cm,4pt)$)     -- ($(n\i3.east)+(\stub cm,4pt)$);
    \draw[mig,westcol] ($(n\i3.east)+(\stub cm,-4pt)$) -- ($(n\i3.east)+(0cm,-4pt)$);
    \draw[mig,westcol] ($(n\i0.west)+(-\stub cm,4pt)$) -- ($(n\i0.west)+(0cm,4pt)$);
    \draw[mig,westcol] ($(n\i0.west)+(0cm,-4pt)$)     -- ($(n\i0.west)+(-\stub cm,-4pt)$);
  }

  \foreach \j in {0,1,2,3}{
    \draw[mig,northcol] ($(n0\j.north)+(4pt,0cm)$)     -- ($(n0\j.north)+(4pt,\stub cm)$);
    \draw[mig,northcol] ($(n0\j.north)+(-4pt,\stub cm)$) -- ($(n0\j.north)+(-4pt,0cm)$);
    \draw[mig,northcol] ($(n3\j.south)+(4pt,-\stub cm)$) -- ($(n3\j.south)+(4pt,0cm)$);
    \draw[mig,northcol] ($(n3\j.south)+(-4pt,0cm)$)     -- ($(n3\j.south)+(-4pt,-\stub cm)$);
  }

\end{tikzpicture}%
}

\begin{figure*}[t]
    \centering

    \setlength{\localUpdatesColumnGap}{0.03\linewidth}
    \setlength{\localUpdatesLeftColumnWidth}{0.55\linewidth}
    \setlength{\localUpdatesRightColumnWidth}{\linewidth}
    \addtolength{\localUpdatesRightColumnWidth}{-\localUpdatesLeftColumnWidth}
    \addtolength{\localUpdatesRightColumnWidth}{-\localUpdatesColumnGap}

    \sbox{\localUpdatesLeftColumnBox}{%
    \begin{minipage}[c]{\localUpdatesLeftColumnWidth}
        \centering
        \begin{subfigure}[b]{\linewidth}
            \centering
            \phantomsubcaption
            \label{fig:local-reactions}

            \begin{tikzpicture}[
                >={Stealth[length=2.2mm]},
                site/.style={
                    circle,
                    draw,
                    minimum size=11mm,
                    inner sep=1pt,
                    font=\small
                },
                lbl/.style={font=\footnotesize},
                inloop/.style={font=\scriptsize},
                birth/.style={->,thick,black},
                predation/.style={->,thick,black},
                death/.style={->,thick,black}
            ]

            \node[
                overlay,
                anchor=east,
                font=\small
            ] at (-3.0,0) {(\thesubfigure)};

            \node[site] (c) at (0,0) {\(r,f\)};

            \draw[birth] (c) to[out=160,in=110,looseness=9] (c);
            \node[lbl,align=center] at (-2.25,1.15)
                {rabbit birth\\\(r\!\to\!r\!+\!1\)};
            \node[inloop,yshift=\inlineBGlyphHeight]
                at (-0.78,0.63)
                {\(b_r r\)};

            \draw[predation] (c) to[out=-160,in=-110,looseness=9] (c);
            \node[
                lbl,
                align=center,
                xshift=-\inlineSterileAnnotationShift
            ] at (-2.20,-1.32)
                {sterile predation\\
                 \((r,f)\!\to\!(r\!-\!1,f)\)};

            \node[
                inloop,
                align=center,
                xshift=\inlineSterileFactorXShift,
                yshift=-\inlineSterileFactorYShift
            ] at (-0.83,-0.43)
                {\scalebox{0.82}{%
                    \(\begin{gathered}
                        d_p\times\\[-\inlineFactorRowReduction]
                        (1-\eta)
                    \end{gathered}\)
                }};

            \draw[death] (c) to[out=20,in=70,looseness=9] (c);
            \node[lbl,align=center] at (2.0,1.15)
                {fox death\\
                 \hspace*{\dimexpr 2\inlinePeriodGlyphWidth\relax}%
                 \(f\!\to\!f\!-\!1\)};
            \node[inloop,yshift=\inlineBGlyphHeight]
                at (0.78,0.63)
                {\(m_f f\)};

            \draw[predation] (c) to[out=-20,in=-70,looseness=9] (c);
            \node[
                lbl,
                align=center,
                xshift=\inlineFGlyphWidth
            ] at (2.0,-1.32)
                {\hspace*{\dimexpr 2\inlineFertileTitleShift\relax}%
                 fertile predation\\
                 \(\quad\,\,(r,f)\!\to\!(r\!-\!1,f\!+\!1)\)};

            \node[
                inloop,
                xshift=-\inlineHalfEtaGlyphWidth,
                yshift=-\inlineDownShift
            ] at (0.86,-0.43)
                {\(\eta\,d_p\)};

            \end{tikzpicture}
        \end{subfigure}

        \par\medskip

        \begin{subfigure}[b]{\linewidth}
            \centering
            \phantomsubcaption
            \label{fig:local-migration}

            \begin{tikzpicture}[
                >={Stealth[length=2.2mm]},
                site/.style={
                    circle,
                    draw,
                    minimum size=11mm,
                    inner sep=1pt,
                    font=\small
                },
                lbl/.style={font=\footnotesize},
                ndiff/.style={->,thick,black,dashed}
            ]

            \node[
                overlay,
                anchor=east,
                font=\small
            ] at (-2.25,0) {(\thesubfigure)};

            \node[site] (ni) at (-1.4,0) {\(n_i\)};
            \node[site] (nj) at (1.4,0) {\(n_j\)};

            \draw[ndiff]
                (ni)
                to[bend left=18]
                node[lbl,above,yshift=2pt,align=center]
                {\phantom{migration}%
                 \llap{\raisebox{\inlineTGlyphHeight}[0pt][0pt]{migration}}\\
                 \(n_{i\to j}\sim\mathrm{Bin}(n_i,p)\)}
                (nj);

            \draw[ndiff]
                (nj)
                to[bend left=18]
                node[lbl,below,yshift=-1pt,align=center]
                {\(n_{j\to i}\sim\mathrm{Bin}(n_j,p)\)}
                (ni);

            \end{tikzpicture}
        \end{subfigure}
    \end{minipage}%
    }
    \setlength{\localUpdatesLeftColumnHeight}{\ht\localUpdatesLeftColumnBox}
    \addtolength{\localUpdatesLeftColumnHeight}{\dp\localUpdatesLeftColumnBox}

    \usebox{\localUpdatesLeftColumnBox}%
    \hspace{\localUpdatesColumnGap}%
    \begin{minipage}[c]{\localUpdatesRightColumnWidth}
        \centering
        \begin{subfigure}[b]{\linewidth}
            \centering
            \phantomsubcaption
            \label{fig:local-torus}

            \sbox{\localUpdatesTorusBox}{\localUpdatesTorusPicture{1}}
            \setlength{\localUpdatesTorusNaturalWidth}{\wd\localUpdatesTorusBox}
            \setlength{\localUpdatesTorusNaturalHeight}{\ht\localUpdatesTorusBox}
            \addtolength{\localUpdatesTorusNaturalHeight}{\dp\localUpdatesTorusBox}
            \pgfmathsetmacro{\localUpdatesTorusScale}{%
                min(\localUpdatesLeftColumnHeight/\localUpdatesTorusNaturalHeight,%
                    \localUpdatesRightColumnWidth/\localUpdatesTorusNaturalWidth)}
            \pgfmathsetmacro{\localUpdatesTorusInverseScale}{1/\localUpdatesTorusScale}
            \pgfmathsetlengthmacro{\localUpdatesTorusLabelShift}{%
                \localUpdatesTorusLabelGap*\localUpdatesTorusInverseScale}
            \sbox{\localUpdatesTorusBox}{\localUpdatesTorusPicture{\localUpdatesTorusInverseScale}}
            \scalebox{\localUpdatesTorusScale}{\usebox{\localUpdatesTorusBox}}
        \end{subfigure}
    \end{minipage}

\caption{Local probabilistic updates in the predator-prey ecosystem of Subsection~\ref{subsec:Eco}. Two species, rabbits and foxes, are initially concentrated in the center of a two-dimensional grid. The ecosystem evolves as the species interact and diffuse across the environment through local stochastic updates. (\subref{fig:local-reactions}) The on-site reaction process~\eqref{eq:Reaction} takes place at each site
  once per time-step, updating the occupancy $(r,f)$. (\subref{fig:local-migration}) The two-site migration process~\eqref{eq:Migration} is executed several times per site per time-step. (\subref{fig:local-torus}) In this scheme the two-site migrations proceed for each species independently in parallel across each of four dimer tilings,
  \textcolor{northcol}{N}, \textcolor{eastcol}{E}, \textcolor{southcol}{S}, \textcolor{westcol}{W}. The animals carry out the whole process depicted in (\subref{fig:local-torus}) twice per time-step.}
    \label{fig:local-updates}
\end{figure*}

\endgroup